\documentclass[useAMS,usenatbib]{mn2e}

\usepackage{natbib,amssymb,color}
\usepackage{times}

\newcommand{\changeb}[1]{#1}

\newcommand{\reva}[1]{#1}

\usepackage{hyperref}
\usepackage{url}
\usepackage{breakurl,sidecap}

\usepackage{amsmath,graphics,graphicx}

\usepackage{epsfig}
\graphicspath{{figures/}}

\newcommand{\sclustera}{J0000-5748}
\newcommand{\sclusterb}{J0102-4915}
\newcommand{\sclusterc}{J0533-5005}
\newcommand{\sclusterd}{J0546-5345}
\newcommand{\sclustere}{J0559-5249}
\newcommand{\sclusterf}{J0615-5746}
\newcommand{\sclusterg}{J2040-5726}
\newcommand{\sclusterh}{J2106-5844}
\newcommand{\sclusteri}{J2331-5051}

\newcommand{\sclusterdemo}{J2337-5942}
\newcommand{\sclusterk}{J2341-5119}
\newcommand{\sclusterl}{J2342-5411}
\newcommand{\sclusterm}{J2359-5009}

\newcommand{\sfclustera}{$J$0000$-$5748}
\newcommand{\sfclusterb}{$J$0102$-$4915}
\newcommand{\sfclusterc}{$J$0533$-$5005}
\newcommand{\sfclusterd}{$J$0546$-$5345}
\newcommand{\sfclustere}{$J$0559$-$5249}
\newcommand{\sfclusterf}{$J$0615$-$5746}
\newcommand{\sfclusterg}{$J$2040$-$5725}
\newcommand{\sfclusterh}{$J$2106$-$5844}
\newcommand{\sfclusteri}{$J$2331$-$5051}
\newcommand{\sfclusterj}{$J$2337$-$5942}
\newcommand{\sfclusterdemo}{$J$2337$-$5942}
\newcommand{\sfclusterk}{$J$2341$-$5119}
\newcommand{\sfclusterl}{$J$2342$-$5411}
\newcommand{\sfclusterm}{$J$2359$-$5009}

\newcommand{\clustera}{SPT-CL\thinspace\sfclustera}
\newcommand{\clusterb}{SPT-CL\thinspace\sfclusterb}
\newcommand{\clusterc}{SPT-CL\thinspace\sfclusterc}
\newcommand{\clusterd}{SPT-CL\thinspace\sfclusterd}
\newcommand{\clustere}{SPT-CL\thinspace\sfclustere}
\newcommand{\clusterf}{SPT-CL\thinspace\sfclusterf}
\newcommand{\clusterg}{SPT-CL\thinspace\sfclusterg}
\newcommand{\clusterh}{SPT-CL\thinspace\sfclusterh}
\newcommand{\clusteri}{SPT-CL\thinspace\sfclusteri}
\newcommand{\clusterdemo}{SPT-CL\thinspace\sfclusterdemo}
\newcommand{\clusterj}{SPT-CL\thinspace\sfclusterj}
\newcommand{\clusterk}{SPT-CL\thinspace\sfclusterk}
\newcommand{\clusterl}{SPT-CL\thinspace\sfclusterl}
\newcommand{\clusterm}{SPT-CL\thinspace\sfclusterm}

\newcommand{\ltsima}{$\; \buildrel < \over \sim \;$}
\newcommand{\ltsim}{\lower.5ex\hbox{\ltsima}}

\newcommand{\be}{\begin{equation}}
\newcommand{\ee}{\end{equation}}
\newcommand{\bea}{\begin{eqnarray}}
\newcommand{\eea}{\end{eqnarray}}

\usepackage{lmodern}
\DeclareMathAlphabet{\mathbfsf}{\encodingdefault}{\sfdefault}{bx}{sl}

\renewcommand{\vec}[1]{\boldsymbol{#1}}

\newcommand{\altaffilmark}[1]{$^{#1}$}

\def\Bonn{1}
\def\StanfordKIPAC{2}
\def\StanfordPhysics{3}
\def\KICPChicago{4}
\def\Munich{5}
\def\ExcellenceCluster{6}
\def\Leiden{7}
\def\ANL{8}
\def\UFlorida{9}
\def\DARK{10}
\def\StonyBrook{11}
\def\MIT{12}
\def\Washington{13}
\def\SLAC{14}
\def\Harvard{15}
\def\CfA{16}
\def\Colby{17}
\def\FNAL{18}
\def\AAUChicago{19}
\def\PhysicsUChicago{20}
\def\ASIAA{21}
\def\Hyderabad{22}
\def\UCStCruz{23}
\def\McGill{24}
\def\Berkeley{25}
\def\Durham{26}
\def\MPE{27}
\def\Melbourne{28}
\def\CTIO{29}

\defcitealias{schrabback2010}{S10}
\defcitealias{massey2014}{M14}
\defcitealias{skelton14}{S14}
\defcitealias{becker11}{BK11}

\begin{document}

\title[Weak Lensing Analysis of 13 Distant SPT Clusters]{Cluster Mass Calibration at High
  Redshift: HST Weak Lensing Analysis of 13 Distant Galaxy Clusters from the South
  Pole Telescope Sunyaev-Zel'dovich Survey}

\author[T.~Schrabback et al.]
{T.~Schrabback\altaffilmark{\Bonn,\StanfordKIPAC,\StanfordPhysics}\thanks{E-mail: schrabba@astro.uni-bonn.de},
D.~Applegate\altaffilmark{\Bonn,\KICPChicago},
J.~P.~Dietrich\altaffilmark{\Munich,\ExcellenceCluster},
H.~Hoekstra\altaffilmark{\Leiden},
S.~Bocquet\altaffilmark{\KICPChicago,\ANL,\Munich,\ExcellenceCluster},
\newauthor
A.~H.~Gonzalez\altaffilmark{\UFlorida},
A.~von der Linden\altaffilmark{\StanfordKIPAC,\StanfordPhysics,\DARK,\StonyBrook},
M.~McDonald\altaffilmark{\MIT},
C.~B.~Morrison\altaffilmark{\Bonn,\Washington},
\newauthor
S.~F.~Raihan\altaffilmark{\Bonn},
S.~W.~Allen\altaffilmark{\StanfordKIPAC,\StanfordPhysics,\SLAC},
M.~Bayliss\altaffilmark{\Harvard,\CfA,\Colby},
 B.~A.~Benson\altaffilmark{\FNAL,\AAUChicago,\KICPChicago},
\newauthor
L.~E.~Bleem\altaffilmark{\KICPChicago,\PhysicsUChicago,\ANL},
I.~Chiu\altaffilmark{\Munich,\ExcellenceCluster,\ASIAA},
S.~Desai\altaffilmark{\Munich,\ExcellenceCluster,\Hyderabad},
R.~J.~Foley\altaffilmark{\UCStCruz},
T.~de~Haan\altaffilmark{\McGill,\Berkeley},
\newauthor
 F.~W.~High\altaffilmark{\KICPChicago,\AAUChicago},
S.~Hilbert\altaffilmark{\Munich,\ExcellenceCluster},
A.~B.~Mantz\altaffilmark{\StanfordKIPAC,\StanfordPhysics},
R.~Massey\altaffilmark{\Durham},
J.~Mohr\altaffilmark{\Munich,\ExcellenceCluster,\MPE},
\newauthor
C.~L.~Reichardt\altaffilmark{\Melbourne},
A.~Saro\altaffilmark{\Munich,\ExcellenceCluster},
P.~Simon\altaffilmark{\Bonn},
C.~Stern\altaffilmark{\Munich,\ExcellenceCluster},
 C.~W.~Stubbs\altaffilmark{\Harvard,\CfA},
A.~Zenteno\altaffilmark{\CTIO}
\\
\vspace{0.4cm}\\
\parbox{\textwidth}{\large Author affiliations are listed at the end of this paper.\\}
}

\maketitle

\begin{abstract}
We
present an HST/ACS weak
gravitational
lensing
 analysis of 13
massive high-redshift (\mbox{$z_\mathrm{median}=0.88$}) galaxy clusters
discovered in the South Pole Telescope (SPT) Sunyaev-Zel'dovich Survey.
This study is part of a larger campaign that aims to robustly calibrate mass-observable scaling relations over a  wide range in redshift
 to enable
improved cosmological constraints from the SPT cluster sample.
We introduce \changeb{new strategies} to ensure that
 systematics in the lensing analysis  do not degrade constraints
on cluster scaling relations significantly.
First, we efficiently remove cluster members
from the source sample
by selecting very blue galaxies in  \mbox{$V-I$}  colour.
Our estimate of the source redshift distribution is based on  CANDELS data,
where we carefully
 mimic
the
source
selection criteria
of the cluster fields.
We apply a  statistical correction  for systematic photometric redshift
errors as derived from {\it Hubble} Ultra Deep Field data and verified through
 spatial
 cross-correlations.
We
account for the impact of  lensing
magnification on the source redshift distribution,
finding that this is particularly relevant for shallower surveys.
Finally, we account for
 biases in the mass modelling caused  by miscentring and uncertainties in
the \reva{concentration--mass} relation using simulations.
\changeb{In combination with  temperature estimates from  \textit{Chandra}
we constrain  the normalisation of the
mass--temperature
scaling relation
$\ln\left(E(z) M_\mathrm{500c}/10^{14}\mathrm{M}_\odot \right)=A+1.5\ln\left(kT/7.2\mathrm{keV}\right)$
to \mbox{$A=1.81^{+0.24}_{-0.14}(\mathrm{stat.}){\pm}0.09(\mathrm{sys.})$},
consistent with self-similar
 redshift evolution when compared to
 lower redshift samples.}
Additionally, the lensing data
constrain the average concentration of the clusters to
\mbox{$c_\mathrm{200c}=5.6^{+3.7}_{-1.8}$}.

\end{abstract}


\defcitealias{vanderlinde10}{V10}
\defcitealias{reichardt13}{R13}
\defcitealias{williamson11}{W11}
\defcitealias{bleem15}{B15}

\section{Introduction}
\label{sec:intro}

Constraints on
the number density of clusters as a function of their mass and redshift
 probe the growth of structure in
the Universe, therefore  holding great promise to
constrain cosmological models
\citep[e.g.][]{haiman01,allen11,weinberg13}.
Previous studies using samples of at most a few
hundred clusters
have delivered some of the tightest cosmological
constraints currently available on
  dark energy properties, theories of modified gravity, and the
  species-summed neutrino mass
(e.g.\thinspace\citealt{vikhlinin09}; \citealt{rapetti09,rapetti13}; \citealt{schmidt09}; \citealt{mantz10,mantz15}; \citealt{bocquet15a}; \citealt{dehaan16}).
Recently, CMB experiments have begun to substantially increase the number of massive, high-redshift clusters found with well-characterised selection functions, detected via their Sunyaev-Zel'dovich
\citep[SZ,][]{sunyaev70,sunyaev72} signature from inverse Compton scattering
off the electrons in the hot cluster plasma
\citep{hasselfield13,bleem15,planck15sz}.
Upcoming experiments such as SPT-3G \citep{benson14} and
eROSITA \citep{merloni12} are expected to soon provide samples of
$10^4$--$10^5$ massive clusters with well-characterised selection functions, yielding a statistical constraining power
that may mark the transition between ``Stage III'' and ``Stage IV'' dark energy constraints \citep[see][]{albrecht06}
from clusters if systematic uncertainties are well controlled.

Cluster observables such as  X-ray
luminosity,
SZ signal,
or  optical/NIR richness and luminosity
have been shown to scale with mass
\citep[e.g.][]{reiprich02,lin04a,andersson11}.
In order to adequately exploit the statistical constraining power of
large cluster surveys,
an accurate and precise calibration of the scaling relations
between such mass proxies and mass is needed.
Already for current surveys  cosmological constraints are primarily limited
by uncertainties in the calibration of  mass--observable
scaling relations
  \citep[e.g.][]{rozo10,sehgal11,benson13,vonderlinden14b,mantz15,planck15szconstraints}.
It is therefore imperative to improve this calibration empirically.
In this context our work
focuses especially on calibrating mass--observable relations at high
redshifts, which
together with low-redshift measurements, provides constraints on their
redshift evolution.
Particularly for constraints on dark energy properties,
which are primarily derived from the redshift evolution of the cluster mass
function,
 it is critical to ensure that
systematic errors in the evolution of mass--observable scaling relations
do not mimic
the signature of dark energy.
Most previous cosmological cluster studies had to rely on priors for the redshift evolution
derived from    numerical cluster simulations \citep[e.g.][]{vikhlinin09,benson13,dehaan16}.
It is crucial to test the assumed models of cluster astrophysics
in these simulations
by
comparing their predictions to observational
constraints on the scaling
relations
\citep[e.g.][]{lebrun14}, and to
shrink the uncertainties
on the scaling relation parameters.

Progress in the field critically requires improvements in the cluster
mass calibration through large multi-wavelength follow-up campaigns.
For example, high-resolution X-ray observations
provide mass proxies with low intrinsic scatter,
which can be used to constrain
the relative masses of clusters
\citep[e.g.][]{vikhlinin09b,reichert11,andersson11}.
On the other hand, weak gravitational lensing  has been recognised as the
most direct technique for the absolute calibration of the normalisation
of cluster mass observable relations
 \citep[][]{allen11,hoekstra13,applegate14,mantz15}.
The main observable is the weak lensing reduced shear, a tangential distortion  caused by the projected tidal gravitational field of the foreground mass distribution.
It is directly related to the differential projected cluster mass distribution, and can be estimated from the
observed shapes of background galaxies
\citep[e.g.][]{bartelmann01,schneider06}.

To date, the majority of cluster weak lensing mass
estimates  have been
obtained for
 lower redshift clusters (\mbox{$z\lesssim 0.6$--$0.7$})
using ground-based observations
\citep[e.g.][]{high12,israel12,oguri12,applegate14,gruen14,umetsu14,hoekstra15,ford15,kettula15,battaglia16,lieu16,vanuitert16,okabe16,simet17,melchior17}.
To constrain the evolution of cluster mass-observable scaling relations,
these measurements need to be complimented with constraints for higher
redshift clusters.
Here, ground-based measurements suffer from low densities of sufficiently
resolved background galaxies with robust shape measurements.
This can be overcome using high-resolution {\it Hubble Space Telescope}
(HST) images,
where so far \citet{jee11} present the only weak lensing constraints for the
cluster mass calibration of a large sample of
massive  high-redshift (\mbox{$0.83\le z \le 1.46$}) clusters,
which were drawn from optically, NIR, and X-ray-selected samples.
Interestingly, their results
suggest a possible evolution in the $M_\mathrm{2500c}-T_\mathrm{X}$ scaling relation
in comparison to self-similar extrapolations from low redshifts, with lower masses at the \mbox{$20-30\%$} level.
HST weak lensing measurements have also been used to constrain
mass-observable scaling relations for lower \citep{leauthaud10} and intermediate mass
 clusters \citep{hoekstra11b}.

This paper is part of a larger effort to obtain improved observational constraints on the calibration of cluster masses as function of redshift.
Here we analyse new HST observations of 13 massive high-$z$ clusters
detected by the  South Pole Telescope \citep{carlstrom11} via the
SZ effect.
This constitutes the first high-$z$  sample of clusters with HST weak
lensing observations which were drawn from a single, well-characterised survey
selection function.
As a major part of this paper,
we carefully investigate and account for the
relevant
 sources of systematic
uncertainty in the weak lensing mass analysis,
and
discuss their relevance for future studies of larger samples.

The primary technical challenges for weak lensing studies are accurate
measurements of galaxy shapes from noisy data in the presence of instrumental distortions, and the need for an accurate knowledge of the source redshift
distribution which enters through the geometric lensing efficiency.
Within the weak lensing community
substantial progress has been made on the former issue through the development of
improved
shape measurement algorithms
tested
 using image simulations
\citep[e.g.][]{miller13,hoekstra15,bernstein16,fenechconti17}.
For the latter issue, previous studies have typically estimated the redshift
distribution from photometric redshifts (photo-$z$s)
given the incompleteness of spectroscopic redshift samples  (spec-$z$s) at the relevant
magnitudes,  requiring that the photo-$z$-based estimates are
sufficiently accurate.
If sufficient wavelength coverage is available,
photo-$z$s can be estimated directly for the weak lensing survey fields of
interest
\citep[used in the cluster context e.g. by][]{leauthaud10,applegate14,ford15}.
Otherwise, photo-$z$s can be used from external reference deep fields,
requiring that statistically consistent and sufficiently representative galaxy populations are selected
in both the survey and reference fields.
For cluster weak lensing studies both approaches are complicated
by the fact that the presence of a cluster means that the corresponding line-of-sight is over-dense at the cluster redshift, while both the default priors of
photo-$z$ codes and the reference deep fields ought to be representative for
the cosmic mean distribution.
Previous studies employing reference fields have typically dealt with this issue
 by applying colour selections (``colour cuts'') that remove galaxies
at the cluster redshift \citep[e.g.][]{high12,hoekstra12,okabe16}.
In case of  incomplete removal the approach
can  be complemented by a statistical
correction for the residual  cluster member contamination if that can be
estimated sufficiently well \citep[e.g.][]{hoekstra15}.
For cluster weak lensing studies a further complication arises when
parametric models are fitted to the measured tangential reduced shear
profiles, as issues such as miscentring \citep[e.g.][]{johnston07,george12} or uncertainties regarding
assumed cluster concentrations can lead to non-negligible biases,
introducing the need
for calibrations using simulations \citep[e.g.][]{becker11}.

This paper is organised as follows:
 Sect.\thinspace\ref{sec:sheartheory} summarises relevant aspects of weak lensing theory.
This is followed by a description of our cluster sample in
Sect.\thinspace\ref{sec:clusters} and a description of the analysed data and
image processing in Sect.\thinspace\ref{sec:data}.
Sect.\thinspace\ref{sec:shear} details on the weak lensing shape measurements and a new test for signatures of potential residuals of charge-transfer inefficiency in the weak lensing catalogues.
In Sect.\thinspace\ref{sec:phot} we describe in detail our approach to
remove cluster galaxies via colour cuts
 and reliably estimate the source redshift distribution using
 data from the CANDELS fields.
In Sect.\thinspace\ref{sec:wlmasses}
 we present our weak lensing shear profile analysis, mass reconstructions, and mass estimates,
which we use in Sect.\thinspace\ref{se:mtx} to constrain the mass--temperature scaling relation.
Finally, we
discuss our
\changeb{findings}
in Sect.\thinspace\ref{sec:discussion}
and conclude in Sect.\thinspace\ref{sec:conclusions}.

Throughout this paper we assume a standard flat $\Lambda$CDM
cosmology characterised by $\Omega_\mathrm{m}=0.3$,  $\Omega_\Lambda=0.7$, and
$H_0=70h_{70}$ km/s/Mpc with $h_{70}=1$,
as approximately consistent with recent CMB
constraints \citep[][]{hinshaw13,planck15cosmo}.
For the computation of large-scale structure noise on the weak lensing
estimates
\reva{and the concentration--mass relation according to \cite{diemer15}}
we furthermore assume \mbox{$\sigma_8=0.8$}, \mbox{$\Omega_\mathrm{b}=0.046$}, and \mbox{$n_\mathrm{s}=0.96$}.
All magnitudes are in the AB system and are corrected for extinction according to \citet{schlegel98}.

\section{Summary of Relevant Weak Lensing Theory}
\label{sec:sheartheory}
The images of distant background galaxies are distorted by the tidal
gravitational field of a foreground mass concentration,
see e.g. the reviews by \citet{bartelmann01,schneider06}, as well as \citet{hoekstra13} in the context of galaxy clusters.
In the weak lensing regime the size of a source is much smaller
than the characteristic scale on which variations in the tidal field occur.
In this case the lens mapping as function of observed position $\vec{\theta}$
can be described using the reduced shear $g(\vec{\theta})$
and
the convergence $\kappa(\vec{\theta})=\Sigma(\vec{\theta})/\Sigma_\mathrm{crit}$, which is the
ratio of the
surface mass density
$\Sigma(\vec{\theta})$
and the
critical surface mass
density
\begin{equation}
  \label{eqn:sigmacrit}
  \Sigma_{\mathrm{crit}} = \frac{c^2}{4\pi G}\frac{1}{D_{\mathrm{l}} \beta },
\end{equation}
with the speed of light $c$, the gravitational constant $G$, and
the geometric lensing efficiency
\begin{equation}
\beta=\mathrm{max}\left[0,\frac{D_\mathrm{ls}}{D_\mathrm{s}}\right] \, ,
\end{equation}
where $D_\mathrm{s}$, $D_\mathrm{l}$, and $D_\mathrm{ls}$ indicate the angular diameter distances to the source, to the lens, and between lens and source, respectively.
The reduced shear
\begin{equation}
g(\vec{\theta})=\frac{\gamma(\vec{\theta})}{1-\kappa(\vec{\theta})}
\end{equation}
describes the observable anisotropic shape distortion due to weak lensing. It is a two component
quantity, conveniently written as a complex number
\begin{equation}
g=g_1+\mathrm{i}g_2=|g|\mathrm{e}^{2\mathrm{i}\varphi} \,,
\end{equation}
where $|g|$ constitutes the strength of the distortion and $\varphi$ its
orientation with respect to the coordinate system.
The reduced  shear $g(\vec{\theta})$ is a rescaled version of the unobservable shear
$\gamma(\vec{\theta})$, and
can be estimated from the ensemble-averaged PSF-corrected ellipticities \mbox{$\epsilon=\epsilon_1+\mathrm{i}\epsilon_2$} of
background galaxies (see Sect.\thinspace\ref{sec:shear}), with the expectation value
\begin{equation}
\langle\epsilon\rangle = g \, .
\end{equation}
Due to noise from the intrinsic galaxy shape distribution and measurement noise we need to
average the ellipticities of a large ensemble of galaxies
\begin{equation}
\langle \epsilon_\alpha \rangle = \frac{\sum  \epsilon_{\alpha,i} w_i}{\sum w_i}
\end{equation}
to obtain useful
constraints, where $\alpha\in\{1,2\}$ indicates the two ellipticity components
and $i$ indicates galaxy $i$. The shape weights \mbox{$w_i=1/\sigma_{\epsilon,i}^2$}
are included to improve the measurement signal-to-noise ratio, where $\sigma_{\epsilon,i}$ contains
contributions both from the measurement noise and the intrinsic shape distribution (see Appendix  \ref{app:shapes_candels}, where we constrain both contributions empirically using CANDELS data).

It is often useful to decompose the shear, reduced shear, and the ellipticity into
their tangential components, e.g. $g_\mathrm{t}$, and  cross components, e.g. $g_\times$, with respect to
the centre of a mass distribution as
\begin{eqnarray}
g_\mathrm{t} & = & - g_1 \cos{2 \phi} - g_2 \sin{2\phi}\label{eq:gt}
\\
g_\times & = & + g_1 \sin{2\phi} - g_2 \cos{2 \phi}  \,,
\label{eq:gx}
\end{eqnarray}
where $\phi$ is the azimuthal angle with respect to the centre.
The azimuthal average
of the  tangential  shear $\gamma_\mathrm{t}$
at a radius $r$ around the centre of the mass distribution is linked to the mean convergence $\bar{\kappa}(<r)$ inside $r$ and  $\bar{\kappa}(r)$ at  $r$ via
\begin{equation}
\langle \gamma_\mathrm{t} \rangle (r) = \bar{\kappa}(<r) - \bar{\kappa}(r)\, .
\end{equation}
The weak lensing convergence and shear scale for an
individual source galaxy at redshift \mbox{$z_i$} with the geometric lensing
efficiency $\beta(z_i)$, which is often conveniently
written as
\begin{equation}
\gamma=\beta_s(z_i)\gamma_\infty \, , \, \kappa=\beta_s(z_i)\kappa_\infty \,,
\end{equation}
where  $\kappa_\infty$ and $\gamma_\infty$ correspond to the values for a
source at infinite redshift, and \mbox{$\beta_s(z_i)=\beta(z_i)/\beta_\infty$}.
In practise, we average the ellipticities of an ensemble of galaxies
distributed in redshift, providing an estimate for
\begin{equation}
\langle g \rangle=\left\langle
  \frac{\beta_s(z_i)\gamma_\infty}{1-\beta_s(z_i)\kappa_\infty} \right\rangle \,.
\end{equation}
While one could in principle compute the exact model prediction for this from the
 source redshift distribution weighted by the lensing weights,
a sufficiently accurate approximation is provided in \citet{hoekstra00}:
\begin{equation}
g^\mathrm{model}\simeq \left[ 1+ \left(
      \frac{\langle\beta_s^2\rangle}{\langle\beta_s\rangle^2} -1
    \right)\langle\beta_s\rangle\kappa_\infty^\mathrm{model} \right]
  \frac{\langle\beta_s\rangle\gamma_\infty^\mathrm{model}}{1-\langle\beta_s\rangle\kappa_\infty^\mathrm{model}}
 \label{eq:gbeta2corrected}
\end{equation}
 \citep[see also][]{seitz97,applegate14}, where
\begin{equation}
\langle\beta_s\rangle=\frac{\sum  \beta_s(z_i) w_i}{\sum w_i} \, ,\, \langle\beta_s^2\rangle=\frac{\sum  \beta_s^2(z_i) w_i}{\sum w_i}\,
\end{equation}
need to be computed from the estimated source redshift distribution, taking the shape weights into account.

When the signal of lenses at different redshifts is compared or stacked,
it can be useful to conduct the analysis in terms of the differential surface mass density
\begin{equation}
\Delta \Sigma(r)=\frac{\sum_{i} w_i \left(\epsilon_\mathrm{t}\Sigma_\mathrm{crit}\right)_i}{\sum_i w_i}\,
 \label{eq:deltaSigma}
\end{equation}
to compensate for the redshift dependence of the signal,
where the the summation is conducted  over sources in a separation interval around $r$.

Gravitational lensing leaves the surface brightness invariant.
Accordingly, a relative change in the observed flux of a source due to
lensing is solely given by the relative magnification of the source
\begin{equation}
\mu =
\frac{1}{(1-\kappa)^2-|\gamma|^2} \,.
\label{eq:magnification_correct}
\end{equation}
Together with the change in solid angle this
also changes the observed density of background sources and their redshift distribution, as investigated in Sect.\thinspace\ref{sec:magnification_model}.

\section{The Cluster Sample}
\label{sec:clusters}

\begin{figure}
  \includegraphics[width=0.99\columnwidth]{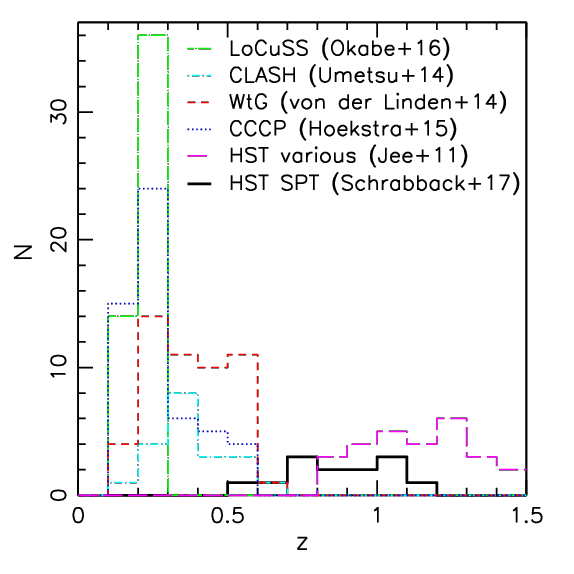}
\caption{Comparison of the cluster redshift distribution of our sample with
  several recent independent studies, plus the larger high-redshift sample
  from   \citet{jee11}, which includes a combination of optically, NIR, and X-ray-selected clusters.
\label{fi:zdist_cluster_surveys}}
\end{figure}

We study a total of 13 distant galaxy clusters detected by the SPT
in the redshift range \mbox{$0.57\le z \le 1.13$} via the
SZ
effect; see
Table \ref{tab:clusters} for details and
Fig.\thinspace\ref{fi:zdist_cluster_surveys} for a comparison of the
cluster redshift distribution to recent large weak lensing cluster samples from
the Canadian Cluster Comparison Project \citep[CCCP;][]{hoekstra15}, Weighing the Giants
\citep[WtG;][]{vonderlinden14}, the Cluster Lensing And Supernova survey with Hubble \citep[CLASH;][]{umetsu14},
the Local Cluster Substructure Survey
\citep[LoCuSS;][]{okabe16},
and the analysis of HST observations of X-ray, optically, and NIR selected
high-redshift clusters by \citet{jee11}.

The SPT clusters were observed in HST Cycles 18 and 19.
At the time of the target selection, the SPT cluster follow-up campaign was
still incomplete.
From the clusters with measured spectroscopic redshifts prior to the corresponding cycle,
we selected the most massive SPT-SZ clusters at \mbox{$0.6\lesssim z\lesssim 1.0$} for
the  Cycles 18 programme, and the  most massive clusters at \mbox{$z\gtrsim
  0.9$} for the Cycle 19 programme.
Nine clusters in our overall sample originate from the first 178 deg$^2$ of the sky
surveyed by SPT  \citep[][hereafter
\citetalias{vanderlinde10}]{vanderlinde10}.
Using updated estimates of the SZ detection significance  $\xi$
from
the
 cluster catalogue for the full 2,500 deg$^2$ SPT-SZ survey
\citep[][hereafter \citetalias{bleem15}]{bleem15},
our selection of clusters from the \citetalias{vanderlinde10} sample includes all
 clusters from the first 178 deg$^2$ at \mbox{$z\ge 0.57$} with
\mbox{$\xi\ge 8$} plus all
clusters at \mbox{$z\ge 0.70$} with
\mbox{$\xi\ge6.6$} (see Table \ref{tab:clusters}), except for SPT-CL{\thinspace}$J$0540$-$5744
(\mbox{$\xi=6.74$}).
Additionally, our sample includes all clusters at \mbox{$z\ge 0.70$} from
\citet[][henceforth \citetalias{williamson11}]{williamson11}, who present a catalogue of the 26 most significant SZ
cluster detections
in the full 2500 deg$^2$ SPT survey region. This adds three clusters in
addition to SPT-CL{\thinspace}$J$2337$-$5942, which is part of both samples.
Finally, with SPT-CL{\thinspace}$J$2040$-$5725 a single further cluster is included  from
\citet[][hereafter \citetalias{reichardt13}]{reichardt13}, who present the
cluster sample constructed from the first 720 deg$^2$ of the SPT cluster survey.
In addition to the aforementioned sample papers, more detailed studies of
individual clusters were published for
SPT-CL{\thinspace}J0546$-$5345 \citep{brodwin10} and SPT-CL{\thinspace}J2106$-$5844 \citep{foley11}.
Spectroscopic cluster redshift measurements are described in \citet{ruel14} and \cite{bayliss16}.
In Table \ref{tab:clusters} we also list X-ray centroids as estimated from
the available {\it Chandra} or XMM-Newton data \citep[detailed in][see also Sect.\thinspace\ref{se:mtx}]{andersson11,benson13,mcdonald13,chiu16}, and BCG positions from \citet{chiu16}.

\begin{table*}
\caption{The cluster sample.
\label{tab:clusters}}
\begin{center}
\addtolength{\tabcolsep}{-1pt}
\begin{tabular}{clccccccccc}
\hline
\hline
Cluster name & $z_l$ & $\xi$ & \multicolumn{6}{c}{Coordinates centres [deg J2000]} & $M_\mathrm{500c,SZ}$ &Sample\\
& & & SZ $\alpha$ & SZ $\delta$  & X-ray $\alpha$ & X-ray $\delta$ & BCG
$\alpha$ & BCG $\delta$ & [$10^{14} \mathrm{M}_\odot h_{70}^{-1}$] &\\
\hline
SPT-CL{\thinspace}$J$0000$-$5748 & 0.702 & 8.49 & 0.2499 & $-57.8064$ & 0.2518
& $-57.8094$ & 0.2502  &  $-57.8093$ & $4.56\pm0.80$ & \citetalias{vanderlinde10}\\
SPT-CL{\thinspace}$J$0102$-$4915 & 0.870 & 39.91 & 15.7294 & $-49.2611$ & 15.7350 & $-49.2667$ & 15.7407 &   $-49.2720$ & $14.43\pm2.10$ & \citetalias{williamson11}\\
SPT-CL{\thinspace}$J$0533$-$5005 & 0.881 & 7.08 & 83.4009 & $-50.0901$ &
83.4018 & $-50.0969$ & 83.4144  &  $-50.0845$ &$3.79\pm0.73$ &
\citetalias{vanderlinde10}\\
SPT-CL{\thinspace}$J$0546$-$5345 & 1.066 & 10.76 & 86.6525 & $-53.7625$ &
86.6532 & $-53.7604$ &  86.6569  &  $-53.7586$ & $5.05\pm0.82$ & \citetalias{vanderlinde10}\\
SPT-CL{\thinspace}$J$0559$-$5249 & 0.609 & 10.64 & 89.9251 & $-52.8260$ &
89.9357 & $-52.8253$ & 89.9301 & $-52.8241$ & $5.78\pm0.95$ & \citetalias{vanderlinde10}\\
SPT-CL{\thinspace}$J$0615$-$5746 & 0.972 & 26.42 & 93.9650 & $-57.7763$ &
93.9652 & $-57.7788$ & 93.9656  &  $-57.7802$  &$10.53\pm1.55$ & \citetalias{williamson11}\\
SPT-CL{\thinspace}$J$2040$-$5725 & 0.930 & 6.24 & 310.0573 & $-57.4295$ & 310.0631$^*$ & $-57.4287$ & 310.0552 &   $-57.4209$ & $3.36\pm0.70$ & \citetalias{reichardt13}\\
SPT-CL{\thinspace}$J$2106$-$5844 & 1.132 & 22.22 & 316.5206 & $-58.7451$ &
316.5174 & $-58.7426$ & 316.5192  &  $-58.7411$ & $8.35\pm1.24$ & \citetalias{williamson11}\\
SPT-CL{\thinspace}$J$2331$-$5051 & 0.576 & 10.47 & 352.9608 & $-50.8639$ &
352.9610 & $-50.8631$ & 352.9631  &  $-50.8650$ &$5.60\pm0.92$ & \citetalias{vanderlinde10}\\
SPT-CL{\thinspace}$J$2337$-$5942 & 0.775 & 20.35 & 354.3523 & $-59.7049$ &
354.3516 & $-59.7061$ & 354.3650  &  $-59.7013$ & $8.43\pm1.27$ & \citetalias{vanderlinde10}, \citetalias{williamson11}\\
SPT-CL{\thinspace}$J$2341$-$5119 & 1.003 & 12.49 & 355.2991 & $-51.3281$ &
355.3009 & $-51.3285$ & 355.3014 &  $-51.3291$ & $5.59\pm0.89$ & \citetalias{vanderlinde10}\\
SPT-CL{\thinspace}$J$2342$-$5411 & 1.075 & 8.18 & 355.6892 & $-54.1856$ &
355.6904 & $-54.1838$ & 355.6913 &   $-54.1848$   & $3.93\pm0.70$ &
\citetalias{vanderlinde10} \\
SPT-CL{\thinspace}$J$2359$-$5009 & 0.775 & 6.68 & 359.9230 & $-50.1649$ &
359.9321 & $-50.1697$ & 359.9324  &  $-50.1722$ & $3.60\pm0.71$ & \citetalias{vanderlinde10}\\
\hline

\end{tabular}
\addtolength{\tabcolsep}{+1pt}
\end{center}

{\flushleft
Note. --- Basic data from \citet{bleem15} and \citet{chiu16} for the 13 clusters targeted in this
  weak lensing analysis.
{\it Column 1:} Cluster designation.
{\it Column 2:} Spectroscopic cluster redshift.
{\it Column 3:} Peak
signal-to-noise ratio
of the SZ
  detection.
{\it Columns 4--9:} Right ascension $\alpha$ and declination $\delta$ of the
cluster centres used in the weak lensing analysis from the SZ peak, X-ray
centroid, and BCG position. $^*$: X-ray centroid from XMM-Newton data,
otherwise {\it Chandra} (see Sect.\thinspace\ref{se:mtx}).
{\it Column 10:} Mass derived from the SZ-Signal.
{\it Column 11:} SPT parent sample for HST follow-up selection.\\
}
\end{table*}

\section{Data and data reduction}
\label{sec:data}
In this section we provide details on the data analysed in this study and their reduction.
For the SPT clusters we make use of HST observations  (Sect.\thinspace\ref{se:hst_observations}) for shape and colour measurements, as well as VLT observations (Sect.\thinspace\ref{se:data_vlt}) for colour measurements in the outer cluster regions.
To optimise our weak lensing pipeline, and to be able to apply consistent selection criteria to photo-$z$ catalogues from \citet{skelton14},  we also process HST observations of the CANDELS fields (Sect.\thinspace\ref{se:data_hst_candels}).

\subsection{HST/ACS data}
\label{sec:hstdata}

\subsubsection{SPT cluster observations}
\label{se:hst_observations}
We measure weak lensing galaxy shapes from high-resolution {\it Hubble Space
  Telescope} imaging obtained during Cycles 18 and 19 as part of programmes 12246 (PI:
C.~Stubbs) and 12477\footnote{This program also includes observations of
  SPT-CL{\thinspace}J0205$-$5829 (\mbox{$z=1.322$}). However, we do not include it in
the current analysis given its high-redshift, which would require deeper
$z$-band observations for the background selection (see
Sect.\thinspace\ref{sec:phot}) than currently available.} (PI: F.~W.~High),
and observed
between  Sep 29, 2011 and Oct 24, 2012 under
low sky background   conditions.
Each cluster was observed with a $2\times 2$ ACS/WFC mosaic in the
F606W filter,
 where each tile consists of 4 dithered exposures
of 480\thinspace s, adding to a total
exposure time of 1.92\thinspace ks per tile.
These mosaic observations
allow us to probe the cluster weak lensing signal out to
approximately the virial radius.
Additionally, a single tile was observed with ACS in the F814W
filter on the cluster centre (1.92\thinspace ks).
These data are included in our photometric analysis
(Sect.\thinspace\ref{sec:phot}).
For the weak lensing shape measurements we chose observations in the F606W filter as it is the most efficient
ACS filter in terms of weak lensing galaxy source density \citep[see,
e.g.][]{schrabback07}.
However note that our analysis in Appendix \ref{app:606vs814} suggests that future programmes could benefit  from
mosaic observations in both F606W and F814W to simultaneously obtain robust shape measurements and colour estimates.
In fact, a
\mbox{$2\times 2$} F814W ACS mosaic  was obtained
for one of the clusters in our sample, {\clusterf}, through the independent
HST programme
12757 (PI: Mazzotta), with observations conducted Jan 19--22, 2012.
For the current analysis we include these additional data in the colour
measurements but not the shape analysis.

We denote magnitudes measured from the ACS F606W and F814W images as $V_{606}$
and $I_{814}$, respectively. By default these correspond to magnitudes
measured in circular apertures with a diameter 0\farcs7  unless explicitly stated differently.

\subsubsection{HST data reduction}
\label{sec:reduction}

For  basic image reductions we largely employ the standard ACS calibration pipeline
\texttt{CALACS}. The main exception is our use of the \citet[][\citetalias{massey2014} henceforth]{massey2014}
algorithm for the  correction of charge-transfer inefficiency (CTI).
CTI constitutes an
important systematic effect
for HST weak lensing shape analyses
if left uncorrected \citep[e.g.][\citetalias{schrabback2010} henceforth]{rhodes07,schrabback2010}.
It is caused by radiation damage in space.
The resulting CCD defects  act as charge traps during the read-out process, introducing non-linear charge-trails behind objects in the parallel-transfer read-out
direction.
\citetalias{massey2014} updated their time-dependent model of the charge trap densities
by fitting
charge trails behind hot pixels in CANDELS
 ACS/F606W imaging exposures of the COSMOS field \citep{grogin2011}, which were obtained at a similar epoch as our cluster data
(between Dec 06, 2011 and Apr 15, 2012).
Given that we conduct the CTI correction
using the
\citetalias{massey2014} code, we also have to CTI-correct the master dark frames using this pipeline.
As further differences to standard \texttt{CALACS} processing we compute
accurately normalised r.m.s. noise maps as detailed in
\citetalias{schrabback2010} and optimise the bad pixel mask, where we flag
satellite trails and cosmic ray clusters, and unflag the removed CTI trails of hot pixels.

The further data reduction for the individual ACS tiles closely follows \citetalias{schrabback2010}, to which we refer the reader for  details. As the first step, we carefully refine relative shifts and rotations between the exposures
 by matching the positions of compact objects.
We then use \texttt{MultiDrizzle} \citep{koekemoer2003} for the cosmic ray removal and stacking, where we employ the \texttt{lanczos3} kernel at the native pixel scale 0\farcs05 to minimise noise correlations while only introducing a low level of aliasing for ellipticity measurements \citep{jee2007}.
The pipeline also generates correctly scaled r.m.s. noise maps for stacks that are used for the object detection.
We conduct weak lensing shape measurements on these individual stacked ACS tiles (see Sect.\thinspace\ref{sec:shear}).

For the joint photometric analysis with available VLT data (Sect.\thinspace\ref{se:source_sel_scatter} with
details given in Appendix \ref{app:details_fors2color_scatter}) we additionally generate stacks for the \mbox{$2\times 2$} ACS mosaics.
Here we iteratively align neighbouring tiles by first resampling them separately onto a common pixel grid, only stacking the exposures of the corresponding tile.
We then use the differences between the positions of matched objects in the overlapping regions to  compute shifts and rotations, in order to update the astrometry.

\subsubsection{CANDELS HST data}
\label{se:data_hst_candels}
When estimating the redshift distribution of our source sample (see
Sect.\thinspace\ref{sec:phot}) we need to
apply the same selection function (consisting of photometric, shape, and size cuts) to the galaxies in the CANDELS fields,
which act as our reference sample.
To be able to employ consistent weak lensing cuts, we
reduce and analyse
ACS imaging in the CANDELS
fields with the same pipeline as the HST observations of the SPT clusters.
This includes data from the CANDELS \citep[][Proposal IDs 12440, 12064]{grogin2011},
GOODS
\citep[][Proposal IDs 9425, 9583]{giavalisco2004}, GEMS \citep[][Proposal ID
9500]{rix2004}, and AEGIS \citep[][Proposal ID 10134]{davis2007} programmes.
Here we perform a tile-wise analysis, always stacking exposures with good spatial
overlap which add to approximately 1-orbit depth, roughly matching the depth of our
cluster field data (see Appendix \ref{app:candels:data} for additional information).

We use these blank field data also as a calibration sample to derive an
empirical
weak lensing weighting scheme
that is based on the measured ellipticity dispersion as function of logarithmic
signal-to-noise ratio and employed in our cluster lensing analysis (see Appendix \ref{app:shapeweights}).
This analysis also provides updated constraints on the dispersion of the intrinsic galaxy
ellipticities
and allows us to compare the weak lensing performance of the ACS
F606W and F814W filters, aiding the preparation of future weak lensing programmes
(see Appendix \ref{app:606vs814}).

\subsection{VLT/FORS2 data}
\label{se:data_vlt}
For our analysis we make use of  VLT/FORS2 imaging of all of our targets
 taken as part of
programmes
086.A-0741 (PI: Bazin),
088.A-0796 (PI: Bazin),
088.A-0889 (PI: Mohr),
and 089.A-0824 (PI: Mohr)
in the $I_\mathrm{BESS}$ pass-band, which we call $I_\mathrm{FORS2}$.
The FORS2 focal plane is covered with two $2k\times 4k$ MIT CCDs.
The data were taken  with the standard resolution collimator in
$2\times 2$ binning, providing imaging
over a $6\farcm8 \times 6\farcm8$ field-of-view with a pixel scale of
0\farcs25, matching the size of our ACS mosaics well.

\begin{table*}
\caption{The VLT/FORS2 $I_\mathrm{FORS2}$ imaging data.
\label{tab:vltdata}}
\begin{center}
\begin{tabular}{cccccc}
\hline
\hline
Cluster name & $t_\mathrm{exp}$ & $I_\mathrm{lim}$ & IQ &
\multicolumn{2}{c}{Used $V_{606}$ range} \\
& & & & bright cut & faint cut\\
\hline
SPT-CL{\thinspace}$J$0000$-$5748 & 2.1\thinspace ks & 26.0 & 0\farcs65 & 24.0--25.5 & 25.5--26.0 \\

SPT-CL{\thinspace}$J$0102$-$4915 & 2.1\thinspace ks & 25.8 & 0\farcs75 & 24.0--25.0 & 25.0--25.5\\
SPT-CL{\thinspace}$J$0533$-$5005 & 2.1\thinspace ks & 25.8 &  0\farcs73 & 24.0--25.5 & -\\%
SPT-CL{\thinspace}$J$0546$-$5345 & 2.1\thinspace ks & 25.7 & 0\farcs75 & 24.0--25.0 & 25.0--25.5\\
SPT-CL{\thinspace}$J$0559$-$5249 & 1.9\thinspace ks & 25.6 &  0\farcs65 & 24.0--25.0 & 25.0--25.5\\

SPT-CL{\thinspace}$J$0615$-$5746 & 2.5\thinspace ks & 25.6 & 0\farcs93& 24.0--24.5 & 24.5--25.5\\
SPT-CL{\thinspace}$J$2040$-$5725 & 2.9\thinspace ks & 25.7 & 0\farcs70 & 24.0--25.0 & 25.0--25.5\\
SPT-CL{\thinspace}$J$2106$-$5844 & 4.8\thinspace ks & 25.8 & 0\farcs80 &  24.0--25.0 & 25.0--25.5\\%
SPT-CL{\thinspace}$J$2331$-$5051 & 2.4\thinspace ks & 25.9 &  0\farcs83 & 24.0--25.5 & 25.5--26.0\\
SPT-CL{\thinspace}$J$2337$-$5942 & 2.1\thinspace ks & 25.7 & 0\farcs80 & 24.0--25.5 & 25.5--26.0\\
SPT-CL{\thinspace}$J$2341$-$5119 & 2.1\thinspace ks & 25.8 & 0\farcs80 & 24.0--25.5 & 25.5--26.0\\
SPT-CL{\thinspace}$J$2342$-$5411 & 2.1\thinspace ks & 25.7 &  0\farcs93&  24.0--25.0 & 25.0--25.5\\
SPT-CL{\thinspace}$J$2359$-$5009 & 2.1\thinspace ks & 25.9 &  0\farcs68& 24.0--25.5 & 25.5--26.0 \\
\hline
\end{tabular}
\end{center}
{\flushleft
Note. --- Details of the analysed VLT/FORS2 imaging data.
{\it Column 1:} Cluster designation.
{\it Column 2:} Total co-added exposure time.
{\it Column 3:} $5\sigma$-limiting magnitude computed for 1\farcs5 apertures
in the stack from the single pixel noise r.m.s. values of the contributing exposures.
{\it Column 4:} Image Quality defined as $2\times \texttt{FLUX\_RADIUS}$
from \texttt{Source Extractor}.
{\it Column 5:} $V_{606}$ magnitude range with low photometric
colour scatter \mbox{$\sigma_{\Delta(V-I)}<0.2$}, for which the ``bright''
colour cut is applied (see Table \ref{tab:app:colourcuts} in Appendix
\ref{app:details_fors2color_scatter}).
{\it Column 6:} $V_{606}$ magnitude range with increased photometric
colour scatter \mbox{$0.2<\sigma_{\Delta(V-I)}<0.3$}, for which the ``faint''
colour cut is applied (see
Table \ref{tab:app:colourcuts} in Appendix
\ref{app:details_fors2color_scatter}).\\
}
\end{table*}

We reduced the data using \texttt{theli} \citep{erben05,schirmer13}, applying bias and flat-field correction, relative photometric
calibration, and sky background subtraction using \texttt{Source Extractor} \citep{bertin1996}.
We use the object positions in the HST F606W image as astrometric
reference for the distortion correction.
For an initial absolute photometric calibration using the stars located in the central HST $I_{814}$ tile we employ the
relation
\begin{equation}
\label{eq:picklesoffset}
I_\mathrm{FORS2} - I_{814} = -0.052 + 0.0095  (V_{606} - I_{814}) \,,
\end{equation}
which was derived employing the \citet{pickles98} stellar library.
This relation is
valid for \mbox{$V_{606} - I_{814} < 1.7$} and assumes total magnitudes for
the computation of \mbox{$I_\mathrm{FORS2} - I_{814}$}.
We list  total exposure times, limiting magnitudes, and delivered image
quality for the co-added images in Table \ref{tab:vltdata}.
For further details on the data reduction see \citet{chiu16}, who also
analyse observations obtained with  FORS2 in the $B_\mathrm{HIGH}$ and
$z_\mathrm{GUNN}$ pass-bands.
In our analysis we do not include these additional bands.
Our initial testing indicates
that
their inclusion would only yield
a minor  increase in the usable background galaxy source density given
the depth of the different observations and typical colours of the dominant background source population.

\section{Weak lensing galaxy shapes}
\label{sec:shear}

\subsection{Shape measurements}
\label{sec:shapemeasure}
For the generation of weak lensing shape catalogues we employ the pipeline
from \citetalias{schrabback2010}, which was successfully used for
cosmological weak lensing measurements that typically have more stringent requirements on the control of systematics than cluster weak lensing studies.
We refer the reader to this publication for a more detailed pipeline
description. Here we summarise the main steps and provide details on recent changes to our pipeline only.
One of the main changes is the application of the pixel-based CTI correction from \citetalias{massey2014} (Sect.\thinspace\ref{sec:reduction}), which is more accurate than the catalogue-level correction employed in \citetalias{schrabback2010}.
This change has become necessary as we analyse more recent ACS data with stronger CTI degradation.

As the first step in the catalogue generation we use \texttt{Source Extractor} \citep{bertin1996} to detect objects in the F606W stacks and measure basic object properties.
For the ellipticity measurement and correction for the point-spread function
(PSF) we employ  the KSB+ formalism \citep{kaiser1995,luppino1997,hoekstra1998} as implemented by \citet{erben2001} with modifications from \citet{schrabback07} and \citetalias{schrabback2010}.
We interpolate the spatially and temporally varying ACS PSF using a model derived from a principal component analysis of PSF variations in dense stellar fields.
\citetalias{schrabback2010} showed that the dominant contribution to ACS PSF ellipticity variations can be described with a single principal component (related to the HST focus position).
This one-parameter PSF model is sufficiently well constrained by the $\sim 10-20$ high-$S/N$ stars available for PSF measurements in extragalactic ACS pointings.
We obtain a PSF model for each contributing exposure based on stellar ellipticity and size measurements in the image prior to resampling (to minimise noise), from which we compute the combined model for the stack.
For the current work we recalibrated this algorithm using archival ACS F606W
stellar field observations taken after Servicing Mission 4.
\changeb{We
processed these data} with the same CTI correction method as our cluster field data.

Following \citetalias{schrabback2010} we select galaxies in terms of  their half-light radius \mbox{$r_\mathrm{h}>1.2  r_\mathrm{h}^{*,\mathrm{max}}$}, where $r_\mathrm{h}^{*,\mathrm{max}}$ is the upper limit of the $0.25$ pixel wide stellar locus,
 and ``pre-seeing'' shear polarisability tensor \mbox{$P^g$} with   \mbox{$\mathrm{Tr}[P^g]/2>0.1$}.
Deviating from \citetalias{schrabback2010} we exclude very extended galaxies with \mbox{$r_\mathrm{h}>7$} pixels, as they are poorly covered
by the employed postage stamps.
As done in \citetalias{schrabback2010} we mask galaxies close to the image boundaries, large
galaxies, or bright stars.

\citetalias{schrabback2010} introduced
an empirical correction for noise bias in the ellipticity measurement as a
function of the KSB signal-to-noise ratio from \citet{erben2001}. \citetalias{schrabback2010} calibrated this correction using simulated images of ground-based
weak lensing observations from STEP2 \citep{massey2007}, and verified that
the same correction robustly corrects simulated high-resolution ACS-like
weak lensing data with less than 2\% residual multiplicative ellipticity
bias ($0.8\%$ on average).
However, as recently shown by \citet{hoekstra15}, the STEP2 image simulations lack sources at the faint end, affecting the derived bias calibration \citep[see also][]{hoekstra17}.
Also, deviations in the assumed intrinsic galaxy shape distribution
influence the noise-bias correction \citep[e.g.][]{viola2014}.
To minimise the impact of such uncertainties
we apply a more conservative galaxy selection requiring
\mbox{$S/N=(\mathrm{Flux}/\mathrm{Fluxerr})_\mathrm{auto}>10$} from
\texttt{Source Extractor}\footnote{This cut is more conservative than  the cut \mbox{$S/N_\mathrm{KSB}>2$} from S10, which is based on the \citet{erben2001}
signal-to-noise ratio definition that includes a radial weak lensing weight
function. \mbox{$S/N_\mathrm{KSB}>2$} approximately corresponds to
\mbox{$S/N=(\mathrm{Flux}/\mathrm{Fluxerr})_\mathrm{auto}\gtrsim 6.5$} for
our typical source galaxies, but note that there is a significant scatter
between both estimates due to the different radial weighting.}.
To be conservative, we additionally
double the
systematic uncertainty for the shear calibration
in the error-budget of our current cluster study (4\%), which is comparable
to the mean shear calibration correction of the galaxies passing our
cuts (average factor 1.05).
In the context of cluster weak lensing studies a relevant question is also
if the image simulations probe the relevant range of shears sufficiently
well. We expect that this is not a major concern for our study given that
\mbox{$\langle g_\mathrm{t}\rangle\lesssim 0.1-0.15$} for all of our
clusters within the radial range used for the mass constraints (see
Sect.\thinspace\ref{sec:wlmasses}).
For comparison, the basic KSB+ implementation used in our analysis was tested in
\citet{heymans06} using shears up to \mbox{$g=0.1$}, where no
indications were found for significant quadratic shear bias terms that would result in an
inaccurate correction using our linear correction scheme.

We apply the same shape measurement pipeline to the CANDELS data discussed in Sect.\thinspace\ref{se:data_hst_candels}.
When mimicking our cluster field selection in these
catalogues and assigning weights, we
rescale the \mbox{$S/N$} values prior to the
\mbox{$S/N$} cut to account for slight differences in depth.
Hence, if a CANDELS tile is slightly shallower (deeper) compared to the
cluster tile considered, we will apply a correspondingly slightly lower
(higher) \mbox{$S/N$} cut in the CANDELS tile to select consistent galaxy
samples.
On average the depth of our CANDELS stacks agrees well with the depth of the
cluster field stacks (to 0.065 mag).
Together with the fact that $\langle\beta\rangle$ depends only  weakly
on $V_{606}$ for our colour-selected sample at the faint end (see
Sect.\thinspace\ref{se:analysis_in_magbins}),
we therefore ignore second-order effects such as incompleteness differences
between the CANDELS and cluster field catalogues.

 \subsection{Test for residual CTI signatures in the ACS cluster data}
\label{sec:ctitests}

CTI generates charge-trails behind objects dominantly in the
parallel-transfer readout direction.
For raw ACS images this corresponds to  the $y$-direction, and this is
approximately also the case for distortion-corrected images if
\texttt{MultiDrizzle} is run
using the native detector orientation.
\citetalias{massey2014} test the performance of their pixel-based
CTI correction
by averaging the PSF-corrected ellipticity estimates of galaxies in blank
field CANDELS data.
Images without CTI correction show a prominent alignment with the
$y$-axis (\mbox{$\langle \epsilon_1 \rangle < 0$}),
where the magnitude of the effect increases with the $y$-separation relative to the readout amplifiers.
In contrast, this alignment is undetected if the
correction is applied.

\begin{figure}
\begin{center}
  \includegraphics[width=0.95\columnwidth]{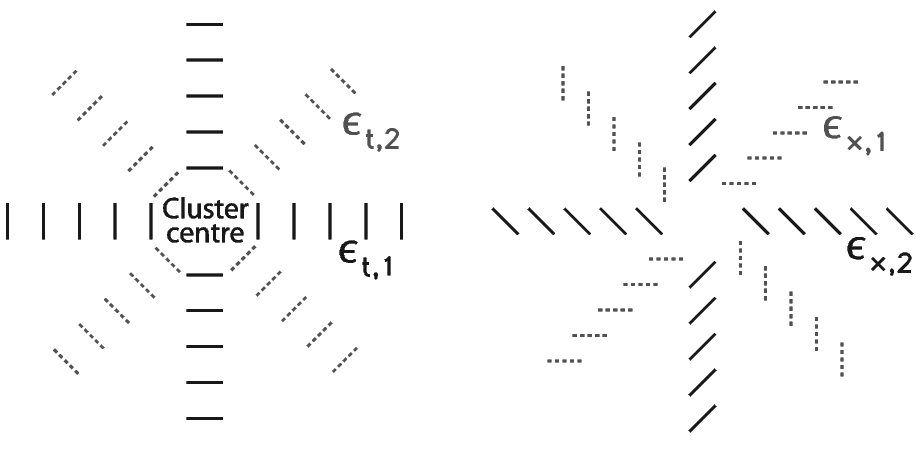}
\vspace{0.3cm}

  \includegraphics[width=0.99\columnwidth]{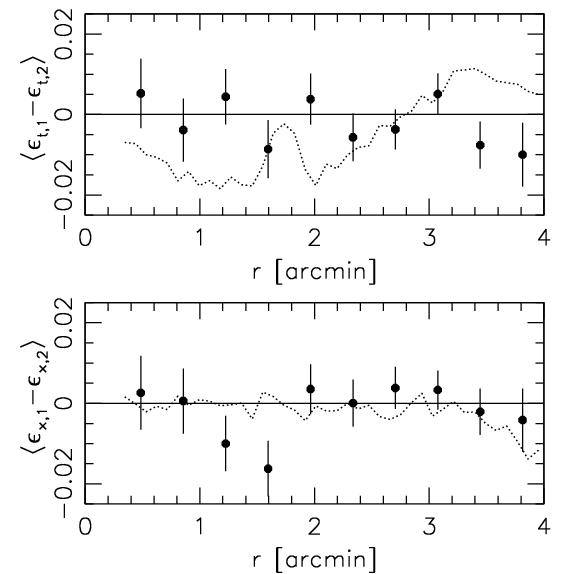}
\end{center}
\caption{Testing for residual CTI systematics in the cluster fields:
{\it Top:} Illustration for the separation of the tangential and cross components of the ellipticity into components affected by CTI ($\epsilon_{\mathrm{t},1}$, $\epsilon_{\times,1}$), and those unaffected by CTI  ($\epsilon_{\mathrm{t},2}$, $\epsilon_{\times,2}$).
The {\it middle} ({\it bottom}) panel shows the difference in the tangential (cross)  ellipticity component with respect to the cluster centre as estimated from the
CTI-affected and the
CTI-unaffected
 components.
Here we combine the signal from all galaxies passing the shape cuts with \mbox{$24<V_\mathrm{606,auto}<26.7$} in
all cluster fields.
The points are consistent with zero (\mbox{$\chi^2/\mathrm{d.o.f.}=0.96$})
suggesting that the CTI has been fully corrected within the statistical
precision of the data. For comparison, the dotted curve shows the signal
which would be measured from an uncorrected CTI saw-tooth ellipticity pattern
with \mbox{$\langle e_1 \rangle=-0.05$},
where small wiggles are caused by the sampling at the galaxy positions and
the masks
applied.
\label{fi:ctitest}}
\end{figure}

We cannot apply the same test to our ACS data of the cluster fields given the
presence of massive clusters, which are always located at the same position
within the mosaics, and whose weak gravitational lensing shear would add to
the saw-tooth CTI signature.
However, we can make use of the fact that CTI primarily affects  the
$\epsilon_1$ ellipticity component (measured along the image axes) but
not the $\epsilon_2$ ellipticity component (measured along the field
diagonals).
The tangential and cross
components of the ellipticity
with respect to  the
cluster centre
\begin{eqnarray}
\epsilon_\mathrm{t} & = & \epsilon_{\mathrm{t},1} + \epsilon_{\mathrm{t},2}\\
\epsilon_\times & = & \epsilon_{{\times},1} + \epsilon_{\times,2}
\end{eqnarray}
 (compare Equations\thinspace\ref{eq:gt} and \ref{eq:gx}) receive contributions  from both ellipticity components
with
\begin{eqnarray}
\epsilon_{\mathrm{t},1} & = & - \epsilon_1 \cos{2 \phi}\\
\epsilon_{\mathrm{t},2} & = & - \epsilon_2 \sin{2\phi}\\
\epsilon_{\times,1}  & = & + \epsilon_1 \sin{2\phi} \\
\epsilon_{\times,2}  & = &  - \epsilon_2 \cos{2 \phi} \,,
\end{eqnarray}
see the sketch in the top panel of Fig.\thinspace\ref{fi:ctitest} for an illustration of these components.
In our test we stack the signal from all clusters.
Here we expect that any
anisotropy in the reduced shear pattern due to cluster halo ellipticity will
average out leading to an approximately circularly symmetric shear field.
Accordingly, in the absence of residual systematics  we expect that \mbox{$\langle
\epsilon_{\mathrm{t},1}-\epsilon_{\mathrm{t},2}\rangle$} and \mbox{$\langle
\epsilon_{\times,1}-\epsilon_{\times,2}\rangle$} are consistent with zero
when averaged azimuthally.
Fig.\thinspace\ref{fi:ctitest} shows that this is indeed the case for our
data (\mbox{$\chi^2/\mathrm{d.o.f.}=0.96$}), confirming the success of the CTI correction within the statistical precision of the data.
For comparison, the dotted line in  Fig.\thinspace\ref{fi:ctitest} shows the
signal that would be caused by a typical uncorrected CTI ellipticity saw-tooth pattern
with \mbox{$\langle
\epsilon_1\rangle=-0.05$}\footnote{\citetalias{massey2014} measure an average uncorrected
CTI-induced galaxy ellipticity  at \mbox{$V\sim 26.5$} of  \mbox{$\langle
\epsilon_1\rangle\simeq -0.04$}
 from CANDELS/COSMOS F606W images, which were observed at a similar epoch
 but have higher background levels than our data, and thus  weaker
 CTI signals.}.

\section{Cluster member removal and estimation of the source redshift distribution}
\label{sec:phot}

Robust weak lensing mass measurements require accurate knowledge of the  mean geometric lensing efficiency $\langle \beta \rangle$ of the source sample and its variance  $\langle \beta^2 \rangle$ (see Sect.\thinspace\ref{sec:sheartheory}).
For a  given cosmological model these depend only on the source redshift distribution and cluster redshift.
Surveys with
sufficiently deep
imaging in
sufficiently
many
 bands can attempt to estimate the probability distribution of source
redshifts directly via photo-$z$s \citep[e.g.][]{applegate14}.
However, such data are not available for our cluster fields.
Hence, we have to rely on an estimate of the redshift distribution from
external reference fields.
Here we use photometric redshift estimates  for the CANDELS fields  from  the 3D-HST team \citep{skelton14}
as primary data set (see Sect.\thinspace\ref{se:photo_ref_cats}).
Additionally, we use spectroscopic and grism redshift estimates for galaxies in the CANDELS fields, as well as
much deeper data from the {\it Hubble} Ultra Deep field (HUDF)
to investigate and statistically correct for systematic features
in the CANDELS photo-$z$s
(Sect.\thinspace\ref{sec:correct_sys_photoz}).

Given that our cluster fields are over-dense at the cluster redshift we have
to apply a colour selection that robustly removes galaxies at the cluster redshift
both in the reference catalogue and our actual cluster field catalogues.
Here we use colour estimates from the HST/ACS F606W
and F814W images in the inner regions
(``ACS-only'' selection, Sect.\thinspace\ref{se:photo_color_select_acs}),
and we use
 VLT/FORS2 $I$-band imaging for the cluster outskirts
(``ACS+FORS2'' selection, Sect.\thinspace\ref{se:source_sel_scatter} with
details given in Appendix \ref{app:details_fors2color_scatter}).
As discussed in
 Appendix \ref{app:why_not_boost} we also explored
a different analysis scheme which
substitutes the colour selection with a statistical correction for cluster
member contamination, but
 we found that we could not control the systematics of the correction to the needed level due to the limited
radial range probed by the F606W images.
We
optimise the analysis by splitting the colour-selected sources into
magnitude bins
(Sect.\thinspace\ref{se:analysis_in_magbins}),
investigate the influence of
line-of-sight variations (Sect.\thinspace\ref{sec:beta_los_variation}),
and account for weak lensing magnification
(Sect.\thinspace\ref{sec:magnification_model}).
Sect.\thinspace\ref{se:photo_number_density_tests} presents consistency
checks for our analysis based on the source number density measured as
function of magnitude and cluster-centric distance.

\subsection{CANDELS photometric redshift reference catalogues from 3D-HST}
\label{se:photo_ref_cats}

We make use of photometric redshift catalogues computed by
the 3D-HST team \citep[][hereafter
\citetalias{skelton14}]{brammer12,skelton14} for the CANDELS fields
\citep{grogin2011}, which
consist of five independent lines-of-sight (AEGIS, COSMOS, GOODS-North, GOODS-South,
UDS).
Hence, their combination efficiently suppresses the impact of sampling variance.
All  CANDELS field were observed by HST with ACS and WFC3,
including ACS  F606W and
F814W\footnote{For the GOODS-North field we estimate the $I_{814}$ magnitudes
  from the \citetalias{skelton14} flux measurements in the F775W and F850LP
  filters. When conducting selections  or binning in $V_{606}$ based on the
  \citetalias{skelton14} photometry we undo their correction for total
  magnitudes in order to employ aperture magnitudes that are consistent with
  our cluster field measurements.}
imaging mosaics that have at least the depth of our cluster field observations \citep[see][]{koekemoer11}.
This includes observations from the CANDELS program \citep{grogin2011}  and
earlier projects \citep[][]{giavalisco2004,rix2004,davis2007,scoville07}.
The \citetalias{skelton14} catalogues are based on detections from combined
HST/WFC3 NIR F125W+F140W+F160W images, and include photometric measurements
from a total of 147 distinct imaging data sets from HST, {\it Spitzer}, and
ground-based facilities with a broad wavelength coverage from
\mbox{$0.3-8\mu\mathrm{m}$} (\mbox{$18-44$} data sets per field).
\citetalias{skelton14} compute photometric redshifts using \textsc{EAZY} \citep{brammer08}, which fits the observed SED constraints of each object with a linear combination of galaxy templates.

We have matched the  \citetalias{skelton14} catalogues with our F606W-detected
shape catalogues of the CANDELS fields (see Sect.\thinspace\ref{sec:shear}).
After applying weak lensing cuts, accounting for masks, and restricting the analysis to the overlap
region of  the ACS and WFC3 mosaics, we find
that
\mbox{$\sim 97.6\%$}
of the galaxies in the shape catalogues with
\mbox{$24<V_{606}<26.5$}
have a direct match within 0\farcs5 in the \citetalias{skelton14} catalogues,
showing that they are nearly complete within our employed magnitude range
(see Appendix \ref{se:app:non_matches} for an investigation of the \mbox{$\sim 2.4\%$} of non-matching
galaxies which shows that they have a negligible impact).

\subsection{Source selection using ACS-only colours}
\label{se:photo_color_select_acs}

\begin{figure}
  \includegraphics[width=0.99\columnwidth]{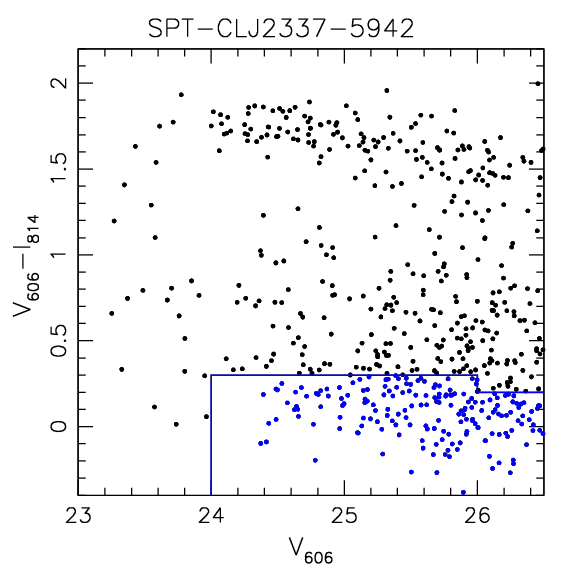}
\caption{Measured  $V_{606}-I_{814}$ colours as function of $V_{606}$ for
  galaxies in the field of SPT-CL{\thinspace}$J$2337$-$5942 that pass our weak
  lensing shape cuts, and that are located within the central $I_{814}$ ACS
  tile. The blue lines indicate the region of blue  galaxies that pass our
 colour selection.
The cluster red sequence is clearly visible at
  \mbox{$V_{606}-I_{814}\sim 1.7$}.
  \label{fig:v_vi}}
\end{figure}

\begin{figure}
  \includegraphics[width=1.03\columnwidth]{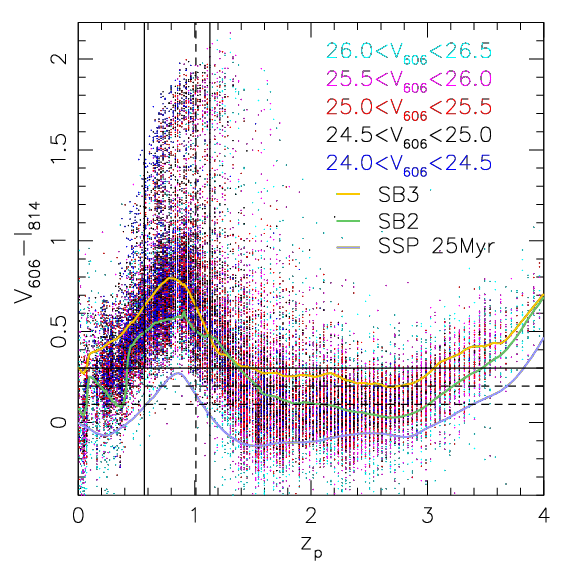}
\caption{$V_{606}-I_{814}$ colours of galaxies in the CANDELS
  fields as function of the peak photometric redshift $z_\mathrm{p}$
  from \citetalias{skelton14}. The colour coding splits the galaxies into our
  different magnitude bins. The horizontal lines mark our different colour
  cuts
(dependent on cluster redshift and galaxy magnitude, see Sect.\thinspace\ref{se:photo_color_select_acs}),
while the vertical lines indicate the cluster
  redshift range \mbox{$0.57\le z \le 1.13$} (solid), as well as
  \mbox{$z=1.01$} (dashed),
  at which cluster redshift the colour cuts change.
The curves indicate synthetic $V_{606}-I_{814}$  colours of galaxy SED
templates from \citet{coe06}.
\label{fig:z_vi}}
\end{figure}

\begin{figure*}
  \includegraphics[width=0.99\columnwidth]{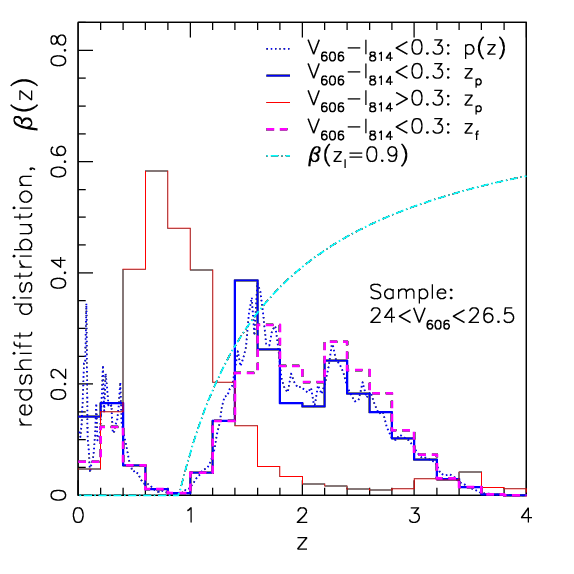}
  \includegraphics[width=0.99\columnwidth]{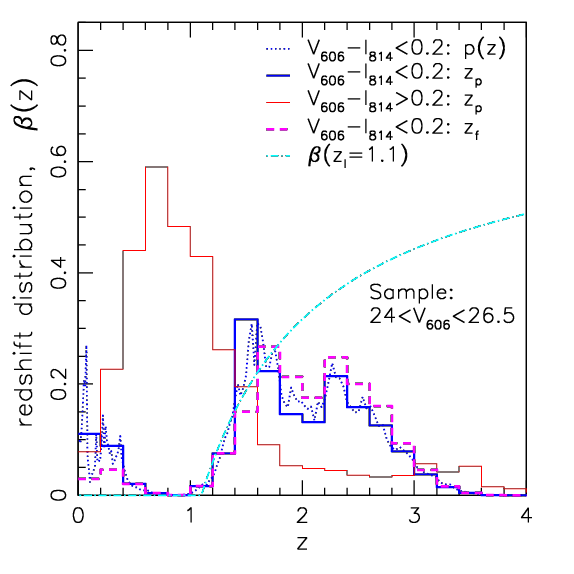}
  \includegraphics[width=0.99\columnwidth]{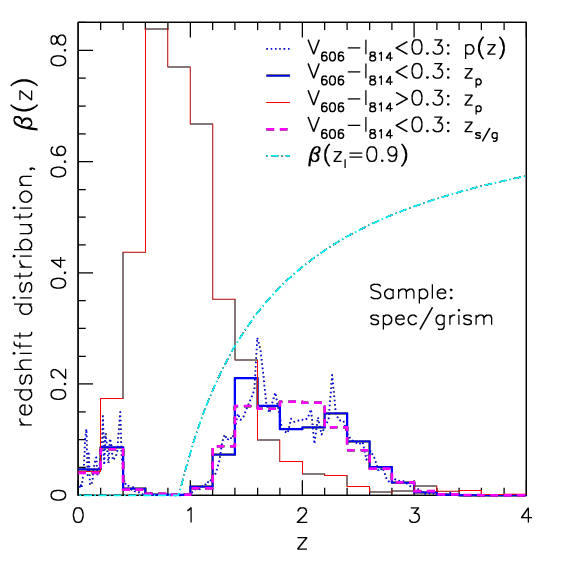}
  \includegraphics[width=0.99\columnwidth]{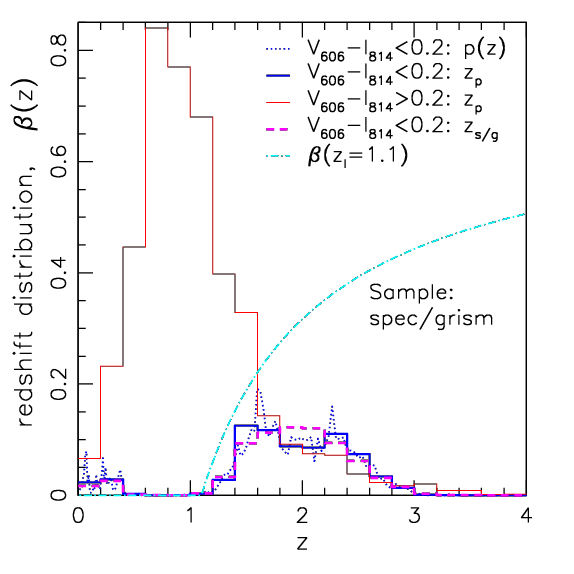}
\caption{Redshift distribution of different galaxy samples in CANDELS:
The {\it top} panels show the full photometric sample of galaxies which have
\mbox{$24.0<V_{606}<26.5$} and
 pass the
shape cuts,
whereas the sample is further reduced to contain only those galaxies with robust spec-$z$s or grism-$z$s in the {\it bottom} panels.
In the {\it left} ({\it right}) panels, a colour cut  \mbox{$V_{606}-I_{814}<0.3$}
(\mbox{$V_{606}-I_{814}<0.2$}) is used to separate the source sample (solid thick photo-$z$ histogram and thin dotted  averaged $p(z)$ in blue) from redder galaxies (thin solid red photo-$z$ histogram) that contain most galaxies at the corresponding cluster redshifts.
The magenta dashed histogram shows the distribution of spec-$z$s or
grism-$z$s in the {\it bottom} panels, and the distribution of photo-$z$s
after the statistical correction based on the HUDF analysis in the {\it top}
panels.
The histograms are normalised according to the total number of galaxies in
the corresponding spectroscopic or photometric sample prior to the colour
selection.
The cyan dashed-dotted curve shows the geometric lensing efficiency $\beta$
for
 clusters at redshift \mbox{$z_l=0.9$} ({\it left}) and \mbox{$z_l=1.1$}
 ({\it right}).
The presence of foreground galaxies in the source sample is not
a concern as long as it is modelled accurately.
\label{fig:zdist_f814w}}
\end{figure*}

In the inner cluster regions we apply a colour selection (indicated in Fig.\thinspace\ref{fig:v_vi}) using our ACS F606W
and F814W images, selecting only
galaxies that are   bluer than
nearly all galaxies at the cluster redshift.
This is illustrated in  Fig.\thinspace\ref{fig:z_vi}, where we
plot the  \textsc{EAZY}
 peak photometric redshift $z_\mathrm{p}$
for the CANDELS galaxies as function of \mbox{$V_{606}-I_{814}$} colour from \citetalias{skelton14}  (measured with the same 0\farcs7 aperture diameter as employed for our ACS colour measurements).
Figures \ref{fig:z_vi} and \ref{fig:zdist_f814w} illustrate that  the selection of blue
galaxies in \mbox{$V_{606}-I_{814}$} colour in CANDELS is very effective in removing
galaxies at our cluster redshifts, while
it
selects the majority of the
\mbox{$z_\mathrm{p}\gtrsim 1.4$} background galaxies.
The latter are high-redshift star-forming galaxies observed at rest-frame UV
wavelength with very blue spectral slopes.
 In contrast, nearly all galaxies at the cluster redshifts show a redder
\mbox{$V_{606}-I_{814}$} colour, as they contain either the 4000\AA \,break
(early type galaxies, see the cluster red sequence in Fig.\thinspace\ref{fig:v_vi}) or the Balmer break (late type galaxies) within the
filter pair.

We note that our approach rejects both red and blue cluster members.
It is therefore more conservative and robust than redder colour cuts
that some studies have used to remove red sequence cluster members only
\citep[e.g.][]{jee11}.
Note that, in contrast, \citet{okabe13} select only galaxies that are
redder than the  red sequence.
This is a useful approach for the low-redshift
clusters targeted in their study, but less effective for the high-redshift clusters studied here, as
most of the \mbox{$z_\mathrm{p}\gtrsim 1.4$} background galaxies are blue at optical wavelengths
(see Fig.\thinspace\ref{fig:zdist_f814w}).
Likewise, some studies of lower redshift clusters have used combinations of
blue and red regions in colour space to minimise cluster member
contamination \citep[e.g.][]{medezinski10,high12,umetsu14}.
\reva{It is evident from Fig.\thinspace\ref{fig:z_vi} that a selection of blue
  galaxies in $V-I$ colour is inefficient for clusters at low redshifts
  \mbox{$z\lesssim 0.4$},
as it would either require extremely blue cuts that drastically shrink the source sample,
or
lead to a  larger residual contamination by galaxies at the cluster
redshift.
Similar
results were
found by \citet{ziparo16}, who conclude that optical observations alone are
not sufficient to reduce the cluster member contamination below the per-cent level
for blue source
samples and clusters at \mbox{$z\sim 0.2$}.
}

For clusters at \mbox{$z<1.01$} we select source galaxies with
\mbox{$V_{606}-I_{814}<0.3$}.
This maximises the background galaxy density
while at the same time
removing
\mbox{$98.5\%$}  of the CANDELS galaxies at
\mbox{$0.6< z_\mathrm{p}<1$} that pass the other weak lensing cuts,
see the top left panel of Fig.\thinspace\ref{fig:zdist_f814w}.
For the higher redshift clusters we apply a
more stringent cut \mbox{$V_{606}-I_{814}<0.2$}
which still yields a  \mbox{$97.6\%$}
suppression of galaxies at \mbox{$1<z_\mathrm{p}<1.13$},
at the expense of a slightly lower source density  (top right panel of
Fig.\thinspace\ref{fig:zdist_f814w}).
When conducting the  analysis for our cluster fields
we apply slightly more conservative colour cuts that are bluer by
0.1\thinspace mag for the faintest sources in our analysis,
as they show the largest
photometric scatter.
As a result, we obtain a similar fraction of removed galaxies at the cluster
redshifts when taking photometric scatter into account (see
Sect.\thinspace\ref{se:source_sel_scatter} and Appendix \ref{app:scatter}).

In Fig.\thinspace\ref{fig:z_vi} we also over-plot synthetic
\mbox{$V_{606}-I_{814}$} colours
of redshifted SED templates for star forming galaxies employed in the  Bayesian  Photometric  Redshift
(\texttt{BPZ})
algorithm \citep{benitez00}.
This includes the
SB3 and SB2 star burst templates from \citet{kinney96} as recalibrated by
\citet{benitez04}.
We additionally include a young star burst model (SSP 25Myr), which
is one of the templates introduced by \citet{coe06} into \texttt{BPZ}
to
improve photometric redshift estimates for very blue galaxies in the HUDF.
The shown SED corresponds to a simple stellar population (SSP) model with an
age of 25 Myr and metallicity \mbox{$Z=0.08$} \citep{bruzual03}.
At the cluster redshifts, the colours of the SB3 and SB2 templates
approximately describe the range of colours of typical blue cloud galaxies,
which are well removed by our colour selection.
In contrast,  while  the colour of the SSP 25 Myr model appears to be
representative for a considerable fraction of the  \mbox{$z\gtrsim
  1.4$} background galaxies, it approximately marks the
location of the most extreme blue outliers at the cluster redshifts, which
are not fully removed by our colour selection scheme.
If the clusters contain a substantial fraction of such extremely blue
galaxies, this might introduce some residual cluster member
contamination in our lensing catalogue.
We investigate this issue in Appendix \ref{se:test_extremely_blue}, concluding
that such galaxies have a negligible impact for our analysis despite the
physical over-density of galaxies in clusters.
We also present
empirical tests for residual contamination by cluster galaxies in Sect.\thinspace\ref{se:photo_number_density_tests}.

\subsection{Statistical correction for systematic features in the photometric redshift distribution}
\label{sec:correct_sys_photoz}
\begin{figure*}
  \includegraphics[width=0.68\columnwidth]{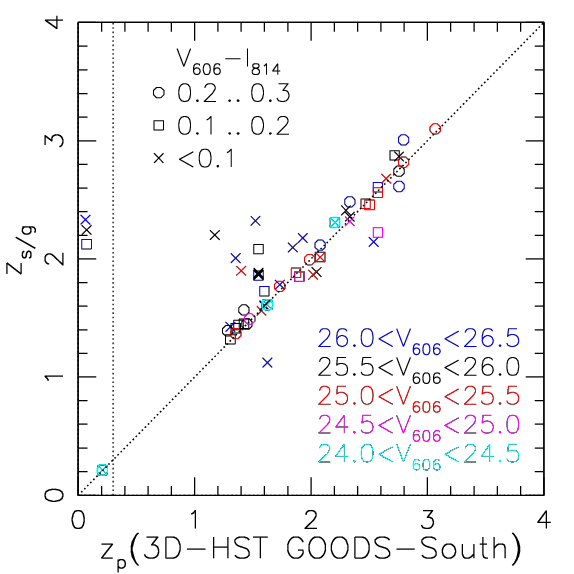}
  \includegraphics[width=0.68\columnwidth]{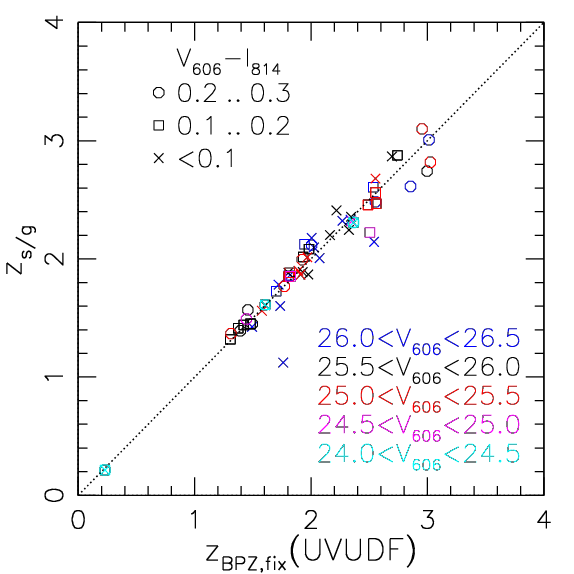}
  \includegraphics[width=0.68\columnwidth]{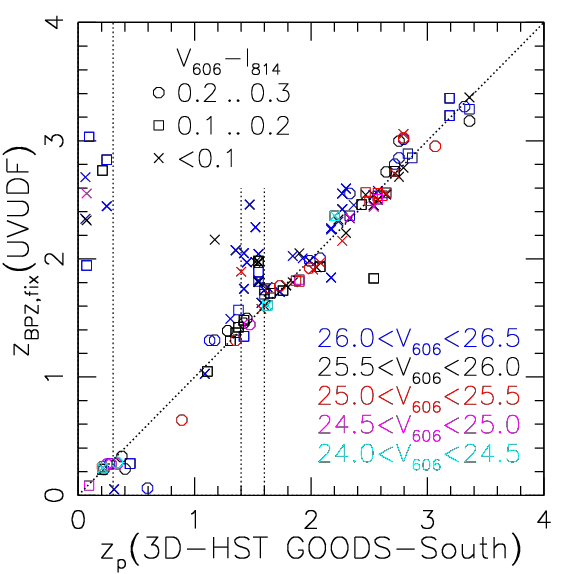}
 \caption{Comparison of redshift estimates in the HUDF including the peak
   photometric redshift from \textsc{EAZY} $z_\mathrm{p}$ estimated by the 3D-HST
   team in the GOODS-South field, the BPZ photometric redshift from  the
   UVUDF project  $z_\mathrm{BPZ,fix}$ (with small bias corrections applied,
   see text), and a combined sample of spectroscopic and grism redshifts
   $z_\mathrm{s/g}$.
We regard the latter as a true but incomplete reference sample, which
reveals the presence of significant outliers for the  $z_\mathrm{p}$  but
not the  $z_\mathrm{BPZ,fix}$ photo-$z$s.
We therefore use the  $z_\mathrm{BPZ,fix}$ photo-$z$s, which do not suffer
from incompleteness at the relevant depth, to derive a statistical
correction for the $z_\mathrm{p}$ photo-$z$s.
 The symbols split the galaxies according to
   \mbox{$V_{606}-I_{814}$} colour and the different colours indicate
   different magnitude bins (based on the 3D-HST photometry).
Galaxies are only included if they pass our weak lensing selection and if
they  are
located
within the area covered by the WFC3 UVIS and IR observations.
In the {\it right} panel the vertical lines indicate the $z_\mathrm{p}$ ranges of our statistical correction for the redshift distribution.
\label{fig:udf_correction}}
\end{figure*}

We
 base our estimate of the source redshift distribution on the CANDELS photo-$z$ catalogues because of their high completeness
at the depth of our SPT ACS observations (Sect.\thinspace\ref{se:photo_ref_cats}), allowing us to select galaxies that are representative for the galaxies used in our lensing analysis.
However, it is important to realise that such photo-$z$ estimates may contain
systematic features (e.g. catastrophic outliers) that can bias the inferred
redshift distribution and accordingly the lensing results.
As an example, the cosmological weak lensing analysis of COSMOS data by \citetalias{schrabback2010} suggests that
the majority
of faint galaxies in the COSMOS-30 photometric redshift catalogue  \citep{ilbert2009} that have a primary peak in their  posterior redshift probability distribution $p(z)$ at low redshifts but also a secondary peak at high redshifts,  are truly at high redshift.
Likewise, the galaxy-galaxy lensing analysis of CFHTLenS data by \citet{heymans12}
indicates that a significant fraction of galaxies with an assigned
photometric redshift \mbox{$z_\mathrm{photo}<0.2$} are truly at high
redshift.
In the following subsections we exploit additional data sets to check the accuracy of the CANDELS photo-$z$s and implement a statistical correction for relevant systematic features.

\subsubsection{Tests and statistical correction based on HUDF data}
\label{se:sub:catastrophic_outliers}

The {\it Hubble} Ultra Deep Field
(HUDF) is located
within one of the CANDELS fields (GOODS-South).
The very deep multi-wavelength observations conducted in the HUDF can therefore be used for  cross-checks of the CANDELS photo-$z$s.

As first data set  we use a combination of high-fidelity spectroscopic redshifts (``spec-$z$s'',
$z_\mathrm{s}$) compiled  by  \citet{rafelski15}\footnote{\citet{rafelski15}
  note that the object 10157 in their catalogue is
  problematic as it consists of a blend of two galaxies at different
  redshifts. We therefore exclude it from the spec-$z$/grism-$z$ sample used
  in our analysis.}, and redshift estimates
extracted
by the 3D-HST team \citep{brammer12,brammer13} from the combination of  deep HST WFC3/IR slitless  grism spectroscopy and very
deep HST optical/NIR imaging.
These ``grism-$z$s'' ($z_\mathrm{g}$)
significantly enlarge the sample of high-$z$ (\mbox{$z>1$}) galaxies with
high quality redshift estimates, where typical errors of the grism-$z$s are
\mbox{$\sigma_z\approx 0.003\times (1+z)$} \citep{brammer12,momcheva16}.

We compare the CANDELS photo-$z$s
 to the HUDF $z_\mathrm{s/g}$ estimates  in the left panel of Fig.\thinspace\ref{fig:udf_correction}.
The majority of the data points closely follow
the diagonal, suggesting that the 3D-HST photo-$z$s are overall well calibrated as
needed for unbiased estimates of the redshift distribution.
However, we note the presence of two relevant systematic features:
first, there are
 three catastrophic outliers that are  at  high
\mbox{$z_\mathrm{s/g}\simeq 2.2$}, but are assigned
a low \mbox{$z_\mathrm{p}\simeq 0.07$}.
Second, there is an increased, asymmetric scatter at \mbox{$1.2\lesssim z_\mathrm{p}\lesssim 1.7$}. Most notably, many galaxies with an assigned photometric redshift  \mbox{$1.4\lesssim z_\mathrm{p}\lesssim 1.6$} are actually at higher redshift.
This is likely the result of  redshift focusing effects \citep[e.g.][]{wolf09} caused by the broad band HST filters.
While this comparison allows us to identify these issues, the matched
catalogue is insufficient to derive a robust statistical correction for our
full photometric sample given the incompleteness of the
\mbox{$z_\mathrm{s/g}$} sample.

To overcome this limitation of incompleteness, we
use
deep photometric redshifts computed
 by \citet{rafelski15} using HUDF data as a second
comparison sample.
Compared to the CANDELS photo-$z$s they benefit from much deeper
HST
optical \citep{beckwith06} and NIR imaging \citep{koekemoer13}, and
additionally incorporate new
HST/UVIS Near UV imaging from the UVUDF project \citep{teplitz13}
 taken in the
F225W, F275W, and F336W filters.
These bands probe the
Lyman break in the
redshift range \mbox{$1.2\lesssim z\lesssim 2.7$}, which contains most
of our weak lensing source galaxies.
At these redshifts, the NIR imaging additionally probes the location of the
Balmer/4000\AA\, break.
Hence, we expect that
the resulting  photo-$z$ should be highly  robust
against catastrophic outliers.
We test this by comparing them to the  $z_\mathrm{s/g}$ redshifts in
the middle panel of Fig.\thinspace\ref{fig:udf_correction}.
Here we use the photo-$z$ estimates $z_\mathrm{BPZ}$ obtained by
\citet{rafelski15}
using \texttt{BPZ} as it
yields the highest
robustness against catastrophic outliers in their analysis.
Note that the  comparison of  $z_\mathrm{BPZ}$ and  $z_\mathrm{s/g}$
suggests that  $z_\mathrm{BPZ}$ slightly overestimates the redshifts for
the  colour-selected sample
in the redshift intervals
\mbox{$1.0\lesssim z_\mathrm{BPZ}\lesssim 1.7$} and \mbox{$2.6\lesssim z_\mathrm{BPZ}\lesssim 3.7$},
with median redshift offsets of 0.071 and 0.171, respectively.
We have therefore subtracted these offsets in the corresponding redshift
intervals, yielding  $z_\mathrm{BPZ,fix}$, which is shown in Fig.\thinspace\ref{fig:udf_correction}.
As visible in the middle panel of Fig.\thinspace\ref{fig:udf_correction},
$z_\mathrm{BPZ,fix}$ correlates tightly with $z_\mathrm{s/g}$.
In particular, the three catastrophic outliers from the left panel are now
correctly placed at high redshifts. Likewise, the redshift focusing effects
are basically removed.
The remaining scatter with  one  moderate outlier
has negligible impact on our results. For example,
\mbox{$\langle\beta\rangle$} agrees to 0.4\% between $z_\mathrm{BPZ,fix}$
and $z_\mathrm{s/g}$ for the matched catalogue and clusters at
\mbox{$z_\mathrm{l}=1.0$} (we include this in the systematic error budget
of Sect.\thinspace\ref{se:zdist_fix_uncertainty}).
This suggests that $z_\mathrm{BPZ,fix}$ provides a
sufficiently accurate approximation for the true redshift.
Hence, we use $z_\mathrm{BPZ,fix}$
as a reference to obtain a statistical
correction for the systematic features of the CANDELS photo-$z$s.

We compare the 3D-HST   photo-$z$s $z_\mathrm{p}$ in the HUDF to
  $z_\mathrm{BPZ,fix}$ in the right panel   of
Fig.\thinspace\ref{fig:udf_correction}, again showing the previously
identified catastrophic outliers at  \mbox{$z_\mathrm{p}<0.3$} and redshift
focusing effects at \mbox{$1.4\lesssim z_\mathrm{p}\lesssim 1.6$}, but now at
the full depth of our photometric sample.
The catastrophic outliers with \mbox{$z_\mathrm{p}<0.3$} are dominated by
blue \mbox{$V_{606}-I_{814}<0.2$} galaxies, for which
 9 out of 12
galaxies appear to
be truly at high redshifts.
In order to implement a statistical correction for these outliers for the
full CANDELS catalogue, we note
the
12
redshift offsets
\mbox{$(z_\mathrm{BPZ,fix}-z_\mathrm{p})_i$}.
We bootstrap this empirically defined distribution to define the correction:
for each CANDELS galaxy
with
\mbox{$z_\mathrm{p}<0.3$} and
\mbox{$V_{606}-I_{814}<0.2$}
we
add a randomly drawn offset to its $z_\mathrm{p}$.
Likewise, we apply a
statistical correction for the redshift focusing within the redshift range
\mbox{$1.4\le z_\mathrm{p} \le 1.6$} for galaxies with
\mbox{$V_{606}-I_{814}<0.1$} (which are most strongly affected, see Fig.\thinspace\ref{fig:udf_correction}),
again randomly sampling from the corresponding
\mbox{$(z_\mathrm{BPZ,fix}-z_\mathrm{p})_i$} offsets in the HUDF.
For the latter correction we split the galaxies into two magnitude ranges
(\mbox{$24<V_{606}<25.5$} and \mbox{$25.5<V_{606}<26.5$}) given that the
fainter galaxies appear to suffer from the redshift focusing effects more
strongly.
We show the resulting distribution of statistically corrected redshifts
$z_\mathrm{f}$ as magenta dashed histograms in the top panels of
Fig.\thinspace\ref{fig:zdist_f814w}.
As expected, it has a lower fraction of low-$z$ galaxies compared to the
uncorrected $z_\mathrm{p}$ distribution, as well as a reduction of the redshift
focusing peak at \mbox{$1.4\le z_\mathrm{p} \le 1.6$}.
Both effects  are
compensated by a higher fraction of high-$z$ galaxies, where we also note
that the local minimum at \mbox{$z_\mathrm{p}\simeq 2$}, which likely
results from the redshift focusing (see also Sect.\thinspace\ref{se:zdist_candels_checks}), is reduced.

Averaged over our full cluster sample, and accounting for the
magnitude-dependent effects  explained in the following sections (e.g. shape weights),
 the application of this correction scheme
leads to a 12\% decrease of the resulting cluster masses.
Of this, 10\% originate from
the correction for catastrophic outliers, and 2\% from the correction for
redshift focusing.

\subsubsection{Uncertainty of the statistical correction of the
  redshift distribution}
 \label{se:zdist_fix_uncertainty}

The statistical correction of the redshift distribution explained in Sect.\thinspace\ref{se:sub:catastrophic_outliers} has a
non-negligible impact on our analysis. Therefore it is important to quantify
its uncertainty.
We consider a number of effects that affect the uncertainty:
first, we estimate the  statistical uncertainty originating from
the limited size of the HUDF catalogue
by generating bootstrapped versions of it, which are then used to generate
the
\mbox{$(z_\mathrm{BPZ,fix}-z_\mathrm{p})_i$} offset samples.
This yields a small, 0.5\% uncertainty regarding the average masses.
Second, our correction scheme assumes that the relative effects seen in the HUDF are
representative for the full CANDELS area. However,
some previous studies suggest that the GOODS-South field, which contains the
HUDF, could be somewhat under-dense at lower redshifts
compared to the cosmic mean \citep[e.g.][]{schrabback07,hartlap09}.
To obtain a worst case estimate of the impact this could have,
we assume
that the GOODS-South field could be under-dense at low redshifts by a factor
3
compared to the cosmic mean.
Hence, we artificially boost the number of HUDF galaxies with
\mbox{$z_\mathrm{p}<0.3$} that are truly at low-$z$ by a factor 3 for the
generation of the offset pool.
On average this leads to a
3\% increase of the
cluster masses.
Third, we note that our correction for redshift focusing incorporates most
but not all of the corresponding outliers in the right panel of
Fig.\thinspace\ref{fig:udf_correction}.
We assume a  conservative 50\% relative uncertainty on the 2\% correction,
corresponding to an absolute 1\% uncertainty.
Adding all individual systematic uncertainties identified here and in
Sect.\thinspace\ref{se:sub:catastrophic_outliers}
in quadrature yields a combined systematic uncertainty for the systematic corrections to the
photometric redshifts
of 3.3\% in the average cluster  mass.

\subsubsection{Consistency checks using spectroscopic and grism redshifts in
  the CANDELS fields}
\label{se:zdist_candels_checks}
In Sect.\thinspace\ref{se:sub:catastrophic_outliers} we obtained a statistical correction for systematic
features in the CANDELS photo-$z$s using very deep data available in the HUDF.
Here we present cross-checks for this correction using
the
CANDELS redshift catalogue from \citet{momcheva16}, which combines a
compilation of high fidelity spectroscopic redshifts
from  \citetalias{skelton14}
with redshift estimates derived from their joint analysis of slitless
WFC3/NIR grism spectra from the 3D-HST project and the  \citetalias{skelton14} photometric
catalogues.
These grism data are shallower than those available in the HUDF (see Sect.\thinspace\ref{se:sub:catastrophic_outliers}) but cover a much wider area.
We restrict the use of these  grism-$z$s to
relatively bright galaxies  (NIR magnitude \mbox{$JH_\mathrm{IR}<24$}).
These galaxies were individually inspected by the 3D-HST team,
 allowing us to  select  galaxies
 classified to have
robust redshift estimates.
For these relatively bright galaxies the continuum emission is  comfortably detected in the grism data, yielding high-quality redshift estimates with a typical  redshift error of \mbox{$\sigma_z\approx 0.003\times (1+z)$} \citep{momcheva16}, which we can neglect compared to the photo-$z$ uncertainties.

For the combined sample of galaxies with spec-$z$s and grism-$z$s we compare the colour-selected   histogram of spec-$z$s/grism-$z$s ($z_\mathrm{s/g}$, using $z_\mathrm{s}$ in case both are available) to the histogram of their photo-$z$s in the bottom panels of
Fig.\thinspace\ref{fig:zdist_f814w}.
Here we note two points:
First, the  spec-$z$s/grism-$z$s  confirm that the colour selection indeed provides a very efficient removal of galaxies at our targeted cluster redshifts.
Second, the high-$z$ galaxies are distributed in a relatively symmetric,
unimodal peak that has a maximum at \mbox{$z\simeq 1.9$} according to
spec-$z$s/grism-$z$s. In contrast, the photo-$z$ histogram shows two
slight peaks (\mbox{$z\simeq 1.5$} and \mbox{$z\simeq 2.3$}).
This is consistent with the conclusion from
Sect.\thinspace\ref{se:sub:catastrophic_outliers}
that the peaks in  the photo-$z$ histogram of the full photometric sample (top panels of
Fig.\thinspace\ref{fig:zdist_f814w}) at these redshifts are a result of
redshift focusing effects and not true large-scale structure peaks in the
galaxy distribution.

As a further cross-check
we reconstruct the redshift distribution of the photometric sample by exploiting its spatial cross-correlation with the
 spec-$z$s/grism-$z$ sample, applying the technique developed by
\citet{newman08,schmidt13,menard13}.
Specifically, we use the implementation in
\textsc{The-wiZZ}\footnote{Available at \url{https://github.com/morriscb/The-wiZZ}} redshift recovery code \citep{morrison17}.
We provide the details of this analysis in Appendix \ref{app:candels_x}, showing that it  independently confirms the presence of the catastrophic redshift outliers and redshift focusing effects.

\subsubsection{Limitations of the averaged posterior
probability distribution}
\label{se:posterior}

Past weak lensing studies suggest that a better approximation of the true
source redshift distribution may be given by the average  photometric redshift
posterior
probability distribution  $p(z)$  of all sources compared to a
histogram of the best-fit (or peak) photometric redshifts \citep[see e.g.][]{heymans12,benjamin13,bonnett15}.
To test this we recompute the  $p(z)$   using  \textsc{EAZY} from the
 \citetalias{skelton14} photometric catalogues, which is
necessary as the $p(z)$ are not reported in the  \citetalias{skelton14}
catalogues.

As visible in
 Fig.\thinspace\ref{fig:zdist_f814w},
 the redshift distribution
inferred from the averaged $p(z)$
is relatively
 similar
to the normalised histogram of the
peak photometric redshifts $z_\mathrm{p}$.
We note that the redshift focusing peak at \mbox{$z_\mathrm{p}\simeq 1.5$}
and local minimum at \mbox{$z_\mathrm{p}\simeq 2$}
are slightly less pronounced in the  averaged $p(z)$, but they do not
reach the level suggested by the corrected $z_\mathrm{f}$ histogram.
More severely, the averaged $p(z)$ over predicts the fraction of
low-$z$ galaxies compared to the $z_\mathrm{f}$ distribution similarly to
the $z_\mathrm{p}$ histogram.
We therefore conclude that the use of the averaged $p(z)$ instead of the
$z_\mathrm{p}$ histogram is insufficient to account for the systematic
features identified in
Sect.\thinspace\ref{se:sub:catastrophic_outliers}.

\begin{figure}
 \includegraphics[width=0.99\columnwidth]{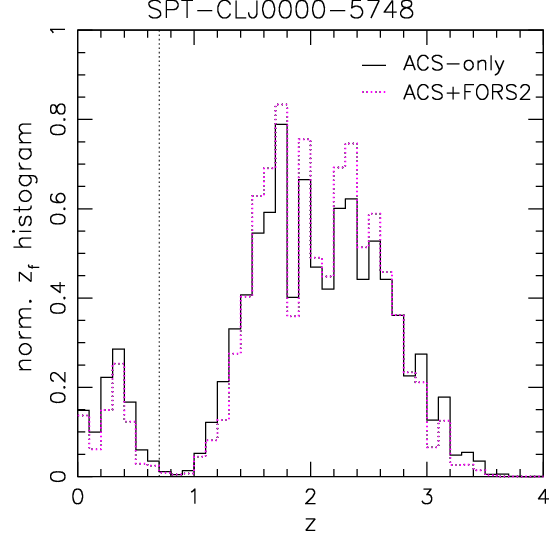}
\caption{
Normalised histogram of the statistically corrected  photometric redshift
estimates $z_\mathrm{f}$
for all galaxies
in our CANDELS catalogues that pass the weak lensing cuts and the colour selection after adding noise to
mimic the properties of the SPT-CL{\thinspace}$J$0000$-$5748 data,
both for the ACS+FORS2 (magenta dotted) and the ACS-only (black solid)
selection.
The vertical dotted line indicates the cluster redshift, at which both selections achieve an efficient suppression also in the presence of noise.
\label{fig:zdist_noisy_minimalistic}}
\end{figure}

\subsection{Source selection in the presence of photometric scatter}
\label{se:source_sel_scatter}
Outside the area of the central F814W ACS tile we only have single band
F606W observations from HST. For the colour selection we therefore have to
combine the F606W  data with the  VLT/FORS2 $I$-band imaging (see Sect.\thinspace\ref{se:data_vlt}).
We measure colours between these
 images as
 described in Appendix \ref{app:details_fors2color}.
 However, VLT/FORS2 $I$-band observations are not available in all CANDELS fields. We therefore need to accurately map the selection in the ACS+FORS2-based $V_\mathrm{606,con}-I_\mathrm{FORS2}$ colour to the \mbox{$V_{606}-I_{814}$} colour
 available in CANDELS.
We empirically obtain this mapping through the comparison of both colour estimates in the inner cluster regions, where both are available (see Appendix
\ref{app:tiecolor}).

As described in Appendix \ref{app:scatter} we  add photometric scatter
to the catalogues from the CANDELS fields to mimic the noise properties of
the cluster fields for the colour selection.
In particular, we apply an empirical model for the (non-Gaussian)
scatter between the ACS-only and the ACS+FORS2 colours derived
from the comparison of the colour measurements in the inner cluster regions.
The ACS-only colour selection has higher signal-to-noise, allowing
us to include galaxies with \mbox{$V_{606}<26.5$} in the analysis.
In contrast, the ACS+FORS2 colour selection is more noisy, which is why we
have to employ shallower magnitude limits (dependent on the depth of the VLT
data, see Table \ref{tab:vltdata}) and
more stringent colour cuts (see Table \ref{tab:app:colourcuts} in Appendix
\ref{app:details_fors2color_scatter}).
 Fig.\thinspace\ref{fig:zdist_noisy_minimalistic} demonstrates that
this approach leads to a robust removal of  galaxies at the cluster redshift
despite the presence of noise.
Here we show the histogram of the statistically corrected redshift estimates $z_\mathrm{f}$
for the CANDELS galaxies passing the colour selection for
SPT-CL{\thinspace}$J$0000$-$5748
 after application of
the photometric scatter.
Averaged over the cluster sample we find that 98.9\% (98.1\%)   of
the CANDELS galaxies with \mbox{$|z_\mathrm{f}-z_\mathrm{l}|\le 0.025$} are
removed by the ACS+FORS2 (ACS-only) colour selection  scheme when the noise
is taken into account.
As shown in  Appendix \ref{se:test_extremely_blue} this translates into a
negligible expected cluster member contamination in the weak lensing analysis.
In addition, we will show in
Sect.\thinspace\ref{se:photo_number_density_tests} that the total source
density and the source density profiles provide limits on the
residual cluster member contamination,
which are consistent with no
contamination.

\subsection{Analysis in magnitude bins}
\label{se:analysis_in_magbins}
\begin{figure}
 \includegraphics[width=1\columnwidth]{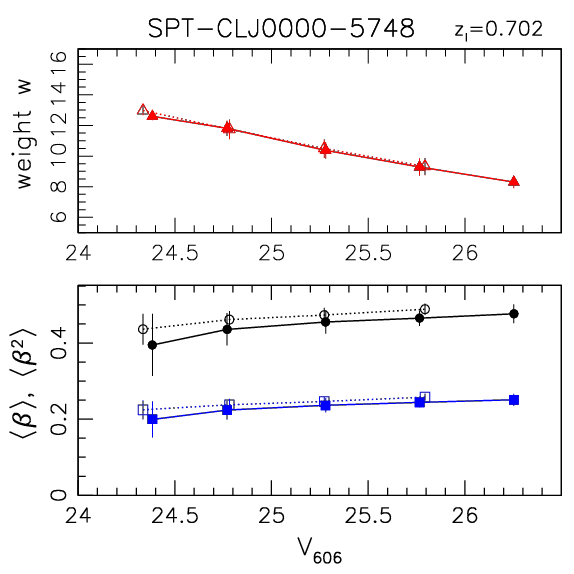}
\caption{Analysis of
SPT-CL{\thinspace}J0000$-$5748
 as function of $V_{606}$
  magnitude, where the solid (open) symbols correspond to the ACS-only (ACS+FORS2) analysis.
{\it Top:} Mean weak lensing shape weight $w$ with error-bars indicating the
dispersion from all selected galaxies in the magnitude bin.
{\it Bottom:}
$\langle\beta \rangle$ (circles) and $\langle\beta^2 \rangle$ (squares) with
 error-bars showing the dispersion of their estimates computed from all CANDELS sub-patches (see Sect.\thinspace\ref{sec:beta_los_variation}), thus indicating the expected line-of-sight variations for the field sizes of our cluster observations.
\label{fig:beta_beta2_w}}
\end{figure}

As shown  in Fig.\thinspace\ref{fig:beta_beta2_w}, $\langle\beta\rangle$
increases moderately within
the magnitude range
\mbox{$24<V_{606}<26.5$}, which is due to a larger fraction of high-redshift
galaxies passing the colour selection at fainter magnitudes.
We only include galaxies with \mbox{$V_{606}>24$} in our analysis
as brighter galaxies contain only a low fraction of background sources.
We split the source galaxies into subsets according to $V_{606}$ magnitude
(0.5\thinspace mag-wide bins) in order to optimise the
$S/N$ of our measurement. This allows us to adequately weight the bins
in the analysis not only accounting for the shape weight $w$, but also the geometric lensing efficiency.

\subsection{Accounting for line-of-sight variations}
\label{sec:beta_los_variation}

There is  statistical
uncertainty on how well we can estimate the cosmic mean $\langle \beta
\rangle$ in a magnitude bin (given our lensing and colour selection) due to
sampling variance and the finite sky-coverage of CANDELS.
Furthermore, the actual redshift distribution along the line-of-sight to
each of our clusters will be randomly sampled from this cosmic mean
distribution, leading to additional statistical scatter, see
e.g.~\citet{hoekstra11}, who show that this is particularly relevant for
high-$z$ clusters.

To account for the  statistical scatter in our weak lensing mass analysis
(Sect.\thinspace\ref{sec:wlmasses}), we subdivide the
CANDELS fields into sub-patches that match the size of our cluster field
observations (single ACS tiles for the ACS-only colour selection and
\mbox{$2\times2$} mosaics for the ACS+FORS2 selection) and compute $\langle \beta
\rangle_i$  and $\langle \beta^2
\rangle_i$ from the  redshift
distribution of each sub-patch $i$.
From the scatter of these quantities between all
sub-patches we compute the resulting scatter in the mass constraints in Sect.\thinspace\ref{sec:profiles}.

Furthermore, we need to investigate if the uncertainty on the estimate of the cosmic mean $\langle \beta
\rangle$ due to the finite sky-coverage of CANDELS adds a significant
systematic uncertainty in our error-budget.
For this, we first compute the uncertainty on the mean  $\langle \beta
\rangle$ from the variance of the  $\langle \beta
\rangle_i$. Assuming all $N$ sub-patches were statistically independent, we find
a very small relative uncertainty   \mbox{$\frac{\Delta \langle \beta
\rangle}{\langle \beta
\rangle}=\sigma_{\langle \beta
\rangle_i}/(\langle \beta
\rangle\sqrt{N-1})=0.3\%$} ($0.6\%$) for our lowest-redshift cluster
SPT-CL{\thinspace}J2331$-$5051 at \mbox{$z_\mathrm{l}=0.576$} and $0.4\%$
($1.1\%$) for our highest-redshift cluster SPT-CL{\thinspace}J2106$-$5844 at \mbox{$z_\mathrm{l}=1.132$} using
the ACS-only (ACS+FORS2) colour selection combining all magnitude bins.
However, due to large-scale structure the $\langle \beta
\rangle_i$ within each CANDELS field will be correlated.
A more conservative estimate can be obtained by computing  $\langle \beta
\rangle_i$ for each CANDELS field (without sub-patches) and estimating  $\frac{\Delta \langle \beta
\rangle}{\langle \beta
\rangle}$ from the variation between the five fields\footnote{Here we
  want to investigate how well we can estimate the cosmic mean redshift
  distribution from CANDELS, for which  sub-patches are not
  needed. The sub-patches are needed to estimate the line-of-sight scatter in
$\langle \beta
\rangle$ between the different cluster fields, as discussed
in the second paragraph of this subsection.}.
This yields \mbox{$\frac{\Delta \langle \beta
\rangle}{\langle \beta
\rangle}=0.6\%$} (0.6\%) for SPT-CL{\thinspace}J2331$-$5051 and $0.4\%$ ($1.0\%$)
 for SPT-CL{\thinspace}J2106$-$5844, again employing the ACS-only (ACS+FORS2)
 colour selection.
This uncertainty is taken into account in our systematic error budget in
Sect.\thinspace\ref{se:systematic_error_budget},
but we note that it is
very small
compared to our statistical errors in all cases.

\subsection{Accounting for magnification}
\label{sec:magnification_model}

\begin{figure*}
  \includegraphics[width=1\columnwidth]{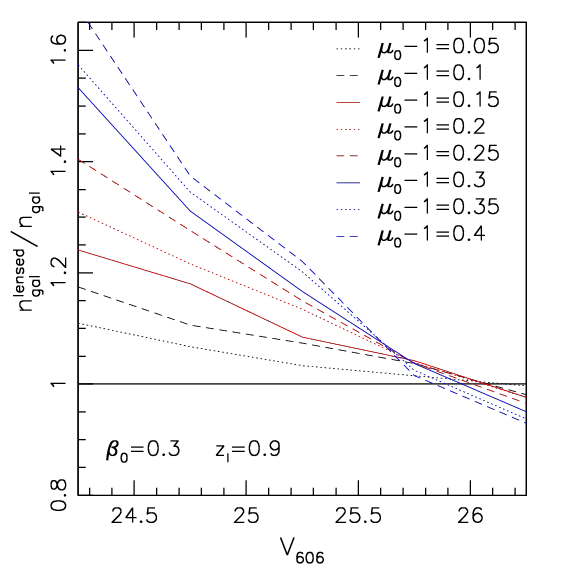}
  \includegraphics[width=1\columnwidth]{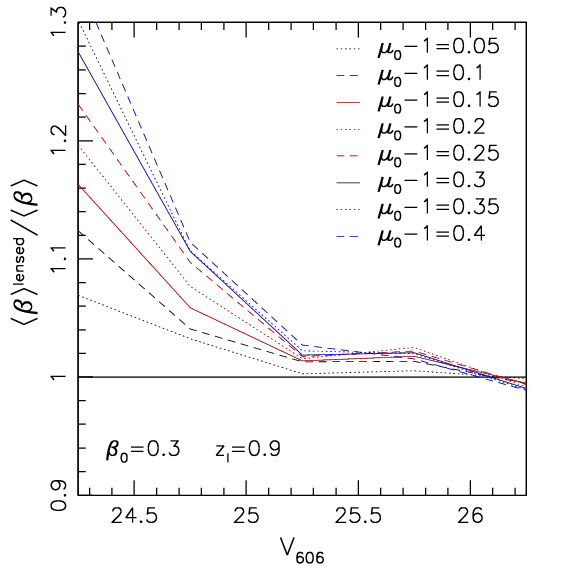}
\caption{Relative change in the source density $n_\mathrm{gal}$ ({\it left}) and
the average geometric lensing efficiency $\langle \beta \rangle$ ({\it right})
for galaxies in the  GOODS-South and GOODS-North fields
as function of the $V_{606}$ aperture magnitude when applying an artificial weak lensing magnification
for \mbox{$z_\mathrm{l}=0.9$}, \mbox{$\beta_0=0.3$}, and a range of $\mu_0$
as indicated in the legend, assuming the linear scaling from Eq.\thinspace(\ref{eq:mu_approximated}).
This analysis uses
the statistically corrected
photometric redshift estimates $z_\mathrm{f}$ for all galaxies in the
\citetalias{skelton14} GOODS-South and GOODS-North catalogues which are located within
the ACS+WFC3 area and pass  our ACS-only colour selection.
\label{fig:magnification}}
\end{figure*}

In addition to the shear, the weak lensing effect of the clusters
magnifies background sources
by a factor $\mu(z)$ given by Eq.\thinspace(\ref{eq:magnification_correct}).
This effectively alters the source redshift distribution, but this effect
has typically been ignored in previous studies.
For our analysis this has three effects: first, it increases the fluxes of sources by a factor $\mu(z)$, which may place them into brighter magnitude bins, thus increasing the total source density by including galaxies which are intrinsically too faint to be included.
Second, it reduces the  source sky area we observe, diluting
the number density of sources by a factor  $\mu(z)$.
Finally, the magnification of object sizes may lead to the inclusion of some small galaxies which would otherwise be excluded by the lensing size cut. However, the large majority of our galaxies are well-resolved with HST, so we will ignore this latter effect (but it may be more relevant for data with lower image quality).

We estimate the impact of the first and second effect
from a colour-selected\footnote{Here we account for the magnitude-dependence of our
  colour cut (see Table \ref{tab:app:colourcuts} in Appendix \ref{app:details_fors2color_scatter}), by basing it on the lensed magnitude.} \citetalias{skelton14} CANDELS catalogue (lensing is
achromatic).
Here we restrict the analysis to  the deeper GOODS fields, initially including
galaxies down to \mbox{$V_{606}=27.5$}.
For this part of the analysis we do not require a matching entry in
our 1 orbit-depth shape catalogue in order to maximise the completeness at
the faint end.
We include the statistical correction for catastrophic
redshift outliers and redshift focusing from
Sect.\thinspace\ref{se:sub:catastrophic_outliers}, where we apply the same scheme
also for one additional
magnitude bin with \mbox{$26.5<V_{606}<27.5$}.
For each cluster redshift we compute $\beta(z_i)$ for each galaxy $i$ (using
\mbox{$z_i=z_{\mathrm{f},i}$})
 in the CANDELS catalogue
and approximate the magnification as
\begin{equation}
\mu(z)-1\simeq \frac{\beta(z)}{\beta_0}\left(\mu_0-1\right) \, ,
\label{eq:mu_approximated}
\end{equation}
where $\mu_0$ indicates the magnification at an arbitrary fiducial $\beta_0$,
for which  we use \mbox{$\beta_0=0.3$} close to the mean  $\beta$
for our higher redshift clusters (compare Table \ref{tab:betastats}).
The
scaling in Eq.\thinspace(\ref{eq:mu_approximated})
is adequate
in the weak lensing limit (\mbox{$|\kappa|\ll 1$},
\mbox{$|\gamma|\ll 1$}),
in which case Eq.\thinspace(\ref{eq:magnification_correct})
simplifies to
\begin{equation}
\mu(z)=\frac{1}{1-2\frac{\beta(z)}{\beta_0}\kappa_0+\left(\frac{\beta(z)}{\beta_0}\right)^2(\kappa_0^2-|\gamma_0|^2)}\simeq
1+2\frac{\beta(z)}{\beta_0}\kappa_0 \, ,
\label{eq:mu_expand}
\end{equation}
where $\kappa_0$ and $\gamma_0$ are the convergence and shear for \mbox{$\beta=\beta_0$}.
In practice we find that the assumed linear scaling with $\beta$ in
Eq.\thinspace (\ref{eq:mu_approximated}) is sufficiently accurate for
all of our clusters within the considered radial range of the tangential reduced shear profile fits
(see Sect.\thinspace\ref{sec:profiles}).

\begin{table*}
\caption{Summary geometric lensing efficiency and source densities.
\label{tab:betastats}}
\begin{center}
\begin{tabular}{cccccc}
\hline\hline
Cluster & $\langle\beta\rangle$ & $\langle\beta^2\rangle$ & $\sigma_{\langle
  \beta \rangle_i}/\langle \beta\rangle$ & \multicolumn{2}{c}{$n_\mathrm{gal} [\mathrm{arcmin^{-2}}]$}\\
& & & & ACS-only & ACS+FORS2\\
\hline
SPT-CL\thinspace$J$0000$-$5748 & 0.466 & 0.243 & 0.053 & 18.2 & 7.2\\
SPT-CL\thinspace$J$0102$-$4915 & 0.374 & 0.163 & 0.068 & 20.4 & 3.6\\
SPT-CL\thinspace$J$0533$-$5005 & 0.368 & 0.159 & 0.062 & 19.7 & 5.4\\
SPT-CL\thinspace$J$0546$-$5345 & 0.303 & 0.107 & 0.083 & 13.1 & 2.9\\
SPT-CL\thinspace$J$0559$-$5249 & 0.505 & 0.288 & 0.064 & 18.2 & 4.0\\
SPT-CL\thinspace$J$0615$-$5746 & 0.334 & 0.132 & 0.075 & 18.0 & 2.3\\
SPT-CL\thinspace$J$2040$-$5725 & 0.344 & 0.141 & 0.077 & 16.2 & 3.5\\
SPT-CL\thinspace$J$2106$-$5844 & 0.282 & 0.093 & 0.087 & 9.2 & 2.0\\
SPT-CL\thinspace$J$2331$-$5051 & 0.522 & 0.308 & 0.059 & 16.2 & 8.3\\
SPT-CL\thinspace$J$2337$-$5942 & 0.425 & 0.205 & 0.059 & 18.3 & 7.6\\
SPT-CL\thinspace$J$2341$-$5119 & 0.320 & 0.122 & 0.067 & 19.1 & 9.3\\
SPT-CL\thinspace$J$2342$-$5411 & 0.300 & 0.105 & 0.082 & 15.8 & 2.5\\
SPT-CL\thinspace$J$2359$-$5009 & 0.423 & 0.204 & 0.055 & 16.6 & 8.7\\
\hline

\end{tabular}
\end{center}
{\flushleft
Note. ---
{\it Column 1:} Cluster designation.
{\it Columns 2--4:}  $\langle\beta\rangle$,  $\langle\beta^2\rangle$, and $\sigma_{\langle
  \beta \rangle_i}/\langle \beta\rangle$ averaged over both colour selection
schemes and all
magnitude bins that are included in the NFW fits according to their
corresponding shape weight sum.
{\it Columns 5--6:} Density of selected sources in the cluster fields for the ACS-only and the
  ACS+FORS2 colour selection schemes, respectively.\\
}
\end{table*}

\begin{figure}
  \includegraphics[width=1.0\columnwidth]{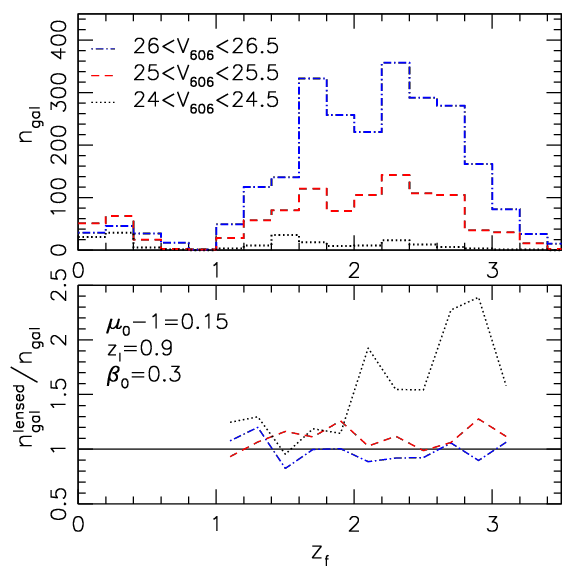}
\caption{{\it Top:} Distribution
 of the statistically corrected photometric redshifts $z_\mathrm{f}$
for galaxies in the
  \citetalias{skelton14} GOODS-South and GOODS-North catalogues located in
  the ACS+WFC3 area when  applying our ACS-only colour selection.
The different  histograms correspond to three different  $V_{606}$ magnitude
bins.
{\it Bottom:} Relative change in those  redshift distributions after application of
weak lensing magnification for a lens at \mbox{$z_l=0.9$} with
\mbox{$\mu_0-1=0.15$}
and \mbox{$\beta_0=0.3$}.
\label{fig:magnification_zdist}}
\end{figure}

For each galaxy in the CANDELS catalogue we compute  $\mu(z_i)$ for a range of $\mu_0$.
We then estimate the magnified magnitude $V_{606,i}^\mathrm{lensed}=V_{606,i}-2.5\log_{10}\mu(z_i)$ for each galaxy,
and keep track of the reduced sky area through a
weight $W_i=1/\mu(z_i)$.
By binning in $V_{606,i}^\mathrm{lensed}$ we then compute the lensed number density
 \begin{equation}
n_\mathrm{gal}^\mathrm{lensed}=\sum_\mathrm{galaxies} W_i/\mathrm{area}
\end{equation}
and the mean lensed geometric lensing efficiency
\begin{equation}
\langle \beta \rangle^\mathrm{lensed}=\sum_\mathrm{galaxies} W_i \beta_i/\sum_\mathrm{galaxies} W_i \, ,
\end{equation}
where the summations are over all galaxies with lensed magnitudes falling into the corresponding bin.
In Fig.\thinspace\ref{fig:magnification} we plot the ratio of these
quantities to their not-lensed counterparts
\mbox{$n_\mathrm{gal}$} and \mbox{$\langle \beta \rangle$} computed in $V_{606}$ bins with uniform weight\footnote{When computing the {\it relative} impact of magnification on the number density and mean lensing efficiency we deliberately do not include the shape weights, as we would  otherwise need to account for the increase in $S/N$ and thus $w$ due to the magnification. Since we perform the full analysis in magnitude bins, with very little variation in $w$ within a bin, our approach constitutes a very good approximation.
}.
This shows that magnification
has only a minor net effect at magnitudes \mbox{$V_{606}\simeq 25.5$--$26$},
which contain a large fraction of our source galaxies.
In contrast, it
significantly boosts both quantities for brighter magnitudes
\mbox{$V_{606}\lesssim 25$}.
The net impact of magnification on high-$z$ cluster mass estimates therefore strongly
depends on the depth of the observations.
For illustration we also
show the redshift distributions within three magnitude bins and their relative
change after applying magnification with
\mbox{$z_\mathrm{l}=0.9$}
and
\mbox{$\mu_0-1=0.15$}
in Fig.\thinspace\ref{fig:magnification_zdist}.

\reva{Previous weak lensing magnification studies have made the simplifying assumptions that
  sources are located at a single redshift and that the source counts can be described
  as a power law. Under these assumptions the ratio of the lensed and non-lensed cumulative source
  densities above a magnitude  $m_\mathrm{cut}$
\begin{equation}
\frac{n(<m_\mathrm{cut})}{n_0(<m_\mathrm{cut})}=\mu^{2.5s-1}
\end{equation}
depends only on the magnification $\mu$ and the slope of the  logarithmic cumulative number counts
\begin{equation}
s=
\frac{\mathrm{d}\log_{10}n(<m)}{\mathrm{d}m}
\end{equation}
\citep[e.g.][]{broadhurst95,chiu16b},
where slopes \mbox{$s>0.4$} (\mbox{$s<0.4$}) lead to a net increase
(decrease) of the counts.
As an illustration we estimate \mbox{$s$} from our colour (\mbox{$V_{606}-I_{814}<0.3$}) and
shape-selected GOODS-South and GOODS-North catalogue, finding that it can
approximately be described as
\begin{equation}
s(V_{606})\simeq +0.88\pm 0.03 -(0.15\pm 0.02)(V_{606}-24)\,
\end{equation}
for
\mbox{$24<V_{606}<26.5$}.
Under these simplifying assumptions we therefore expect a significant boost
in the source density at bright magnitudes (\mbox{$V_{606}\simeq 24$--$25$})
where the slope of the number counts is steep, and only a small boost towards
 fainter magnitudes (\mbox{$V_{606}\simeq 26.5$}), where the slope of the
 number counts is shallower.
This roughly agrees with the more accurate results shown in
Fig.\thinspace\ref{fig:magnification},
but there are noticeable differences, such as the slight net decrease in the
source
density at \mbox{$V_{606}\simeq 26.5$} in Fig.\thinspace\ref{fig:magnification}.
As our sources are not at a single redshift, the simplifying assumptions are
clearly not met, which is why we base our analysis on the more accurate approach
described above.
}

When fitting the reduced cluster shear profiles with NFW models in
Sect.\thinspace\ref{sec:wlmasses}, we compute a $\mu(\langle\beta\rangle_j,r)$
profile for magnitude bin $j$ and a given mass from the
NFW model predictions for both $\kappa(\langle\beta\rangle_j,r)$ and
$\gamma(\langle\beta\rangle_j,r)$ according to
Eq.\thinspace(\ref{eq:magnification_correct}).
Employing Eq.\thinspace(\ref{eq:mu_approximated}) with
\mbox{$\beta=\langle\beta\rangle_j$} we compute the corresponding $(\mu_0-1)(r)$ profile and obtain
radius-dependent corrections $\langle\beta\rangle^\mathrm{lensed}_j/\langle\beta\rangle_j(r)$ and  $n_{\mathrm{gal},j}^\mathrm{lensed}/n_{\mathrm{gal},j}(r)$
by
interpolating our CANDELS-based estimates
(Fig.\thinspace\ref{fig:magnification}) between the  discrete $\mu_0$
values.
The fact that we compute the magnification in the NFW prediction from both
$\kappa$ and $\gamma$ is our primary motivation to conduct the interpolation
in terms of $\mu_0$ and not $\kappa_0$.
This  provides a more accurate correction than if the shear contribution is ignored,
even though we assume the linear scaling in $\beta$ in
Eq.\thinspace(\ref{eq:mu_approximated}) to simplify the CANDELS analysis.

\begin{figure}
  \includegraphics[width=1.0\columnwidth]{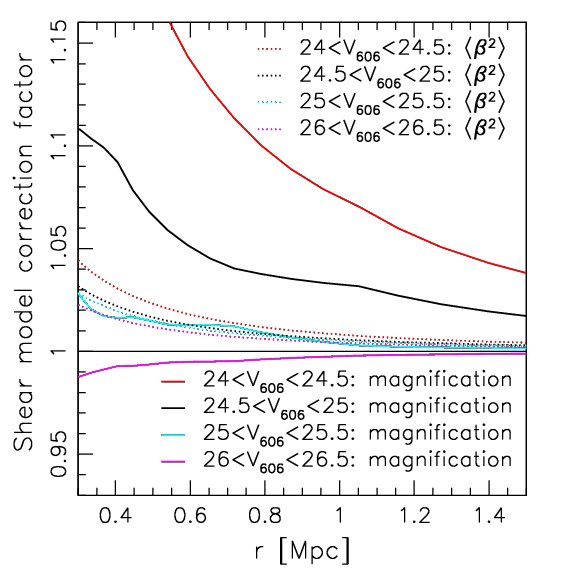}
\caption{Correction factors to the reduced shear profile model of a
  $M_\mathrm{200c}=7\times 10^{14}\mathrm{M}_\odot$ galaxy cluster   at
\mbox{$z_\mathrm{l}=0.87$}
 due to the
  impact of magnification on the source redshift distribution (solid curves)
and the finite width of the redshift distribution ($\langle \beta^2\rangle$, see
Eq.\thinspace\ref{eq:gbeta2corrected}, dotted).
The different colours correspond to different bins in the $V_{606}$ aperture magnitude.
\label{fig:magnification_shear}}
\end{figure}

On average the application of the  correction for magnification-induced changes in the redshift distribution reduces our  estimated cluster masses by 3\%.
This net impact is relatively small since the majority of our sources are at
\mbox{$V_{606}>25$}, requiring small corrections.
Also, we exclude the cluster cores, where the
correction is the largest (see Fig.\thinspace\ref{fig:magnification_shear}),
 from our tangential shear profile fits
(see Sect.\thinspace\ref{sec:wlmasses}).
However, we emphasise that weak lensing studies of high-$z$ clusters using shallower data will be affected more strongly and should adequately model this effect.

We note a subtle limitation of our modelling approach for magnification which
results from our choice
to conduct the analysis as function of aperture magnitude.
Here we ignore the fact that the increase in size due to magnification will lead to a larger fraction of the flux being outside the fixed aperture radius than without magnification.
As a test we also conducted the magnification analysis using aperture-corrected magnitudes
from CANDELS\footnote{The 3D-HST CANDELS catalogue provides aperture magnitudes, which we can directly compare to our measurements, plus an aperture correction based on the $H$-band, which is however not available for our cluster fields.}, finding similar  models as in Fig.\thinspace\ref{fig:magnification} but shifted to brighter magnitudes, with \mbox{$\langle\beta\rangle^\mathrm{lensed}/\langle\beta\rangle$} reaching unity at \mbox{$V_{606}^\mathrm{tot}\simeq 25.0$--$25.5$}.
Given the very minor impact magnification has for our data compared to the statistical uncertainties, the described subtle limitation can safely be ignored for the current study.
In the future this can be avoided by computing aperture corrections in the filter used for shape measurements both for CANDELS and the cluster fields.

\subsection{Number density consistency tests}
\label{se:photo_number_density_tests}
The measurements of the total source density and its
radial dependence can be used to test the cluster member removal and
our procedure to consistently select galaxies in the cluster and  CANDELS
fields in the presence of noise
(Sect.\thinspace\ref{se:source_sel_scatter}).
When computing the source density we account for masks and
apply an approximate correction\footnote{Here we approximate the sky area blocked by a galaxy
  through the $N_\mathrm{pix}$ parameter from \texttt{Source
    Extractor}. \citet{hoekstra15} present a more detailed treatment using
  image simulations, finding that obscuration by cluster members is a
  relatively minor effect for their analysis. Our cluster
  galaxies are  at higher redshift and are thus more strongly dimmed, leading to
  an even smaller impact of obscuration by cluster members.
Our pipeline automatically masks the image region around bright and very
extended galaxies.
With this applied we find
that accounting for the sky area blocked by unmasked brighter galaxies via
the  $N_\mathrm{pix}$ parameter leads to
  $\lesssim 1$ per cent changes in the source density even
for the faintest galaxies considered in our analysis.}
for the impact of
obscuration by cluster members \citep{simet15}.
We also
account for the impact of
cluster magnification,
employing the corresponding radius-dependent magnification model for each
cluster from Sect.\thinspace\ref{sec:magnification_model}.

\subsubsection{Total source density}
\label{se:total_ngal}
\begin{figure}
  \includegraphics[width=1\columnwidth]{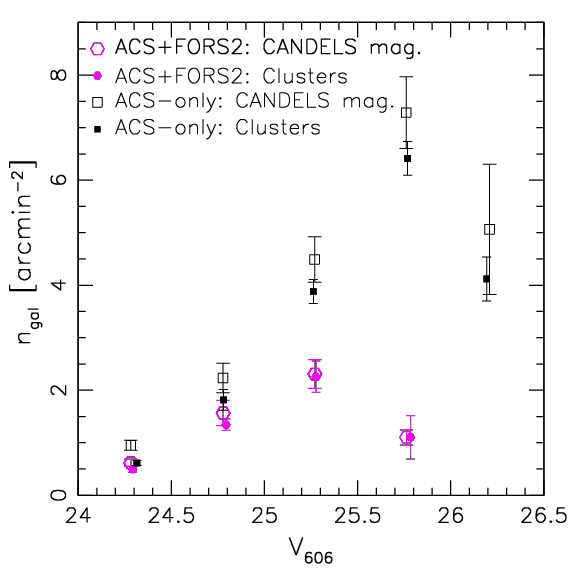}
\caption{Selected source density $n_\mathrm{gal}$ as function of $V_{606}$ accounting for
  masks and averaged over all the cluster fields (small solid
  symbols) and the corresponding  source density averaged over the CANDELS fields
  when mimicking the same selection and accounting for
  photometric scatter and magnification (large open symbols).
Black squares show the ACS-only selection,
  while magenta hexagons correspond to the ACS+FORS2 selection.
We include only galaxies located within the fit range of the shear profiles (see
Sect.\thinspace\ref{sec:profiles})
to avoid limitations of the magnification correction at small radii.
The error-bars show the uncertainty on the mean from the variation between
the contributing cluster fields or the five CANDELS fields, respectively,
assuming Gaussian scatter.
They are correlated between
magnitude bins due to  large-scale structure.
\label{fig:ngal}}
\end{figure}

In Fig.\thinspace\ref{fig:ngal} we compare the average  density of selected
source galaxies in the cluster fields as function
of $V_{606}$ to the corresponding average
 density in the CANDELS
fields corrected for the expected influence of magnification given our best-fit NFW cluster mass models.
There is very good agreement for the
 ACS+FORS2
selection
and reasonable agreement for the ACS-only selection (error-bars are
correlated because of large-scale structure).
Fig.\thinspace\ref{fig:ngal} also
visualises that the ACS-only
analysis (with two ACS bands)
provides a substantially higher average total density of selected sources
in the cluster fields
 of
16.8 galaxies/arcmin$^2$
compared to
 5.2 galaxies/arcmin$^2$
 for the ACS+FORS2 colour selection (see Table
 \ref{tab:betastats} for the total source densities in each field).
This shows that either substantially deeper ground-based imaging or
ACS-based colours for the full imaging area would be required for
the colour selection in order to adequately exploit the full depth of the ACS shape catalogues.

\subsubsection{Source density profile}

\begin{figure}
  \includegraphics[width=1\columnwidth]{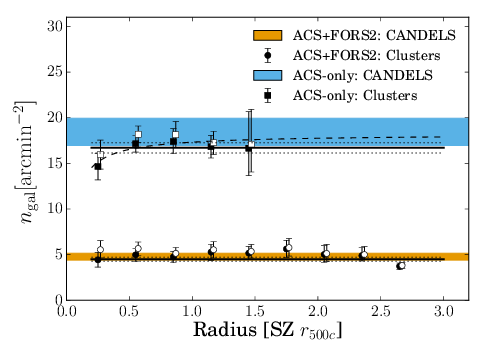}
 \caption{Density of sources $n_\mathrm{gal}$ for the ACS-only and ACS+FORS2
   colour selections as function of cluster
   centric distance around the X-ray centre in units of \mbox{$r_\mathrm{500c}$}, as estimated from
   the SZ signal. The profiles account
   for both masks and  obscuration by cluster members.
Solid (open) symbols include (do not include) the correction for magnification.
 The coloured regions
   indicate the one sigma constraints on the mean background density from
   the five CANDELS fields. Black solid and dotted lines show the maximum likelihood and 68\% uncertainty range for a constant density model. The dashed line shows for the ACS-only selection the maximum likelihood contamination model following a $1/r$ functional form.
\label{fig:ngal_radius}}
\end{figure}

As an additional cross-check for the removal of cluster galaxies and our
magnification model we plot
the radial source density profiles for the ACS-only
and ACS+FORS2 selected samples in Fig.\thinspace\ref{fig:ngal_radius}, averaged over all
clusters, as function  of cluster-centric
distance from the X-ray centre (nearly identical results are obtained when using the
SZ peak location, see Sect.\thinspace\ref{sec:wlmasses}) in units of their corresponding \mbox{$r_\mathrm{500c}$} as
estimated from the SZ signal.
Here we compare the case with and without applying the magnification
correction.
\changeb{
The difference is
small
given that
the magnification is relatively weak for the majority of the
clusters.
Also, most of the source galaxies have faint magnitudes, where the net
impact of magnification is small}
(see Fig.\thinspace\ref{fig:magnification}).
The net difference is
strongest for the inner cluster regions where the magnification is strongest.

\changeb{In the case of a complete cluster galaxy removal and an accurate
  correction for magnification we expect}
to measure
a number density that is consistent with being flat as function of radius.
To test this,  we perform a model comparison test using the cluster sample-averaged number density profiles, assuming that errors were independent and Gaussian distributed.
Each radial bin used in the test is the average of at least three clusters, while uncertainties are determined by bootstrapping the cluster sample.
With these measures, the $\chi^2$ statistic should be a crude yet useful approximation to the true uncertainty distribution, while allowing us to use analytic model quality of fit and comparison tests.

As expected for adequate removal of cluster members and magnification correction, we find that the source
density profiles are consistent with being flat.
For the ACS-only case, the maximum likelihood constant number density model
returns a \mbox{$\chi^2 = 2.54$} with 4 degrees of freedom, while a $1/r$
inverse-r profile with two parameters, the contamination fraction $f_{500}$
at $r_\mathrm{500c}$  and the background number density \citep{hoekstra07},
returns a \mbox{$\chi^2 = 0.74$} with 3 degrees of freedom. Both are
acceptable models at \mbox{$p>0.05$},
\changeb{
where the improvement in $\chi^2$ is consistent with random according to an F-test
(\mbox{$p>0.05$}).
The rather low  $\chi^2$ values
 might be due to our assumption of independent errors between bins, which  neglects the effects of large-scale structure.
For} the ACS+FORS2 selection, the constant number density model returns \mbox{$\chi^2=9.02$} with 8 degrees of freedom, while an exponential model \citep[see Appendix \ref{app:why_not_boost} and][]{applegate14}, which is preferred over the inverse-r model in this case,
returns \mbox{$\chi^2=7.74$} with 7 degrees of freedom, again suggesting that a flat number density model is sufficient (\mbox{$p > 0.05$} from the F-test).
For a general test for the consistency of the combined number density profile being flat we allow for negative  $f_{500}$ \changeb{in these fits}, which could for example
be mimicked by
an incorrect magnification correction.
The maximum likelihood parameter value for the inverse-r contamination model fit to the ACS-only number density profile
\changeb{indeed}
peaks at slightly,
\changeb{but not significantly} negative values, \mbox{$f_{500} = -0.050_{-0.052}^{+0.038}$}.
Fig.\thinspace\ref{fig:ngal_radius} shows the measured number density profiles and maximum-likelihood model fits for both selections.

\section{Weak lensing constraints and mass analysis}
\label{sec:wlmasses}

We reconstruct the projected mass distribution
in our cluster fields in Sect.\thinspace\ref{sec:massmaps}
and constrain the cluster masses
via fits to the
tangential reduced shear profile in Sect.\thinspace\ref{sec:profiles}.
In Sect.\thinspace\ref{se:stack_mc} we compare the stacked shear profiles
from all clusters for the different centres used in the analysis
 and
investigate the consistency of the data with different \reva{concentration--mass} relations.
In Sect.\thinspace\ref{sec:sims} we detail on the simulations used to
calibrate the mass estimates.
We discuss the systematic error budget in Sect.\thinspace\ref{se:systematic_error_budget}.

\subsection{Mass maps}
\label{sec:massmaps}

\begin{figure*}
 \includegraphics[height=7.4cm]{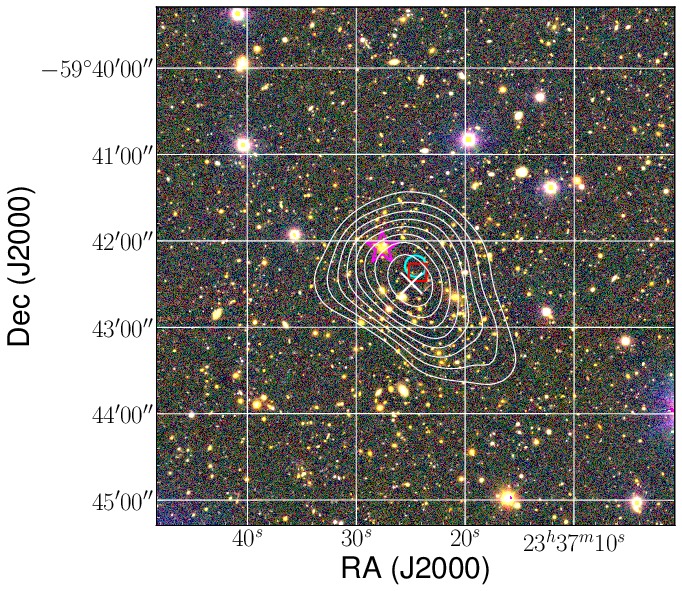}
\includegraphics[height=7.4cm]{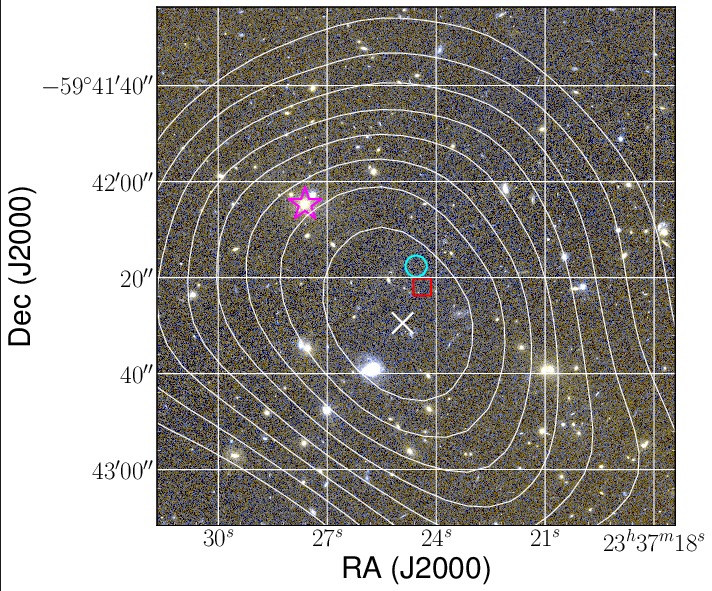}
\caption{Contours of the
signal-to-noise
 map of the Wiener-filtered mass reconstruction for \clusterdemo,
starting at $2\sigma$ in steps of  $0.5\sigma$ with the cross indicating the peak location,
overlaid on
a VLT/FORS2 $BIz$ colour image ({\it left}, \mbox{$6^\prime\times 6^\prime$}),
as well as a  \mbox{$1\farcm8\times 1\farcm8$} cut-out of the ACS imaging
({\it right}, using F606W as blue, F814W as red, and the sum
\mbox{$\mathrm{F606W}+2\times \mathrm{F814W}$} as green channel).
The cyan circle, red square, and magenta star
indicate
the positions of the SZ peak, X-ray centroid, and BCG, respectively.
The corresponding plots of the other clusters are shown in Appendix \ref{se:additional_figs}.
\label{fig:massplotdemo_map}}
\end{figure*}

The weak lensing shear and convergence are linked as they are both
based on second-order derivatives of the lensing potential.
Therefore, a reconstruction of the convergence
field can be obtained from the shear field up to a
constant \citep[][]{kaiser93},
which is the mass-sheet degeneracy \citep{schneider95}.
  Motivated by the different
colour-selected source densities in the inner and outer regions of our
clusters we employ a
Wiener filter for the convergence reconstruction
using  an  implementation described in \citet{mcinnes09} and \citet{simon09}.
This code
estimates the convergence on a grid taking the
spatial variation in the source number density into account; it
applies more smoothing where the number density of sources is lower.
The smoothing in the Wiener filtered map
 employs the shear two-point correlation function $\xi_+(\theta)$
\citep[e.g.][]{schneider06}
as a prior on the angular correlation of the convergence,
which affects the degree of smoothing.
For this, we measure  $\xi_+(\theta)$ in the cluster fields and find that
it is on average approximately described by the fitting function
$\xi_+^\mathrm{fit}(\theta)=0.012(1+\theta/\mathrm{arcmin})^{-2}$.
We fix the mass-sheet degeneracy by setting the average  convergence inside
each cluster  field to zero.
 While
this underestimates the overall convergence for our relatively small
cluster fields, this is irrelevant as we use the reconstructions to
study  positional offsets and the signal-to-noise  ($S/N$)
 ratio of relative mass distributions,
but not to obtain quantitative mass constraints.
To compute the $S/N$ mass maps
 we generate 500 noise maps for each cluster by
randomising the ellipticity phases and repeating the mass reconstruction.
We then define the  $S/N$ mass map as the ratio of the reconstruction from the
actual data and the r.m.s. image of the noise reconstructions.

\begin{table*}
\caption{Locations (\mbox{$\alpha, \delta$}) of the mass map signal-to-noise peaks of the clusters,
  their positional uncertainty (\mbox{$\Delta\alpha, \Delta\delta$}) estimated by bootstrapping the galaxy
  catalogue, and the peak signal-to-noise ratio.
\label{tab:masspeaklocations}}
\begin{center}
\begin{tabular}{lccccccc}
\hline\hline
Cluster & $\alpha$ & $\delta$ & $\Delta\alpha$ & $\Delta\delta$ & $\Delta\alpha$ & $\Delta\delta$ &\mbox{$(S/N)_\mathrm{peak}$}   \\
 &  [deg J2000] &  [deg J2000] & [arcsec] & [arcsec] & [kpc] & [kpc] &\\
 \hline
SPT-CL{\thinspace}$J$0000$-$5748  &   0.25195  & $ -57.80875 $ & 1.9  &  2.2  &  14  &  16  &  5.7 \\
SPT-CL{\thinspace}$J$0102$-$4915  &   15.71743  & $ -49.25458 $ & 7.1  &  7.9  &  55  &  61  &  5.7 \\
SPT-CL{\thinspace}$J$0533$-$5005  &   83.39772  & $ -50.09984 $ & 10.2  &  8.0  &  79  &  62  &  3.0 \\
SPT-CL{\thinspace}$J$0546$-$5345  &   86.65396  & $ -53.75861 $ & 5.1  &  3.7  &  41  &  30  &  3.6 \\
SPT-CL{\thinspace}$J$0559$-$5249  &   89.92875  & $ -52.82297 $ & 4.3  &  3.6  &  29  &  24  &  5.0 \\
SPT-CL{\thinspace}$J$0615$-$5746  &   93.96562  & $ -57.77979 $ & 4.3  &  2.8  &  34  &  23  &  5.1 \\
SPT-CL{\thinspace}$J$2040$-$5725  &   310.06389  & $ -57.42232 $ & 5.0  &  5.1  &  40  &  40  &  3.1 \\
SPT-CL{\thinspace}$J$2106$-$5844  &   316.52210  & $ -58.74336 $ & 7.2  &  4.5  &  59  &  37  &  2.9 \\
SPT-CL{\thinspace}$J$2331$-$5051  &   352.96521  & $ -50.86360 $ & 1.8  &  2.3  &  12  &  15  &  5.1 \\
SPT-CL{\thinspace}$J$2337$-$5942  &   354.35384  & $ -59.70819 $ & 1.8  &  2.5  &  13  &  19  &  6.0 \\
SPT-CL{\thinspace}$J$2341$-$5119  &   355.30057  & $ -51.33015 $ & 5.3  &  5.4  &  42  &  44  &  3.3 \\
SPT-CL{\thinspace}$J$2342$-$5411  &   355.69305  & $ -54.18043 $ & 3.7  &  9.9  &  30  &  80  &  3.1 \\
SPT-CL{\thinspace}$J$2359$-$5009  &   359.93213  & $ -50.16822 $ & 4.7  &  4.8  &  35  &  35  &  5.2 \\
\hline
\end{tabular}
\end{center}
\end{table*}

As an example,
Fig.\thinspace\ref{fig:massplotdemo_map}
shows an overlay of the \mbox{$S/N$} contours of the mass
reconstruction (starting at $2\sigma$ in steps of $0.5\sigma$) for
 \clusterdemo\,
 on a FORS2 $BIz$ colour image (left) as well as a \mbox{$1\farcm8\times 1\farcm8$} cut-out of the ACS imaging (right).
Here we also indicate the locations of the X-ray centroid,
BCG, and SZ peak (see Table \ref{tab:clusters}).
The corresponding figures for the other clusters are shown in Appendix \ref{se:additional_figs}.

For all clusters the weak lensing reconstruction shows a mass peak associated with the cluster, with a peak signal-to-noise ratio \mbox{$(S/N)_\mathrm{peak}$} between $2.9\sigma$ and $6.0\sigma$.
Typically, the mass reconstructions follow the  distribution of red cluster galaxies well, especially for the clusters with \mbox{$(S/N)_\mathrm{peak}\gtrsim 4$}.
Of these clusters, \clustera \,and \clusteri \, show relatively symmetric  mass reconstructions consistent with more relaxed morphologies, while
\clusterb, \clustere, \clusterf, \clusterj, and \clusterm \, show  more elongated or perturbed  morphologies.
In particular, \clusterb\, is known to be an extreme merger \citep{menanteau12}, for which our mass reconstruction separates both main components well \citep[see also the independent weak lensing analysis by][]{jee14}.

\begin{figure}
  \includegraphics[width=1\columnwidth]{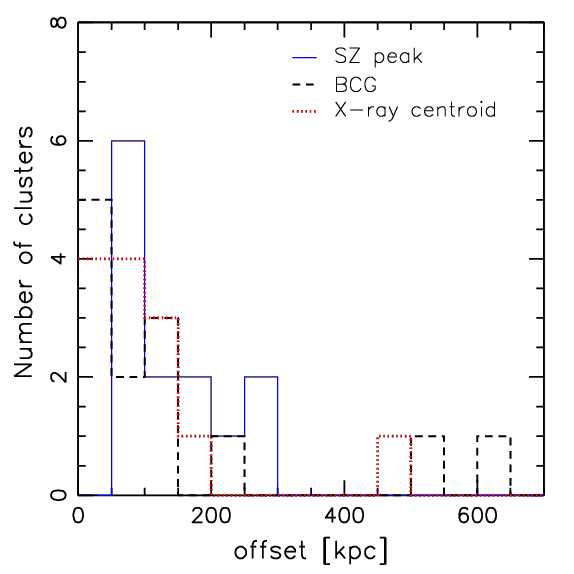}
 \caption{Histograms of spatial offsets between the peak in the mass reconstruction signal-to-noise map and the indicated centres.
\label{fig:massoffset_histo}}
\end{figure}

In the mass signal-to-noise  maps we determine the position of the mass peak of  the corresponding cluster by identifying the pixel with the highest $S/N$ within 90$^{\prime\prime}$ from the SZ peak location.
We report these positions in Table \ref{tab:masspeaklocations} along with estimates of their uncertainty and peak signal-to-noise $(S/N)_\mathrm{peak}$.
The positional uncertainties     are estimated by generating 500 bootstrap samples of the source catalogue for which we repeat the reconstruction and identification of the peak location. The average r.m.s. positional uncertainty (including both directions) for the full sample is 59 kpc.

\cite{dietrich12} investigate the origin of offsets between peaks in weak
lensing mass reconstructions and the projected position of the 3D centre (defined as the minimum of the potential) of
cluster-scale dark matter haloes
in the Millennium Simulation \citep{springel05,hilbert09}.
Without shape noise and smoothing applied in the mass reconstruction they find very
small offsets: their analysis using sources at \mbox{$z=3.06$} is most
similar to the set-up of our study, yielding a 90th percentile offset of
$5.6 h^{-1}$ kpc.
Hence, projection effects and large-scale structure have a negligible impact
for the measured offsets in typical observing scenarios.
\cite{dietrich12} find that smoothing and shape noise increase the offsets substantially,
where the addition of shape noise has the biggest impact
unless unnecessarily large
smoothing kernels are used.
Our bootstrap analysis provides an estimate for the positional uncertainty due to shape
noise.
However, the analysis likely underestimates the true positional uncertainty with respect to the 3D
cluster centre
as it does not explicitly account for the
impact of smoothing.
Nevertheless, we can use the distribution of offsets between the peaks in
the mass signal-to-noise maps
and
different proxies for the cluster centre,
namely
 the X-ray centroid,  SZ peak, and BCG position,
to test if these are similarly good proxies for the true 3D cluster centre position.
Fig.\thinspace\ref{fig:massoffset_histo} shows
a histogram of the corresponding offset distributions.
We also summarise the average, r.m.s., and median of these offset distributions
in Table \ref{tab:offsetsstats}, where
 errors indicate the dispersion of the corresponding values when  bootstrapping the cluster sample.
While the X-ray  centroids yield the smallest average and median offsets,
their r.m.s. offset is similar to the one for the SZ peaks.
The distribution of  offsets between the BCG locations and the mass  signal-to-noise peaks has
the largest r.m.s. offset, resulting from two outliers:
the largest offset  occurs for the merger {\clusterb} (642 kpc),
where  the BCG is located in the
South-Eastern component
 while
the highest signal-to-noise  peak in the mass reconstruction coincides
with the North-Western cluster component, which also shows a strong concentration
of galaxies but has a less bright BCG (see
Fig.\thinspace\ref{fig:massplotb}).
In contrast, both the  SZ peak and the X-ray centroid are located between the
two cluster cores and peaks of the mass reconstruction, resulting in smaller offsets.
{\clusterc} also shows a large (522 kpc) offset between the  BCG
and the mass signal-to-noise  peak, while the latter is broadly consistent with the
SZ peak, X-ray centroid, and strongest galaxy concentration.
This could also be explained with a merger scenario, where a smaller
component hosting a brighter BCG is falling into the main cluster.

\begin{table}
\caption{Average, r.m.s., and median of the offsets [kpc] between the peaks in the mass reconstruction signal-to-noise maps and the
SZ peak, X-ray centroid, and BCG location.
\label{tab:offsetsstats}}
\begin{center}
\begin{tabular}{lccc}
\hline\hline
Centre & average & r.m.s. & median\\
 \hline
SZ peak  & $ 137 \pm 21 $ & $ 158 \pm 23 $ & $ 100 \pm 13   $\\
X-ray centroid  & $ 105 \pm 30 $ & $ 154 \pm 50 $ & $ 64 \pm 44   $\\
BCG location  & $ 159 \pm 52 $ & $ 249 \pm 72 $ & $ 80 \pm 54   $\\
\hline
\end{tabular}
\end{center}
{\flushleft
Note. --- Errors indicate  the  dispersion of the values when  bootstrapping the cluster sample.}
\end{table}

\subsection{Individual shear profile analysis}
\label{sec:profiles}

\begin{figure}
 \includegraphics[width=1\columnwidth]{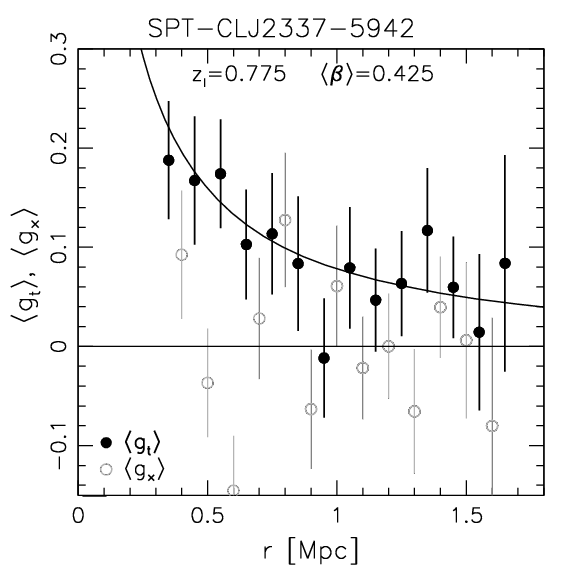}
 \caption{Tangential reduced shear profile (black solid circles) of {\clusterdemo} centred
 on the
X-ray centroid and obtained by combining the profiles of all
 contributing magnitude bins
 of the ACS-only plus the ACS+FORS2 selection
 (see Sect.\thinspace\ref{sec:profiles}). The curve shows the
 correspondingly-combined best-fitting NFW model prediction obtained by fitting the data in the range  \mbox{$500\thinspace \mathrm{kpc}\le r \le 1.5 \thinspace\mathrm{Mpc}$} and using
the \citet{diemer15} \reva{concentration--mass} relation.
The grey open circles indicate the 45 degrees-rotated reduced cross-shear component, which is a test for systematics, shifted by \mbox{$dr=-0.05$\thinspace Mpc}
 for clarity.
The corresponding plots of the other clusters are shown in Appendix \ref{se:additional_figs}.
\label{fig:massplotdemo_profile}}
\end{figure}

We compute profiles of the tangential reduced shear
(Eq.\thinspace\ref{eq:gt}) around the cluster centres in 14
linearly-spaced bins of transverse physical separation between 300\thinspace
kpc and  1.7\thinspace Mpc (100\thinspace kpc-wide bins),
but note that we  restrict the fit range to
 \mbox{$500\thinspace\mathrm{kpc}\le r \le 1.5 \thinspace\mathrm{Mpc}$}
when deriving mass constraints.
Smaller scales are more susceptible to the impact of miscentring, cluster
substructure, uncertainties in the \reva{concentration--mass} relation, and shear
calibration,
while larger scales suffer from an increasingly incomplete  azimuthal
coverage, where 1.5\thinspace Mpc (1.3\thinspace Mpc) equals the largest radius with full azimuthal
coverage at the median (lowest)
redshift of the targeted clusters.
We repeat the analysis for the different proxies for the cluster centre (X-ray
centroid, SZ peak, and BCG position),
but regard the measurements using the
X-ray
centroids as our primary (default) results,
given that they yield the smallest average and median offsets from the peaks
in the mass signal-to-noise maps (Sect.\thinspace\ref{sec:massmaps}).

We compute separate tangential reduced shear profiles for each  magnitude bin and colour selection scheme,
where we use the  ACS-only selection in the inner cluster regions where both ACS bands are available, and the ACS+FORS2 selection in the outer cluster regions.
Each magnitude bin for both colour selection schemes
has a separate value for $\langle \beta \rangle$ and $\langle \beta^2 \rangle$ (see
Sect.\thinspace\ref{se:analysis_in_magbins},
Fig.\thinspace\ref{fig:beta_beta2_w}, and average values reported in Table \ref{tab:betastats}),
which we correct for magnification as a function of cluster centric distance
as described in Sect.\thinspace\ref{sec:magnification_model}.
We fit the profiles from all ACS-only  magnitude bins plus those
ACS+FORS2 bins that have sufficiently low photometric scatter (Sect.\thinspace\ref{se:source_sel_scatter} and Appendix \ref{se:scatter_acsfors2})
jointly with a reduced shear profile model  (see
Eq.\thinspace\ref{eq:gbeta2corrected}) according to \citet{wright00}, assuming spherical mass distributions that follow the
NFW density profile
\citep{navarro97}.
Here we use a  fixed \reva{concentration--mass} (\reva{$c(M)$}) relation,
where we by default employ the \reva{$c(M)$} relation from \citet[][]{diemer15},
but also test the consistency of the data with other relations in Sect.\thinspace\ref{se:stack_mc}.
While the mass distributions in individual clusters may well deviate from an
NFW profile, we account for the net impact on an ensemble of clusters in Sect.\thinspace\ref{sec:sims}.
Due to the fixed \reva{concentration--mass} relation we fit a one-parameter model to each cluster.
Here we perform a $\chi^2$ minimisation using $M_{200\mathrm{c}}$ as free parameter,
comparing the predicted reduced shear values to the measured values in each contributing bin in radius and magnitude.
Given its dominance we employ
a pure shape noise covariance matrix derived from our
 empirical weighting scheme (see Appendix \ref{app:shapes_candels}).
In the fit we also allow for \mbox{$M_{200\mathrm{c}}<0$}, which we model by switching the sign of the tangential reduced shear profile.
For the calibration of scaling relations we make use of the full  likelihood
distribution (see Sect.\thinspace\ref{se:mtx}).
In addition, we identify the maximum likelihood (minimum $\chi^2$) location and the $\Delta\chi^2=1$ points, which we report in  Table \ref{tab:mass}, where
conversions between over-density masses use the assumed \reva{$c(M)$} relation.
Note that the derived mass constraints are expected to be biased due to
effects in the mass modelling such as miscentring, which will be addressed in Sect.\thinspace\ref{sec:sims}.

The statistical errors in
Table \ref{tab:mass}
include two additional minor noise sources.
The first source is given by
 line-of-sight
variations in the redshift distribution, which we estimate from the
dispersion  $\sigma_{\langle
  \beta \rangle_i}$ of the estimates $\langle
  \beta \rangle_i$ from the CANDELS sub-patches (see
  Sect.\thinspace\ref{sec:beta_los_variation}).
In Table \ref{tab:betastats} we report $\sigma_{\langle
  \beta \rangle_i}/\langle \beta\rangle$,
which introduces an additional relative noise in the mass estimates
of  $\sigma_{M,z}/M\simeq 1.5\sigma_{\langle
  \beta \rangle_i}/\langle \beta\rangle$, where \mbox{$M\in
  \left\{M_{200\mathrm{c}}, M_{500\mathrm{c}}\right\}$}.
Further statistical noise is added by projections of uncorrelated
large-scale structure \citep{hoekstra01}.
To estimate it we compute 500 random realisations of the cosmic shear field
per cluster for our reference cosmology and the colour-selected source redshift distribution  as detailed in Appendix B of \citet{simon12},
with the non-linear matter power spectrum estimated following
\citet{takahashi12}\footnote{This approach generates Gaussian random shear
  fields based on the matter power spectrum.
Comparing the resulting scatter in cluster
  mass estimates,
  \citet{hoekstra11} show that approaches using the
 shear power
  spectrum provide good approximations
to more accurate estimates from a  ray-tracing analysis
  through the Millennium Simulation \citep{springel05,hilbert09}.}.
We add these cosmic shear field realisations to the measured shear field in the  SPT cluster fields and repeat the cluster mass analysis for each realisation. The dispersion in the best-fit mass estimates then yields an estimate for the   large-scale structure noise.
We find that it  amounts to
\mbox{$30-50\%$} of the statistical errors from
shape noise.
\reva{Additional scatter between profile-fitted weak lensing mass estimates and
  halo masses defined via  spherical overdensities is caused by  halo triaxiality, variations in cluster density profiles, and correlated large-scale
  structure \citep[e.g.][]{gruen15,umetsu16}.
This scatter typically amounts to
  \mbox{$\sim 20\%$} for massive clusters \citep{becker11} and is not
  explicitly listed in Table \ref{tab:mass}.
Instead, we absorb it in the intrinsic
scatter accounted for in the scaling relation analysis (see
Sect.\thinspace\ref{se:mtx} and Dietrich et al.~in prep.).
}

\begin{table*}
\caption{Weak lensing mass constraints from the NFW fits to the reduced shear
  profiles
using
scales
  \mbox{$500\mathrm{\,kpc}<r<1.5$ Mpc} and the \citet{diemer15} \reva{$c(M)$} relation for two different over-densities
  \mbox{$x \in \{200\mathrm{c}, 500\mathrm{c}\}$}.
Columns 2--5 correspond to the default analysis  centring around
the X-ray centroids, while columns 6--9 list results for the analysis
centring around the SZ peaks.
$M_{x}^\mathrm{biased,ML}$ are the maximum likelihood mass estimates
\reva{in $10^{14}\mathrm{M}_\odot$} {\it without} bias correction applied. All errors are statistical 68\%
uncertainties, listing the contributions from shape noise (asymmetric errors),
uncorrelated
large-scale, and  line-of-sight variations in the redshift
distribution.
Systematic uncertainties are listed in Table \ref{tab:sys}.
The
factor
$b_{x}$ indicates the expected mass bias factor for the
scaling relation analysis when the full likelihood distribution of the mass
constraints is used.
\label{tab:mass}}
\tabcolsep=0.13cm
\begin{center}
\begin{tabular}{crrrrrrrrrrr}
\hline\hline
& \multicolumn{4}{c}{X-ray centres} & \multicolumn{4}{c}{SZ centres}\\
Cluster&  \multicolumn{1}{c}{$M_{200\mathrm{c}}^\mathrm{biased,ML}$} &
$b_{200\mathrm{c}}$ &  \multicolumn{1}{c}{$M_{500\mathrm{c}}^\mathrm{biased,ML}$} &
$b_{500\mathrm{c}}$
&  \multicolumn{1}{c}{$M_{200\mathrm{c}}^\mathrm{biased,ML}$} &
$b_{200\mathrm{c}}$ &  \multicolumn{1}{c}{$M_{500\mathrm{c}}^\mathrm{biased,ML}$} &
$b_{500\mathrm{c}}$
\\
\hline
SPT-CL\thinspace$J$0000$-$5748   & $6.2_{-2.4}^{+2.6} \pm 1.1 \pm 0.5$ & 0.91    & $4.2_{-1.6}^{+1.8} \pm 0.7 \pm 0.3$ & 0.88    & $6.5_{-2.5}^{+2.6} \pm 1.1 \pm 0.5$ & 0.80      & $4.5_{-1.7}^{+1.8} \pm 0.7 \pm 0.3$ & 0.82 \\
SPT-CL\thinspace$J$0102$-$4915   & $11.1_{-2.8}^{+2.9} \pm 1.2 \pm 1.1$ & 0.86   & $7.9_{-2.1}^{+2.2} \pm 0.9 \pm 0.8$ & 0.88    & $14.4_{-2.8}^{+2.8} \pm 1.2 \pm 1.5$ & 0.79     & $10.3_{-2.1}^{+2.1} \pm 0.9 \pm 1.1$ & 0.79 \\
SPT-CL\thinspace$J$0533$-$5005   & $4.3_{-2.4}^{+2.7} \pm 1.0 \pm 0.4$ & 0.88    & $2.9_{-1.6}^{+1.9} \pm 0.7 \pm 0.3$ & 0.87    & $2.4_{-1.8}^{+2.4} \pm 1.0 \pm 0.2$ & 0.80      & $1.6_{-1.2}^{+1.7} \pm 0.7 \pm 0.1$ & 0.81 \\
SPT-CL\thinspace$J$0546$-$5345   & $5.4_{-3.3}^{+3.7} \pm 1.1 \pm 0.7$ & 0.86    & $3.7_{-2.3}^{+2.6} \pm 0.8 \pm 0.5$ & 0.85    & $2.6_{-2.4}^{+3.5} \pm 1.1 \pm 0.3$ & 0.72      & $1.8_{-1.6}^{+2.4} \pm 0.8 \pm 0.2$ & 0.73 \\
SPT-CL\thinspace$J$0559$-$5249   & $8.0_{-2.9}^{+3.1} \pm 1.0 \pm 0.8$ & 0.79    & $5.4_{-2.0}^{+2.2} \pm 0.7 \pm 0.5$ & 0.81    & $4.7_{-2.5}^{+2.9} \pm 1.0 \pm 0.5$ & 0.84      & $3.2_{-1.7}^{+2.0} \pm 0.7 \pm 0.3$ & 0.85 \\
SPT-CL\thinspace$J$0615$-$5746   & $6.8_{-2.6}^{+2.9} \pm 1.0 \pm 0.8$ & 0.88    & $4.7_{-1.8}^{+2.0} \pm 0.7 \pm 0.5$ & 0.85    & $5.8_{-2.5}^{+2.8} \pm 1.0 \pm 0.7$ & 0.82      & $3.9_{-1.7}^{+1.9} \pm 0.7 \pm 0.4$ & 0.80 \\
SPT-CL\thinspace$J$2040$-$5726   & $2.1_{-1.9}^{+2.9} \pm 0.8 \pm 0.2$ & 0.87    & $1.4_{-1.3}^{+2.0} \pm 0.6 \pm 0.2$ & 0.81    & $2.1_{-2.0}^{+2.9} \pm 0.8 \pm 0.2$ & 0.80      & $1.4_{-1.3}^{+2.0} \pm 0.6 \pm 0.2$ & 0.80 \\
SPT-CL\thinspace$J$2106$-$5844   & $8.8_{-4.6}^{+5.0} \pm 1.5 \pm 1.1$ & 0.85    & $6.1_{-3.3}^{+3.7} \pm 1.1 \pm 0.8$ & 0.86    & $8.2_{-4.3}^{+5.0} \pm 1.5 \pm 1.1$ & 0.81      & $5.7_{-3.1}^{+3.6} \pm 1.1 \pm 0.7$ & 0.78 \\
SPT-CL\thinspace$J$2331$-$5051   & $3.8_{-2.1}^{+2.5} \pm 1.1 \pm 0.3$ & 0.85    & $2.6_{-1.4}^{+1.7} \pm 0.7 \pm 0.2$ & 0.92    & $4.0_{-2.1}^{+2.5} \pm 1.1 \pm 0.4$ & 0.85      & $2.7_{-1.4}^{+1.7} \pm 0.7 \pm 0.2$ & 0.87 \\
SPT-CL\thinspace$J$2337$-$5942   & $10.5_{-2.8}^{+2.9} \pm 1.3 \pm 0.9$ & 0.88   & $7.2_{-2.0}^{+2.1} \pm 0.9 \pm 0.7$ & 0.91    & $10.0_{-2.8}^{+2.9} \pm 1.3 \pm 0.9$ & 0.82     & $6.9_{-2.0}^{+2.1} \pm 0.9 \pm 0.6$ & 0.83 \\
SPT-CL\thinspace$J$2341$-$5119   & $2.4_{-1.9}^{+2.5} \pm 1.1 \pm 0.2$ & 0.91    & $1.6_{-1.3}^{+1.7} \pm 0.7 \pm 0.2$ & 0.89    & $2.3_{-1.8}^{+2.5} \pm 1.1 \pm 0.2$ & 0.80      & $1.5_{-1.2}^{+1.7} \pm 0.7 \pm 0.1$ & 0.80 \\
SPT-CL\thinspace$J$2342$-$5411   & $8.6_{-3.5}^{+3.8} \pm 1.4 \pm 1.1$ & 0.87    & $6.0_{-2.5}^{+2.8} \pm 1.0 \pm 0.7$ & 0.84    & $7.0_{-3.4}^{+3.7} \pm 1.4 \pm 0.9$ & 0.79      & $4.8_{-2.4}^{+2.7} \pm 1.0 \pm 0.6$ & 0.81 \\
SPT-CL\thinspace$J$2359$-$5009   & $5.0_{-2.6}^{+3.0} \pm 1.1 \pm 0.4$ & 0.91    & $3.4_{-1.8}^{+2.1} \pm 0.8 \pm 0.3$ & 0.92    & $5.7_{-2.5}^{+2.8} \pm 1.1 \pm 0.5$ & 0.83      & $3.9_{-1.7}^{+1.9} \pm 0.8 \pm 0.3$ & 0.84 \\
\hline
\end{tabular}
\end{center}
\end{table*}

For visualisation we show profiles in
Figs.\thinspace\ref{fig:massplotdemo_profile} and \ref{fig:massplota} to
\ref{fig:massplotm}, where we have combined shear estimates
from the different magnitude bins and colour selections for the analysis
using the
X-ray centroids as centres.
Here we stack all profiles scaled to the same average $\langle \beta \rangle$ of all magnitude bins of the cluster
as
\begin{equation}
\langle g_\mathrm{t} \rangle_\mathrm{comb}(r_k)=\sum_{j \in \mathrm{mag \, bins}}\langle g_\mathrm{t} \rangle_j(r_k)\frac{\langle \beta \rangle}{\langle \beta \rangle_j}W_{j,k}/\sum_{j \in \mathrm{mag \, bins}} W_{j,k} \, ,
\end{equation}
where $k$ indicates the radial bin, $j$ the magnitude bin and colour
selection scheme, and $W_{j,k}=( {\langle \beta \rangle_j}/{\langle \beta
  \rangle})^{2}\sum w_i$ is the rescaled sum of the shape weights of the
contributing galaxies.

\begin{figure}
  \includegraphics[width=1\columnwidth]{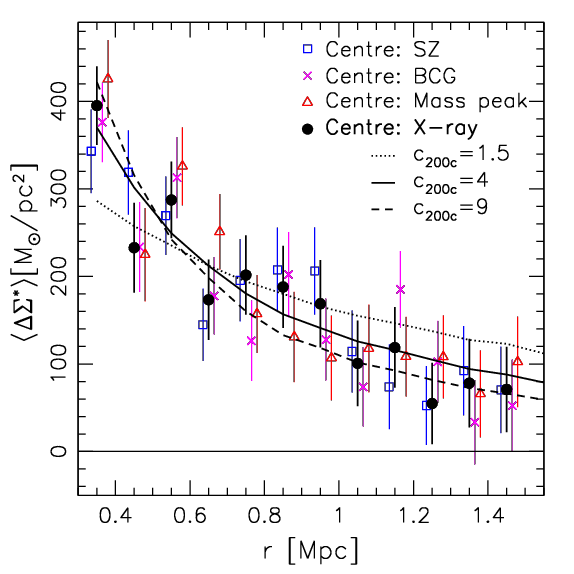}
  \includegraphics[width=0.95\columnwidth]{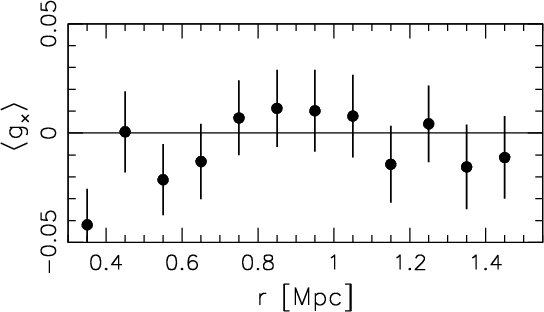}
\caption{\reva{{\it Top}:} Weighted average of the rescaled differential surface mass density
  profiles  from all clusters. The circles, squares, crosses, and triangles
  show the signal measured around the X-ray centroids,  the  SZ peak
  positions, the BCG locations, and the weak lensing mass peaks,
  respectively. The circles showing the signal around the  X-ray centroids are
  displayed at the correct radius, while the other symbols are shown
  with a horizontal  offset for clarity. The curves show the correspondingly
  averaged best-fit model predictions for different fixed concentrations
  for the analysis employing the   X-ray centres and using an extended fit
  range 300\thinspace kpc to 1.5\thinspace Mpc, which increases the
  sensitivity for constraints on the average concentration.
\reva{{\it Bottom}: Profile of the stacked reduced cross-shear component of all
  clusters measured with respect to their X-ray centres.}
\label{fig:shear_profile_stack}}
\end{figure}

\subsection{Stacked signal and constraints on the average cluster concentration}
\label{se:stack_mc}
Miscentring  reduces the shear signal at small radii.
To test if our data
show signs for this, we compare the stacked signal for the different centres
(\reva{top panel of} Fig.\thinspace\ref{fig:shear_profile_stack}).
To stack the signal from clusters at different redshifts and lensing
efficiencies
we employ the differential surface mass density
$\Delta \Sigma(r)$ (see Eq.\thinspace\ref{eq:deltaSigma}), where we
compute $\Sigma_\mathrm{crit}$ using the $\langle\beta\rangle$ of the corresponding magnitude bin and colour selection scheme.
Our clusters span a significant range in mass.
Here we expect higher $\Delta \Sigma(r)$ profiles for the more massive clusters.
Before stacking, we therefore scale them to approximately the same signal amplitude.
For this we compute a theoretical NFW model for the differential surface
mass density $\Delta\Sigma_\mathrm{model}$
for each cluster assuming its mass inferred from the SZ signal
$M_{500\mathrm{c,SZ}}$ \citep{bleem15}\footnote{We weight according to the SZ mass and not
  the lensing-inferred mass. The latter is more noisy and would  give higher
  weight to clusters for which the lensing mass estimate scatters up.} and
a fixed \mbox{$c_\mathrm{200c}=4$},
 and then scale the cluster signal as
\begin{equation}
\Delta \Sigma^*(r)=s \Delta \Sigma(r)\equiv \frac{\langle
  \Delta\Sigma_\mathrm{model}(800\mathrm{kpc})\rangle}{\Delta\Sigma_\mathrm{model}(800\mathrm{kpc})} \Delta \Sigma(r)
\,.
\end{equation}
We evaluate the theoretical model at an intermediate scale
$r=800\mathrm{kpc}$,
but note that the exact choice is not important as we are only interested in
an approximate rescaling to optimise the weighting.
We then compute the weighted average
\begin{equation}
\langle \Delta \Sigma^*(r_j) \rangle = \sum_{i\in
  \mathrm{clusters}} \Delta \Sigma^*_i(r_j)\hat{W_{ij}}/ \sum_{i\in
  \mathrm{clusters}} \hat{W_{ij}} \,,
\end{equation}
with $\hat{W_{ij}}=\left(s\sigma(\Delta \Sigma(r_j))\right)^{-2}$ and
$\sigma(\Delta \Sigma(r_j))$ indicating the $1\sigma$ uncertainty of $\Delta \Sigma(r_j)$.

The results are shown in \reva{in the top panel of} Fig.\thinspace\ref{fig:shear_profile_stack}.
We first note that the stacked profiles are fairly similar for the different
centre definitions.
This is also the case for an analysis centred on the peaks in the weak lensing
mass reconstruction.
Such an analysis should not suffer from miscentring, but is
rather expected to deliver shear estimates that are biased high \citep[see e.g.][]{dietrich12}.
The similarity of the shear profiles  suggests that, for the sample as a whole, miscentring appears to have
relatively minor impact at the radial scales considered in our analysis.

We also fit the reduced shear profiles of all clusters using models with different fixed concentrations.
For three of these fixed concentrations
and the analysis using the X-ray
centres
we show the averaged best-fit models from all clusters in Fig.\thinspace\ref{fig:shear_profile_stack}, using the same scale factors and weights as used for the data.
In these fits we use an extended fit range  300\thinspace kpc to 1.5\thinspace
Mpc to increase the sensitivity of the data for constraints on the
concentration, which are mostly derived from the change in the slope between small
and large radii (compare Fig.\thinspace\ref{fig:shear_profile_stack}).
 Adding the $\chi^2$ from the individual  clusters with equal weights
we compute the total  $\chi^2_\mathrm{tot}$ of the sample
as function of the  fixed concentration,
allowing us to place constraints on the average  concentration of the
sample\footnote{An alternative approach to constrain cluster concentrations
  from weak lensing data is to fit both mass and concentration
  simultaneously for each cluster.
These individual constraints are however very weak due to shape noise, and
they are strongly affected by large-scale structure projections \citep[e.g.][]{hoekstra03}.}
 to
\mbox{$c_\mathrm{200c}=5.6^{+3.7}_{-1.8}$} using the X-ray centres (\mbox{$\chi^2_\mathrm{tot}/\mathrm{d.o.f.}=747.3/744$}),
\mbox{$c_\mathrm{200c}=4.9^{+3.1}_{-1.7}$} using the SZ centres (\mbox{$\chi^2_\mathrm{tot}/\mathrm{d.o.f.}=754.6/712$}),
\mbox{$c_\mathrm{200c}=5.5^{+3.5}_{-1.8}$} using the BCG centres (\mbox{$\chi^2_\mathrm{tot}/\mathrm{d.o.f.}=754.2/774$}),
and
\mbox{$c_\mathrm{200c}=6.5^{+3.6}_{-2.2}$} (\mbox{$\chi^2_\mathrm{tot}/\mathrm{d.o.f.}=749.2/752$}) when centring on the weak lensing
mass peaks.
We stress that the fitting was conducted for each cluster separately (see Sect.\thinspace\ref{sec:profiles}),
and that the stacked signal shown in Fig.\thinspace\ref{fig:shear_profile_stack} is for illustrative
purposes only.
This is important given that the scaling is only approximate, while the
individual analyses account for all effects (e.g. reduced shear).

Due to miscentring the  estimates using the X-ray, SZ, and BCG centres may
be slightly biased low, while the estimate based on the mass peak centre  is likely biased high.
Given that all constraints are well consistent within the uncertainties, we
conclude that miscentring has a negligible impact for the constraints on
concentration at the current statistical precision.
These estimates are consistent with predictions from recent numerical
simulations.
In particular,  the \reva{$c(M)$} relation from  \citet{diemer15}, which corresponds
to our default analysis, yields average
concentrations
\mbox{$3.5\lesssim c_\mathrm{200c}\lesssim 4.6$} (average 3.8) in our mass and redshift
range, fully consistent with our constraints.
Accordingly, it is not surprising that it
provides
similarly good fits to the data as the best-fit fixed concentrations, e.g. we obtain \mbox{$\chi^2_\mathrm{tot}/\mathrm{d.o.f.}=748.8/745$} with the \citet{diemer15}  \reva{$c(M)$} relation for the analysis using the X-ray centres.
For comparison, the  \reva{$c(M)$} relation from \citet{duffy08} yields lower average
concentrations \mbox{$2.4\lesssim c_\mathrm{200c}\lesssim 3.0$} (average
2.7) in our mass and redshift
range, which agrees with  our constraints at the \mbox{$\sim 2\sigma$} level only.

\reva{In the bottom panel of Fig.\thinspace\ref{fig:shear_profile_stack} we
  additionally show the stacked profile of the  reduced cross-shear component of all
  clusters measured with respect to their X-ray centres (computed without
  rescaling). We find that it is
consistent with zero, providing
  another consistency check for our analysis.}

\subsection{Calibration of the mass estimates with simulations and consistency checks in the data}
\label{sec:sims}

We have adopted a simplistic model for the mass distribution in clusters, namely a
spherical NFW halo with a known centre and a concentration fixed by a \reva{concentration--mass} relation.
However, effects such as choosing an improper cluster centre (``miscentring''), variations in cluster density profiles, and noise bias in statistical estimators can introduce
substantial biases in the mass constraints derived from fits of such a  model to cluster
weak lensing shear profiles \citep[e.g.][]{becker11,gruen15}.
To estimate and correct for these biases in our
analysis we apply our measurement procedure to a large sets of  simulated
cluster weak lensing data
based on the Millennium XXL simulation \citep{angulo12}
and the simulations created by \citet[][henceforth  \citetalias{becker11}]{becker11}.
The details of this analysis will be presented in Applegate et al.~(in
prep.). Here we only summarise the most important points relevant to this analysis.

\subsubsection{Simulations}

The two simulations considered for our calibration differ in the redshifts
of the available snapshots, in the cluster mass range, and the input
cosmology.
The difference in cosmology alters the \reva{concentration--mass} relation in the
simulation \citep[e.g.][]{diemer15}, but this is small compared to the range
of  \reva{$c(M)$} relations we consider
(Sect.\thinspace\ref{se:shear_profile_compare_mc}).
Likewise, the calibration
does not depend on mass to a level that is important for this analysis when the full
likelihood distribution of the mass constraints is used (see Applegate et al.~in prep.).
We find that the bias has some dependence on redshift and therefore
interpolate between the two available snapshots that match our observations
best.
For the calibration of both $M_{200\mathrm{c}}$ and $M_{500\mathrm{c}}$ these are the \mbox{$z=0.5$} snapshot of the \citetalias{becker11} simulation
and the \mbox{$z=1$} snapshot of the  Millennium XXL simulation.
Note that both simulations yield consistent bias
calibrations at \mbox{$z=0.25$}, where data are available from both
simulations (see Applegate et al.~in prep.).

For the \citetalias{becker11} \mbox{$z=0.5$} snapshot we include 788 haloes with \mbox{$M_\mathrm{500c}>1.5  \times
   10^{14}\mathrm{M}_\odot/h$}, providing a good match to the SPT cluster mass range.
Since the sample provided to us by \citetalias{becker11} is selected in \mbox{$M_\mathrm{500c}$}, we are only able to measure the bias in \mbox{$M_\mathrm{200c}$} at \mbox{$M_\mathrm{500c} > 4 \times 10^{14}\mathrm{M}_\odot/h$}, above which we are still complete.
For the MXXL \mbox{$z=1$} snapshot we include the 2100 most massive haloes, corresponding to \mbox{$M_\mathrm{200c}>3.5  \times
   10^{14}\mathrm{M}_\odot/h$}.
For the calibration of weak lensing estimates for \mbox{$M_\mathrm{500c}$} this  sample is
complete for
  \mbox{$M_\mathrm{500c}\gtrsim 3.2  \times
   10^{14}\mathrm{M}_\odot/h$},
matching the mass range of the studied SPT clusters well
(compare to Table \ref{tab:clusters}).

The generation of simulated shear fields from the underlying N-body simulations is described in BK11.
In short, all particles within \mbox{$400 h^{-1}$} Mpc along the line-of-sight to each cluster are projected onto a common plane to produce a $\kappa$ map, from which a fast Fourier transform can compute the shear field on a regular grid.
The procedure is similar for MXXL, except that particles within \mbox{$200
  h^{-1}$} Mpc are used, and three orthogonal projection directions are employed.

We create mock observations matching each cluster in our observed sample.
We first select a profile centre location by randomly choosing an offset
from the true cluster centre, which is defined as the position of the
most-bound particle in the simulation, according to different probability
distributions reflecting our assumptions on the miscentring distributions
of SZ and X-ray centres (Sect.\thinspace\ref{se:miscentring_distributions}).
We then bin and azimuthally average the simulated reduced shear grid, matching the binning in the observed shear profile, and add Gaussian random noise to each bin matching the observed noise levels.
We fit the cluster masses from these simulated weak lensing data
as done for the real clusters, calculating scans of $\chi^2$ versus $M_{\mathrm{meas}}$.
To obtain a bias calibration for the scaling relation analysis (see Sect.\thinspace\ref{se:mtx})
we model the ratio $M_{\mathrm{meas}}/M_{\mathrm{true}}$,
where $M_{\mathrm{true}}$ denotes the corresponding halo mass,
 as a log-normal distribution.
 We associate the mean of the log-normal distribution as the inferred average bias and the width of the distribution as the intrinsic scatter from cluster triaxiality, substructure, and line-of-sight projections.
We fit the log-normal distribution to the population of clusters in each
snapshot, marginalising over the statistical uncertainty for each cluster
(see Applegate et al.\, in prep).
While we perform the analysis in bins of true halo mass to check for mass-dependence of the bias, we instead only use one all-encompassing mass bin to determine the bias for this analysis.

We repeat the whole procedure for a number of miscentring distributions and  \reva{$c(M)$} relations.
We list individual bias numbers for each cluster
for the
X-ray and SZ miscentring distributions and the \citet{diemer15}  \reva{$c(M)$} relation
in Table \ref{tab:mass},
and sample-averaged values for a
number of  configurations in
Table~\ref{tab:bias}.
We stress that the quoted  bias numbers are adequate for quantitative analyses that take the full likelihood distribution of the mass constraints into account,  as done in our scaling relation analysis presented in Sect.\thinspace\ref{se:mtx}.
We correct the mass estimates as
\begin{equation}
M_x^\mathrm{WL}=\frac{M_x^\mathrm{biased}}{b_x} \, .
\end{equation}
As an approximation we also apply these bias
correction factors to the maximum likelihood values and confidence
intervals indicated in Figures \ref{fig:mass_corrected_comparison_centres} to \ref{fig:m_tx_sz} in the following sections.
However, note that the  bias factors may differ at some level for the  maximum
likelihood estimates and the fits that use the
full likelihood distribution due to differences in the impact of noise bias.
We plan to investigate this issue further in Applegate et al.~(in prep.).

\subsubsection{Miscentring distributions}
\label{se:miscentring_distributions}

For the SPT clusters we have proxies for the cluster centres, where we in
particular use the  X-ray centroids and SZ peaks for the mass analysis.
These need to be related to the cluster centres defined by halo finding
algorithms used to predict the cluster mass function from simulations.
These offsets will typically
lower the measured shear from the expected NFW signal at small radii \citep[e.g.][]{johnston07,george12}.
To mimic this effect in the BK11 and MXXL N-body simulations, where we have neither mock SZ nor mock X-ray observations, we employ
offset distributions derived from
the Magneticum Pathfinder Simulation (\citealt{dolag16}; see
also \citealt{bocquet16}),
which is a
  large volume, high-resolution cosmological hydrodynamical simulation.
It includes simulated SZ and X-ray observations,
where we make use of
SPT mock catalogues
 (\citealt{saro14}; Gupta et al.~in prep.)
that include  the
full SPT cluster detection procedure.
We find that the most relevant parameter regarding the centring
uncertainty when using the SZ centres is the smoothing scale $\theta_\mathrm{c}$ used for the cluster
detection \citep[see][]{bleem15}.
We therefore use the actual distribution of  $\theta_\mathrm{c}$ values for
our clusters from \citet{bleem15} for the generation of the miscentring
distribution.

\begin{table}
\caption{Mass recovery bias factors for the analysis taking the full likelihood
  distribution into account, averaged
  over all of our clusters,
 for different miscentring
  distributions and \reva{concentration--mass} relations.  The statistical
  uncertainty of the bias correction ranges from $1.5\%$ for our lower
  redshift clusters to $2.5\%$ for our highest
  redshift clusters.
\label{tab:bias}}
\begin{center}
\begin{tabular}{cccc}
\hline\hline
Miscentring & \reva{$c(M)$} rel.  & \mbox{$\langle b_\mathrm{200c} \rangle$} & \mbox{$\langle b_\mathrm{500c} \rangle$} \\
\hline
None & Diemer+15 & 0.95 & 0.96 \\
X-ray-hydro & Diemer+15 & 0.87 & 0.87 \\
SZ-hydro & Diemer+15 & 0.81 & 0.81 \\
SZ-hydro & \mbox{$c_\mathrm{200c}=4$} & 0.79 & 0.81 \\
SZ-hydro & \mbox{$c_\mathrm{200c}=3$} & 0.89 & 0.86 \\
SZ-hydro & \mbox{$c_\mathrm{200c}=5$} & 0.73 & 0.77 \\
\hline
\end{tabular}
\end{center}
\vspace{0.2cm}
\end{table}

\subsubsection{Impact and uncertainty of the miscentring correction}
\label{se:impact_miscentring}
Using the default \reva{$c(M)$} relation \citep{diemer15} and comparing
the analyses using the miscentring distributions from the  hydrodynamical
simulation
to the case
without miscentring, we estimate that miscentring on average
introduces
a moderate
 mass bias of \mbox{$8$--$9\%$} when using the  X-ray centres, and
a more substantial bias of \mbox{$14$--$15\%$}
using the SZ centres (see Table \ref{tab:bias}).  The SZ measurements less accurately determine the cluster centre, which on-average increases the bias correction.  This result is  consistent with the smaller
average offsets from the mass peaks found for the X-ray centres (Sect.\thinspace\ref{sec:massmaps}).

\begin{figure}
\begin{center}
  \includegraphics[width=0.9\columnwidth]{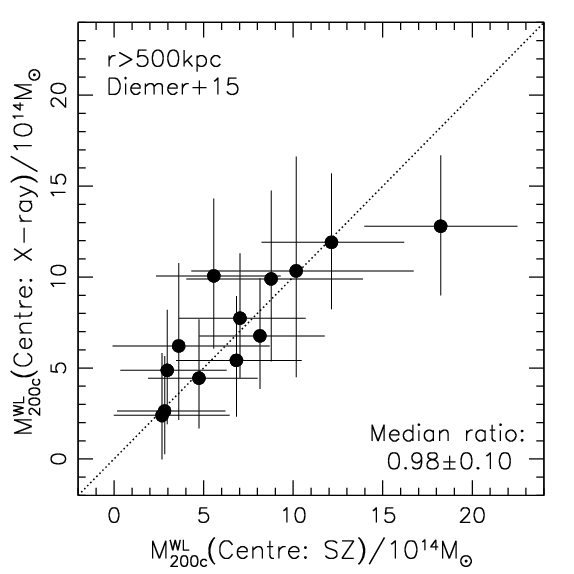}
\end{center}
\caption{Comparison of the
bias-corrected
 weak lensing
  mass estimates
using the X-ray versus the SZ centres.
 The high-mass outlier is the  merger
 SPT-CL{\thinspace}$J$0102$-$4915, for which the location of the SZ peak is closer
 to the centre between the two peaks of the mass reconstruction (see
 Fig.\thinspace\ref{fig:massplotb}), resulting in a higher mass estimate.
\label{fig:mass_corrected_comparison_centres}}
\end{figure}

As a consistency check for the miscentring correction we compare the
bias-corrected
mass estimates
using the  X-ray and SZ centres in
Fig.\thinspace\ref{fig:mass_corrected_comparison_centres}.
Their median ratio \mbox{$0.98\pm 0.10$},
 with an uncertainty
estimated by bootstrapping the clusters, is consistent with unity as
expected in the case of accurate bias correction.
We, however, note that the small sample size leads to a significant
uncertainty of this median ratio, making it not a very stringent test for the
accuracy of the bias correction.

The accurate correction for mass modelling biases such as the
one introduced by miscentring is an
active field of research \citep[e.g.][]{lsst12}.
Our analysis using a miscentring distribution based on a hydrodynamical
simulation is a step forward in this respect, but we acknowledge that it is
still simplistic. In particular, it ignores that
positional offsets are not always in a random direction.
This is prominently demonstrated by the merger  SPT-CL{\thinspace}$J$0102$-$4915, for which the location of the SZ peak is closer
 to the centre between the two peaks of the mass reconstruction (see
 Fig.\thinspace\ref{fig:massplotb}), leading to an increased mass estimate
 (compare Fig.\thinspace\ref{fig:mass_corrected_comparison_centres}).
 Due to this simplification in our current analysis, we conservatively assign a
 large uncertainty for the miscentring correction, which amounts to 50\% of
 the correction, corresponding to a 4\% uncertainty in mass when using the X-ray centres and 7\% when using the SZ centres.
 Future analyses can
 reduce this uncertainty by simulating all observables
including the weak lensing data from the same  hydrodynamical
simulation (see Sect.\thinspace\ref{se:dis_massmodel}).

\subsubsection{Uncertainties in the \reva{concentration--mass} relation}
\label{se:shear_profile_compare_mc}
For the case of SZ miscentring
Table \ref{tab:bias} lists average bias numbers for
the  \reva{$c(M)$} relation from
\citet{diemer15},
as well as
 fixed concentrations
\mbox{$c_\mathrm{200c} \in \{3,4,5\}$}.
Our bias correction procedure effectively maps the \reva{$c(M)$} relation used for the fit to the
observed
\reva{$c(M)$} relations in the simulations that are used for the bias correction (BK11, Millennium XXL).
The remaining question is how well the \reva{$c(M)$} relations in these simulations resemble
the true average \reva{$c(M)$} relation in the Universe, especially regarding the
impact of  baryons.
\citet{duffy10} show that the impact of  baryon physics appears to have only
a relatively minor (\mbox{$\lesssim 10\%$}) influence on the concentrations of
very massive clusters.
 \citet{deboni13} find similar numbers at low redshifts (for complete halo
 samples), and slightly stronger effects at \mbox{$z=1$} (\mbox{$\sim 15$--$20$\%}).
Interpolating between the \mbox{$\langle b_\mathrm{500c} \rangle$} values in
Table \ref{tab:bias}
we estimate that a 10--20\% uncertainty on the concentration around
\mbox{$c_\mathrm{200c}=4$} leads to a \mbox{$\sim 2$--$4\%$} systematic
uncertainty for the constraints on  \mbox{$M_\mathrm{500c}$}, where we
conservatively adopt
the
larger number in our systematic error budget (see Sect.\thinspace\ref{se:systematic_error_budget}).

 \citet{deboni13} note that  differences in the definition of the
 concentration
can lead to shifts in the values measured from  N-body
simulations  of up to 20\%. This is not a concern for our analysis, as
we directly estimate the calibration from the simulated weak lensing data,
and therefore do not rely on concentration measurements in the simulations.

\begin{table*}
  \caption{Systematic error budget for our current study and our expectation for what can be achieved in similar studies in the near future with moderate analysis improvements.
\label{tab:sys}}
\begin{center}
\begin{tabular}{lccccccll}
\hline
\hline
  &      \multicolumn{3}{c}{{\bf Current}} & \multicolumn{3}{c}{{\bf Near future}} & & \\
Source & rel.~error   &  \multicolumn{2}{c}{rel.~error $M_\mathrm{500c}$} & rel.~error  &
\multicolumn{2}{c}{rel.~error $M_\mathrm{500c}$} & Sect./ & Improve via \\
& signal & &  & signal  & &  & App. &\\
\hline
{\bf Shape measurements:} &&&&&&&&\\
Shear calibration & 4\% & \multicolumn{2}{c}{6\%} & 1\% & \multicolumn{2}{c}{1.5\%} &\ref{sec:shear} & Image simulations\\
\hline
{\bf Redshift distribution:} &&&&&&&&\\
$\langle\beta \rangle$ sys.~photo-$z$  & 2.2\% & \multicolumn{2}{c}{3.3\%} & 1.5\% & \multicolumn{2}{c}{2.2\%} & \ref{se:zdist_fix_uncertainty}
& Improved priors + $p(z)$\\
$\langle\beta \rangle$ cosmic variance & 1\% & \multicolumn{2}{c}{1.5\%} & 1\%& \multicolumn{2}{c}{1.5\%} &\ref{sec:beta_los_variation} & More reference fields\\
$\langle\beta \rangle$ deblending & 0.5\% & \multicolumn{2}{c}{0.8\%} & 0\%& \multicolumn{2}{c}{0\%} & \ref{se:app:non_matches} &
F606W-detected photo-$z$s\\
$\langle\beta \rangle$ LCBG contamination & 0.9\% & \multicolumn{2}{c}{1.4\%} & 0.5\%& \multicolumn{2}{c}{0.8\%} & \ref{se:test_extremely_blue} &
Apply model\\
\hline
{\bf Mass model:}  &&& &&&&&\\
Miscentring for X-ray (SZ) centres &  & \multicolumn{2}{c}{4\% (7\%)} & & \multicolumn{2}{c}{2\% (3.5\%)} &  \ref{se:miscentring_distributions} &
 Hydro sims, weak lensing\\
\reva{$c(M)$} relation &  & \multicolumn{2}{c}{4\%} & & \multicolumn{2}{c}{2\%} &  \ref{se:shear_profile_compare_mc} &
 Hydro sims, weak lensing\\
\hline
{\bf Total} for X-ray (SZ) centres {\bf :} &  & \multicolumn{2}{c}{9.2\% (10.8\%)} & & \multicolumn{2}{c}{4.2\% (5.1\%)} &  &
\\
\hline
\end{tabular}
\end{center}
\vspace{0.2cm}
\end{table*}

\subsection{Statistical precision versus systematic uncertainty}
\label{se:systematic_error_budget}

We summarize the identified sources of  systematic uncertainty for our study in
Table \ref{tab:sys},
pointing to their corresponding sections, and listing  their associated relative uncertainties in the
measured weak lensing signal and mass constraints.
Combining all systematic error contributions in quadrature, we estimate an
overall systematic mass uncertainty of 9\% (11\%) for the analysis using the X-ray (SZ) centres.
This
can be compared to the
combined statistical mass signal-to-noise ratio of the sample, which we
approximate as
\begin{equation}
\left(S/N
\right)_\mathrm{mass}^\mathrm{sample}=\sqrt{\sum_\mathrm{clusters}\left(M_{\mathrm{500c},i}/\Delta
    M_{\mathrm{500c},i}^\mathrm{stat.}\right)^2}\simeq 7.3\,,
\end{equation}
which corresponds
to a \mbox{$\sim 14\%$} precision, ignoring the impact of intrinsic scatter,
e.g. from cluster triaxiality.
Accordingly, our total uncertainty is dominated by statistical measurement
noise and not systematic uncertainties.

For the analysis of larger future data sets with improved statistical
precision it will be important to further reduce systematic uncertainties.
When discussing the individual sources of systematic uncertainty we have already suggested strategies how their influence can be reduced in the future.
The largest contributions to the systematic error budget currently come from the shear calibration, miscentring corrections, and uncertainties in the \reva{$c(M)$} relation.
All of these can be reduced
with better simulations. For the latter two issues the weak lensing data
can themselves provide information that help to reduce these uncertainties (see also Sect.\thinspace\ref{se:dis_massmodel}).
As a rough guess we expect that it should be possible to cut the
systematic uncertainties
associated with the mass modelling
by half in the coming years with moderate effort
(compare Table \ref{tab:sys}),
and
note that
 some improved shape measurement techniques have already reached significantly
higher accuracy \citep[e.g.][]{bernstein16,fenechconti17}.
We further discuss the strategies to reduce systematic uncertainties  in Sect.\thinspace\ref{sec:discussion}.

\section{Constraints on the $M$--$T_\mathrm{X}$ scaling relation}
\label{se:mtx}

In the self-similar model \citep[e.g.][]{kaiser86} galaxy clusters form through the gravitational
collapse of the most overdense regions in the early Universe. In this model the
cluster baryons are heated through gravitational processes only, leading to predictions
for cluster scaling relations.
Deviations from
self-similarity, e.g. regarding  the slope of the X-ray
luminosity--temperature relation \citep[e.g.][]{arnaud99},
suggest that non-gravitational effects, such as heating by active galactic
nuclei or radiative cooling, provide non-negligible contributions
to the energy budget of clusters.
However,
the redshift evolution of cluster X-ray observables appears to be consistent with
self-similar predictions \cite[e.g.][]{maughan06}, suggesting that
non-gravitational effects have a similar impact at low and high redshifts.
If this ``weak self-similarity'' \citep[e.g.][]{bower97} also applies to
cluster masses, we expect
a scaling between temperature and mass in the form
\begin{equation}
 \label{eqn:METx}
M_x E(z) \propto T^\alpha \, ,
\end{equation}
\citep[e.g.][]{mathiesen01,boehringer12}, where
\begin{equation}
E(z)=\frac{H(z)}{H_0}=\sqrt{\Omega_\mathrm{m}(1+z)^3+\Omega_\Lambda}
\end{equation}
indicates the redshift dependence of the Hubble parameter, here assuming a flat $\Lambda$CDM cosmology, and
\mbox{$\alpha=3/2$} corresponds to the self-similar prediction for the slope
of the relation.

The main constraints on cluster scaling relations from our sample will be
presented in a
forthcoming paper (Dietrich et al.~in prep.) that combines our measurements with a  complementary
 sample of clusters  at  lower redshifts with Magellan/Megacam observations
 and accounts for the SPT selection function, which is especially important
 when calibrating SZ scaling relations.
However, here we already combine our measurements with core-excised \emph{Chandra} X-ray temperature estimates  $T_\mathrm{X}$ that are available for 12 clusters in our sample.
Details of the specific measurements are provided in \cite{mcdonald13}, with the analysis pipeline adapted based on \cite{vikhlinin06}. In short, \emph{Chandra} ACIS-I data were reduced using \textsc{ciao} v4.7 and \textsc{caldb} v4.7.1. All exposures were initially filtered for flares, before applying the latest calibrations and determining the appropriate epoch-based blank-sky background. Point sources were identified via an automated wavelet decomposition technique \citep{vikhlinin98} and masked. Spectra were extracted in a core-excised region from $(0.15$--$1)\times r_{500c}$ \citep{mcdonald13} and fit over 0.5--10.0\,keV using a combination of an absorbed, optically-thin plasma (\textsc{phabs}$\times$\textsc{apec}), an absorbed hard background component (\textsc{phabs}$\times$\textsc{bremss}), and a soft background (\textsc{apec}), see  \cite{mcdonald13} for details.

\begin{figure}
  \includegraphics[width=1\columnwidth]{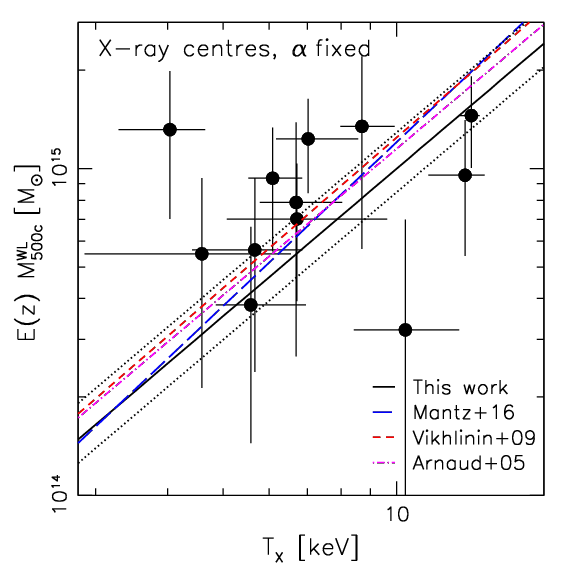}
\caption{
Core-excised X-ray temperatures measured in the range  $(0.15$--$1)\times r_{500c}$ based on {\it Chandra} data versus
   $E(z) M_\mathrm{500c}^\mathrm{WL}$ from the weak lensing  analysis
  using the X-ray centroids and assuming the \reva{$c(M)$} relation from \citet{diemer15}.
The solid
 black line shows our best-fit  estimate of the scaling relation
when assuming a fixed slope
 $\alpha=3/2$.
The dotted lines
correspond to normalisations that are  lower or higher by $1\sigma$,
combining the statistical and systematic uncertainties of our constraints.
 The dashed and dashed-dotted lines indicate  best-fit estimates
derived by   \citet{arnaud05,vikhlinin09b} and \citet{mantz16}.
\label{fig:m_tx}}
\end{figure}

\begin{figure}
  \includegraphics[width=1\columnwidth]{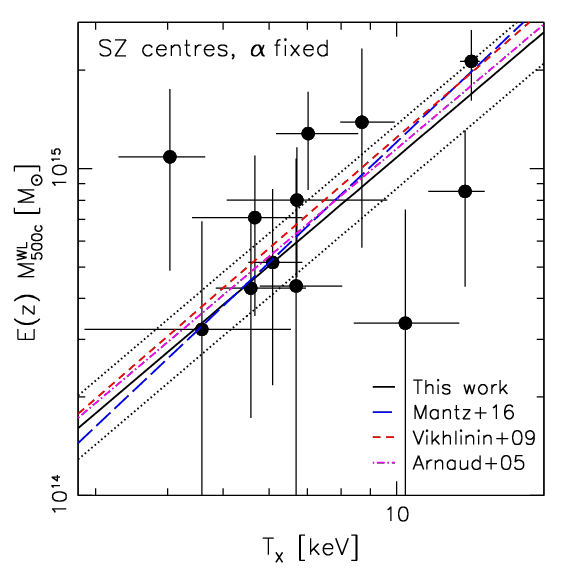}
\caption{As Figure
\ref{fig:m_tx}, but employing the weak lensing results for the SZ centres.
\label{fig:m_tx_sz}}
\end{figure}

Figures \ref{fig:m_tx} and \ref{fig:m_tx_sz}  show the bias-corrected
$M_\mathrm{500c}^\mathrm{WL}E(z)$
using
the
\citet{diemer15}
  \reva{$c(M)$} relation
as function of the core-excised  $T_\mathrm{X}$ estimates (Table
\ref{tab:tx}) for the analyses centring on the X-ray centroids or SZ peaks, respectively.
For comparison we show best-fit estimates for the scaling relation
derived by \citet[][based on their \mbox{$T_\mathrm{X}>3.5$ keV} sample]{arnaud05}, \citet{vikhlinin09b}, and \citet{mantz16}
using clusters at lower and intermediate redshifts ($z \lesssim 0.6$).

To obtain quantitative constraints on the scaling relation,
we assume the functional form
\begin{equation}
\ln \left(E(z) M_\mathrm{500c}/10^{14}\mathrm{M}_\odot \right)=A+\alpha \left[\ln
   \left(kT/7.2\mathrm{keV}\right)\right]\, ,
\end{equation}
where
the temperature  pivot point roughly corresponds to the mean temperature of the sample.
Our fitting method is
based on the approach of \cite{kelly07}, which incorporates measurement errors
in the $x$- and $y$- coordinates and has been extended to include log-normal intrinsic scatter.
The method has been generalized to use the exact likelihood from the lensing analysis, and a two-piece normal approximation to the X-ray likelihood \citep{applegate16}.
For this analysis we use the lensing  likelihood based on the dominant shape noise
only and absorb the minor contributions from large-scale structure
projections and line-of-sight variations in the redshift distribution (see
Sect.\thinspace\ref{sec:profiles}) in the intrinsic scatter
$\sigma_\mathrm{M}$.

\begin{table}
\caption{Core-excised {\it Chandra} X-ray temperatures used for our
  constraints on the $M$--$T_\mathrm{X}$ scaling relation.
\label{tab:tx}}
\begin{center}
\begin{tabular}{lr}
\hline
\hline
Cluster & $T_\mathrm{X}$ [keV]\\
\hline
SPT-CL\,$J$0000$-$5748  & $6.7_{-1.6}^{+2.9}$ \\
SPT-CL\,$J$0102$-$4915  & $13.5_{-0.6}^{+0.5}$\\
SPT-CL\,$J$0533$-$5005  & $4.6_{-1.7}^{+2.0}$\\
SPT-CL\,$J$0546$-$5345  & $6.7_{-0.9}^{+1.4}$\\
SPT-CL\,$J$0559$-$5249  & $6.1_{-0.6}^{+0.8}$\\
SPT-CL\,$J$0615$-$5746  & $13.1_{-1.8}^{+1.1}$\\
SPT-CL\,$J$2106$-$5844  & $8.7_{-0.7}^{+1.2}$\\
SPT-CL\,$J$2331$-$5051  & $5.6_{-0.7}^{+1.4}$\\
SPT-CL\,$J$2337$-$5942  & $7.0_{-0.9}^{+1.6}$\\
SPT-CL\,$J$2341$-$5119  & $10.4_{-1.9}^{+2.5}$\\
SPT-CL\,$J$2342$-$5411  & $4.0_{-0.8}^{+0.6}$\\
SPT-CL\,$J$2359$-$5009  & $5.7_{-1.3}^{+1.2}$\\

\hline
\end{tabular}
\end{center}
\end{table}

We fix the slope of the scaling
relation to the self-similar prediction ($\alpha=3/2$) for the current
analysis, given  the limited sample size and mass range.
\changeb{We then obtain constraints
\mbox{$(A,\sigma_\mathrm{M})=(1.81^{+0.24}_{-0.14},0.05^{+0.32}_{-0.05})$}
for our default analysis using the X-ray centres.
When alternatively using the  SZ peaks as centre for the weak
lensing analysis we obtain  consistent results
\mbox{$(A,\sigma_\mathrm{M})=(1.89^{+0.20}_{-0.19},0.31^{+0.04}_{-0.31})$}.}
In addition to these statistical uncertainties there is a 9\% (11\%) systematic
uncertainty for the analysis using the X-ray (SZ) centres, directly propagating into the normalisation of the scaling relation  (see Sect.\thinspace\ref{se:systematic_error_budget}).
The obtained  constraints are consistent with
 the
aforementioned results from
lower redshift samples when assuming
 self-similar redshift evolution \changeb{within \mbox{$1\sigma$}} (see Figures \ref{fig:m_tx} and \ref{fig:m_tx_sz}).

\citet{jee11} present an HST weak lensing analysis for 27 galaxy clusters at
\mbox{$0.83\le z \le 1.46$}, using a heterogeous  sample that includes optically, NIR, and X-ray-selected clusters.
\changeb{Their analysis suggests
a possible evolution in the $M_\mathrm{2500c}$--$T_\mathrm{X}$ scaling relation
in comparison to self-similar extrapolations from lower redshifts.
For example, at \mbox{$T_\mathrm{X}=5$ keV} their estimated scaling relation
has a lower amplitude by \mbox{$27\pm 7\%$} (statistical uncertainty from
\citet{jee11} only)
compared to the best-fit relation from \\
 \citet{arnaud05}.
We do not find significant indications for a similar evolution for the
$M_\mathrm{500c}$--$T_\mathrm{X}$ scaling relation,
 but note that our statistical uncertainties are significantly larger given
 our smaller sample size and more conservative radial fit range.
There are various additional differences in the analyses, such as different samples for the calibration of the
source redshift distribution, our more conservative removal of cluster
galaxies, and our calibration of modelling biases on simulations, making the
direct comparison difficult.
Importantly, both studies   use different overdensities for the scaling
relation constraints\footnote{We do not report $M_\mathrm{2500c}$ masses as these are not
available in the \citetalias{becker11} simulation, preventing us to compute
accurate bias corrections for this overdensity.}.
Furthermore, \citet{jee11} use X-ray temperature estimates from the
literature that typically do not exclude the core regions.
Including the cores should, on average, reduce  the temperatures in the presence
of cool-core clusters.
This would, however, aggravate the tension between the \citet{jee11} results
and the self-similar extrapolations from lower redshift} samples.

\section{Discussion}
\label{sec:discussion}

In our analysis we have introduced a number of new aspects and systematic investigations for weak lensing studies of high-redshift clusters.
Here we discuss their relevance also in the context of future
weak lensing programmes.
Our study using HST and VLT data
provides a
demonstration for
future weak lensing science investigations that combine  deep
high-resolution space-based
shape measurements, e.g. from \textit{Euclid}  \citep{laureijs11} or WFIRST
\citep{spergel15},
with deep photometry,
e.g. from LSST  \citep{lsst09}.

\subsection{The benefits and challenges of using faint blue galaxies for weak lensing}
For deep  weak lensing surveys conducting shape measurements at optical wavelengths
the majority of the high-redshift (\mbox{$z\sim 1.5$--$3$}) sources
are blue star forming galaxies observed at rest-frame UV wavelengths with
blue observed optical colours (see the top left panel of Fig.\thinspace\ref{fig:zdist_f814w}).
These galaxies are useful as the source sample in weak lensing studies of high-redshift clusters
both because of their high source density and high geometric lensing efficiency,
but also because they can  be readily distinguished from both blue and red
cluster galaxies using optical colours (see
Sect.\thinspace\ref{se:photo_color_select_acs}).
This enables a nearly complete removal of cluster galaxies from the weak lensing
source sample,
which is important both in order to minimise modelling uncertainties
regarding cluster member contamination (Appendix \ref{app:why_not_boost}), and to ensure that intrinsic
alignments of galaxies within the targeted clusters cannot bias mass constraints  \citep[but note
that
this appears to be a negligible effect at the precision of current samples, see][]{sifon15}.

To exploit these benefits, a number of challenges need to be overcome.
Here we first stress that high signal-to-noise optical
photometry is needed to robustly select these galaxies in colour space.
In the case of our study a well-matched colour selection was possible in
areas covered by ACS in both F606W and F814W.
However, outside the F814W footprint we had to rely on the combination of
F606W ACS imaging and VLT $I_\mathrm{FORS2}$ images,
which, despite a good VLT integration time \mbox{$\langle
  t_\mathrm{exp}\rangle=2.4$ ks} and \mbox{$\lesssim 0\farcs8$} seeing,
delivered a density of usable
sources that is only 32\% of the density from the ACS-only
$V_{606}-I_{814}$ selection (Sect.\thinspace\ref{se:total_ngal}).
This highlights that future weak lensing programmes and surveys should carefully
tune the relative depth of their bands (regarding both red and blue filters)
to maximise the science output of their data.

While our analysis is based on simple colour cuts due to the limited data
available in our cluster fields, we expect that similar conclusions apply
for surveys that aim at computing individual photometric redshifts for the weak
lensing source galaxies.
Photometric redshift selections correspond to higher dimensional
cuts in colour-colour space.
However, depending on the survey characteristics, the large population of blue high-$z$
galaxies may only be detected in a few of the bluer optical pass bands,
effectively
reducing photo-$z$ cuts to a selection in a relatively small colour-colour space.
As a result, individual  photometric redshift estimates for  faint blue galaxies have
typically
large uncertainties unless deep photometry is available over a very broad
wavelength range (in particular including deep $u$-band and NIR observations).
For cluster weak lensing studies noise in individual photometric redshifts is not a problem
as
long as cluster galaxies can be removed robustly and
the overall source redshift distribution can be modelled accurately.

\subsection{Robust estimates of the source redshift distribution}

We employ a statistically consistent selection of source galaxies
matched in filter, magnitude, colour, and shape properties
in our
cluster fields and observations of the  CANDELS fields.
This allows us to estimate the average source redshift distribution and its
statistical variation between lines-of-sights using the CANDELS data and
apply this information for the cluster weak lensing analyses.
At depths similar to our data, the CANDELS fields are currently among the extragalactic fields that are best studied both photometrically and spectroscopically.
We have shown that they cover enough sky area to reduce the cosmic variance contribution to the uncertainty on the mean lensing efficiency at our cluster redshifts to the \mbox{$\sim 1\%$} level (Sect.\thinspace\ref{sec:beta_los_variation}), which is much smaller than current statistical weak lensing uncertainties.
Therefore, we expect that the CANDELS fields will remain to be an important calibration sample for estimates of the source redshift distribution in deep weak lensing data in the near future.

As revealed by our comparison to HUDF data
(Sect.\thinspace\ref{se:sub:catastrophic_outliers}) and confirmed via
spatial cross-correlations with spectroscopic/grism redshifts
(Appendix\thinspace\ref{app:candels_x}),
the  3D-HST  CANDELS photo-$z$s
 suffer from catastrophic redshift
outliers (primarily galaxies at \mbox{$2\lesssim z \lesssim 3$} that are assigned a low photometric redshift \mbox{$z_\mathrm{p}<0.3$}) and  redshift
focussing effects at \mbox{$z_\mathrm{p}\simeq 1.5$}.
Together these would on average bias our mass estimates high by 12\% if not accounted for.
For our current study we have implemented an empirical correction for these systematics effects.
We plan to investigate this issue and its causes in detail in a future paper  (Raihan et al.~in prep.).
Given  the high photometric quality, depth, and broad wavelength coverage of the CANDELS data, we speculate that  some other current photometric redshift data sets might suffer from similar effects.
This is supported by the weak lensing analyses of
\citetalias{schrabback2010} and  \citet{heymans12} as discussed in Sect.\thinspace\ref{sec:correct_sys_photoz}.
We therefore expect that also other weak lensing programmes will have to
implement similar correction schemes or improved  photometric redshift
algorithms,
and apply these either to deep field data in case of colour  cut analyses, or
their survey data in case of individual photo-$z$ estimates.
Surveys that obtain  individual photo-$z$s can also attempt to identify and remove galaxies in  problematic \mbox{$z_\mathrm{p}$} ranges at the cost of reduced sensitivity.
We
 stress that the use of the average redshift posterior probability
distribution instead of the peak photometric redshift estimates is not
sufficient to cure the identified issue for the 3D-HST photo-$z$s
(Sect.\thinspace\ref{se:posterior}).

One route to calibrate photo-$z$s is via very deep spectroscopy for
representative
galaxy samples. At present,  such spectroscopic samples are very
incomplete at the depth of our
analysis, which is why we resorted
to the comparison of the CANDELS photo-$z$s to photometric redshifts  for
the HUDF
\citep{rafelski15}, which are  based on deeper data and a broader
wavelength coverage.
We find that this is a viable approach at the precision of current and
near-term high-$z$ cluster samples with weak lensing measurements, but it is likely not of sufficient accuracy for
the calibration of very large future data sets.
To prepare for the analyses of such
data sets it is vital and timely to obtain larger spectroscopic
calibration samples, including both
highly complete deep samples for direct calibration,
but also very large,
potentially shallower and less complete samples \cite[][]{newman15}.
The later can be used to infer information on the redshift distribution via
spatial cross-correlations \citep[e.g.][]{newman08,matthews10,schmidt13,rahman15,rahman16}, for which we provide one of the first practical
applications in the context of weak lensing measurements \citep[see Appendix \ref{app:candels_x} and][]{hildebrandt17}.

As an important ingredient for our modelling of the redshift
distribution we carefully matched the selection criteria and noise properties between our cluster
field data and the CANDELS data to ensure that consistent
galaxy populations are selected between both data sets
(see Sect.\thinspace\ref{se:source_sel_scatter} and Appendix \ref{app:scatter}).
For the colours obtained from the combination of  ACS F606W and
VLT $I_\mathrm{FORS2}$ data we empirically estimated the net scatter distribution by
comparing to the colours estimated in the inner cluster regions from ACS
F606W and F814W data.
We note that systematic effects such as residuals from the PSF
homogenisation can add scatter which may well deviate from
 Poisson noise distributions that are often assumed, e.g. in photometric
 redshift codes.
As we empirically sample from the actual scatter distribution such effects are automatically accounted for in our
analysis.
For future surveys that vary in data quality we recommend to obtain
repeated imaging observations of spectroscopic reference fields that span the full range of varying observing
conditions, in order to generate similar empirical models for the impact of
the actual noise properties.

\subsection{Accounting for magnification}
\label{se:dis:mag}
The impact of weak lensing magnification on the source redshift distribution has typically been ignored in past weak lensing studies.
Our investigation of this effect in
Sect.\thinspace\ref{sec:magnification_model} indicates that the net effect
is small for our study given the depth of our data.
However,
shallower programmes such as DES \citep[][]{des05} or KiDS
\citep{kuijken15}, which aim to calibrate high-$z$ cluster masses by
combining measurements from a large number of clusters, will need to
carefully account for the resulting boost in the average lensing efficiency
\mbox{$\langle \beta \rangle$}.
For example, Fig.\thinspace\ref{fig:magnification_shear} illustrates that
the impact of
magnification on the source redshift distribution has a larger impact on the
reduced shear profile at brighter magnitudes
than the typically applied correction for the finite width of
the source redshift distribution.

We point out that knowledge of the redshift distribution is needed
at fainter magnitudes than the targeted depth limit of a survey in order to be able to
compute the actual correction for the impact of magnification
(Sect.\thinspace\ref{sec:magnification_model}).
Accordingly, it is necessary to obtain spectroscopic redshift samples for
photo-$z$ calibration to greater depth than the targeted survey depth.
The difference in depth depends on the maximum magnification that is
considered, and therefore the magnitude limit, the cluster redshift and
mass, as well as the considered fit range.

We also note that it
is important
to take magnification into account when using the source density and the density profiles as validation tests for the cluster member removal (see Sect.\thinspace\ref{se:photo_number_density_tests}).
Programmes with ground-based resolution will also need to account for the
change in source sizes due to magnification  as function of redshift,
cluster-centric distance, and mass,
as shape cuts could otherwise introduce
redshift- and mass-dependent selection
biases.

\subsection{Shape measurement biases}
\label{se:dis_shape}
Currently the shear calibration uncertainty constitutes the largest
individual contribution to the systematic error budget of our study (4\% for
the shear calibration corresponding to a 6\% mass uncertainty).
This is due to the fact that we base the calibration on simulations from the
STEP project (Sect.\thinspace\ref{sec:shapemeasure}) which lack faint
galaxies that influence the bias calibration \citep{hoekstra15} and do not
probe shears as high as those used in our analysis.
However, this  source of systematics can easily be reduced through image
simulations that resemble real galaxy populations and cluster-regime shears
more accurately, and which can be generated recent tools such as
\textsc{GalSim} \citep{rowe15}.
We therefore expect that shear measurement biases in cluster weak lensing
studies will soon be reduced to the  levels reached in cosmic shear
measurements \citep[e.g.][]{fenechconti17}.
Also see \citet{hoekstra17},
\changeb{whose results suggest}
 that the impact of the higher density of
sources in cluster regions on shape measurement biases should be negligible
for current data.

In addition, additive shape measurement biases can be relevant for cluster
weak lensing in particular for pointed follow-up programmes where the clusters
are always centred at similar detector positions.
An example for such a potential source of bias can be CTI residuals.
However, through a new null test we have shown that our data
show no significant CTI-like residuals within the current statistical uncertainty (Sect.\thinspace\ref{sec:ctitests}).

\subsection{Accounting for biases in the mass modelling}
\label{se:dis_massmodel}
We have calibrated our mass estimates using reduced shear profile
fits to simulated cluster weak lensing data from N-body simulations (see
Sect.\thinspace\ref{sec:sims}).
One important source for bias is miscentring of the reduced shear
profile.
As we do not know the location of the centre of the 3D cluster potential we have to rely on
observable  proxies for the
cluster centre, leading to a suppression of the expected reduced shear signal at small radii.
Based on the work from \cite{dietrich12} we expect that the peaks in the  reconstructed weak lensing mass
maps of the clusters (see Sect.\thinspace\ref{sec:massmaps}) should provide a
tight tracer for the centre of the 3D cluster potential,
but we do not use these centres for our mass constraints in our current analysis as they are expected to yield masses that are biased high.
By studying the offset distributions between  the mass
peaks and the other  proxies for the
cluster centre  we find that the
 X-ray centroids provide the smallest average offsets, closely followed by
 the SZ peak locations.
 Hence, they also provide good proxies for the cluster centre, which is why we
employ them
 as centres for our mass constraints.
To account for the expected remaining bias caused by miscentring,
we  randomly misplace the centre
in the simulated weak lensing data based on offset distributions measured
between the 3D cluster centre and the SZ peak location or X-ray centroid in
hydrodynamical simulations (see Sect.\thinspace\ref{se:miscentring_distributions}).
Future studies could further advance this approach by simulating all
observables including SZ, X-ray, and
weak lensing data  from the same hydrodynamical simulation, in order
to also account for possible covariances between these observables.
Our analysis of the prominent merger SPT-CL\,$J$0102$-$4915 demonstrates that
such covariances exist, as both the X-ray centroid and SZ peak are located
between the two peaks of the mass reconstruction (see
Fig.\thinspace\ref{fig:massplotb}). Hence, the misplacement is not in a
random direction.
To validate the accuracy of the employed simulations, the measured offset
distributions between the mass peaks and the different proxies for the
centre can be compared between the  real data and the simulations.
This approach could be further expanded by explicitly accounting for the
miscentring in the fitted reduced shear profile model
(e.g.~\citealt{johnston07,george12}; also see \citealt{koehlinger15} for the impact of
miscentring in stacked
Stage IV analyses).

A further uncertainty for the mass constraints arises from uncertainties in the assumed \reva{$c(M)$} relation.
The applied calibration procedure essentially maps the measurements onto the  \reva{$c(M)$} relation of the simulation.
Remaining uncertainties reflect our ability to simulate the true \reva{$c(M)$} relation of the Universe, especially with respect to the impact of  baryons.
These uncertainties are expected to shrink
with further advances in simulations, in particular thanks to the recent advent of large hydrodynamical simulations \citep[e.g.][]{dolag16}.
In addition, the weak lensing measurements themselves
can be used to test if the inferred reduced shear profiles are consistent with the simulation-based priors on the \reva{$c(M)$} relation, in particular if information from the inner reduced shear profiles is incorporated.
Using the X-ray centroids  our analysis yields a best-fitting fixed concentration
for the sample of
\mbox{$c_\mathrm{200c}=5.6^{+3.7}_{-1.8}$}
when including
scales \mbox{$>300$ kpc} (Sect.\thinspace\ref{se:stack_mc}).
This is fully consistent with recent results for the \reva{$c(M)$} relation from
simulations \citep[e.g.][]{diemer15}, but higher than earlier results from
\citet{duffy08}, \reva{which, however, are based on a WMAP5 cosmology
  \citep{komatsu09} with lower $\Omega_\mathrm{m}$ and $\sigma_8$, reducing the resulting concentrations}.
We note that future studies that aim to obtain tighter constraints on the
concentration will have to account for the impact of miscentring and
stronger shears in the inner cluster regions, which we could ignore for this
part of our analysis given the statistical uncertainties.

\section{Conclusions}
\label{sec:conclusions}
We have presented a weak gravitational lensing analysis of 13
high-redshift clusters from the SPT-SZ Survey, based on shape measurements in
high resolution HST/ACS data and colour measurements that also incorporate VLT/FORS2
imaging.
We have introduced new
methods for the weak lensing analysis
of high redshift clusters and carefully investigated the impact of
systematic uncertainties
as discussed in Sect.\thinspace\ref{sec:discussion} in the context of future programmes.
In particular, we select blue galaxies in \mbox{$V_{606}-I_{814}$} colour to achieve a nearly complete removal of cluster galaxies, while selecting most of the relevant source galaxies at \mbox{$1.4\lesssim z \lesssim 3$} (see Sect.\thinspace\ref{se:photo_color_select_acs}).
Carefully matching our selection criteria we estimate the source redshift distribution using data from CANDELS, where we apply a statistical correction for photometric redshift outliers.
This correction is derived from the comparison to deep spectroscopic and photometric data from the HUDF (see Sect.\thinspace\ref{sec:correct_sys_photoz}), and checked using spatial cross-correlations (see Appendix \ref{app:candels_x}).
We account for the impact of lensing magnification on the source redshift distribution, which we find is especially important for shallower surveys (see Sect.\thinspace\ref{sec:magnification_model}).
We also introduce a new test for residual contamination of galaxy shape estimates from charge-transfer inefficiency, which is in particular applicable for pointed  cluster follow-up observations (see Sect.\thinspace\ref{sec:ctitests}).
Finally, we account for biases in the mass modelling through simulations (see Sect.\thinspace\ref{sec:sims}).

At present, our weak lensing mass constraints are limited by statistical
uncertainties given the small cluster sample and the limited depth of the data
for the colour selection in the cluster outskirts.
For the current study the total systematic  uncertainty on the cluster mass
scale at high-$z$ is at the
\mbox{$\sim 9\%$} level, where the largest contributions come from the shear calibration and mass modelling.
As discussed in
Sect.\thinspace\ref{se:systematic_error_budget} we have identified
strategies how this can be reduced to the  \mbox{$\sim 4\%$} level  in the
near future based on exisiting calibration data and improved simulations.
This is particularly relevant for near-term studies using larger HST
data sets.

We have used our measurements to derive updated constraints on the
\mbox{$M_\mathrm{500c}$--$T_\mathrm{X}$} scaling relation for massive high-$z$ clusters in
combination with {\it Chandra} observations.
Compared to scaling relations calibrated at lower redshifts we find no indication
for a significant deviation from self-similar redshift evolution at our
\changeb{current
\mbox{$\sim 20\%$} precision} (see Sect.\thinspace\ref{se:mtx}).
Our measurements will additionally be used in companion papers to derive updated
 constraints on additional mass-observable scaling relations,
where we also incorporate weak lensing measurements at lower redshifts from
Magellan/Megacam (Dietrich et al.~in prep.) and the Dark Energy Survey
\citep[DES,][Stern et al.~in prep.]{des05},
 and to derive
 improved cosmological constraints from the SPT-SZ cluster sample.

We  investigate the offset distributions between different proxies for the cluster
centre and the weak lensing mass reconstruction, where we find that the X-ray
centres provide the smallest average offsets
(see Sect.\thinspace\ref{sec:massmaps}).
Our analysis constrains
 the average
 concentration of the cluster sample to
\mbox{$c_\mathrm{200c}=5.6^{+3.7}_{-1.8}$}
(Sect.\thinspace\ref{se:stack_mc}) when
using the  X-ray
centres and including information
from smaller scales ($300\, \mathrm{kpc}<r<500\, \mathrm{kpc}$), which are
excluded for the conservative mass constraints.

With the advent of the next generation of deep cluster surveys
such as SPT-3G \citep{benson14},  the Dark Energy Survey \citep[DES][]{des05}, Hyper-Suprimecam
\citep[HSC][]{miyazaki12}, eROSITA \citep{merloni12}, and
Advanced ACTPol \citep{henderson16}
 it will be vital to further tighten the weak lensing
calibration of cluster masses in order to exploit these surveys for
constraints on cosmology and cluster astrophysics.
At low and intermediate redshifts, weak lensing surveys such as DES, HSC, and KiDS
are expected to soon calibrate cluster masses at the few per cent level,
especially if  large numbers of clusters can be reliably selected down to
lower masses and if their weak lensing signatures are combined statistically
\citep[e.g.][]{rozo11}.
Such surveys will also provide some statistical weak lensing constraints for clusters
out to \mbox{$z\sim 1$} \citep[e.g.][]{vanuitert16}, but it still needs to be demonstrated how reliably
such measurements can be conducted from the ground as most of the
distant background galaxies are poorly resolved.
At such cluster redshifts HST is  currently unique with its capabilities
to measure robust individual cluster masses with good signal-to-noise.
\changeb{Clusters at high redshifts and high masses are very rare.
As a result,} stacking analyses of shallower wide-area surveys cannot compete in terms of precision for their mass calibration with a large HST
program that obtains
pointed follow-up observations for all of them.
Our current study is an important pathfinder towards such a program.
For comparison, stacked analyses tend to be more powerful for lower mass
clusters, which are too numerous to be followed up individually. The
combination of deep pointed follow-up for high-mass clusters and stacked
shallower measurements for lower mass clusters is therefore particularly powerful for
obtaining constraints on the slope of mass-observable scaling relations.
In addition,
good signal-to-noise ratios
\changeb{for individual clusters,}
as provided by deep pointed follow-up, are needed for constraints on
intrinsic scatter.

In the 2020s  weak lensing Stage IV dark energy experiments
such as \textit{Euclid} \citep{laureijs11}, LSST \citep{lsst09}, and WFIRST \citep{spergel15} are expected to provide
a precise calibration of cluster masses over a
wide range in redshift \citep[for a forecast for \textit{Euclid} see][]{koehlinger15}.
To reach their weak lensing science goals they
will require highly
accurate calibrations for the redshift distribution and shear estimation.
Further efforts will be needed to fully exploit these calibrations and weak
lensing data sets for cluster mass
estimation.
For example, the shear calibration needs to be extended towards
stronger shear, and magnification has to be taken into account when estimating
the source redshift distribution (Sect.\thinspace\ref{se:dis:mag}).
We also stress that it will be vital to pair such observational studies with
analyses of large sets of  hydrodynamical
  simulations, in order to accurately calibrate the weak lensing mass
  estimates and account for covariances with other observables (see Sect.\thinspace\ref{se:dis_massmodel}).

LSST and \textit{Euclid}  will still have
signficantly
lower densities of high-redshift background source galaxies compared to HST
observations.
In order to extend the mass calibration for massive clusters out to very high redshifts (\mbox{$z\gtrsim
  1.3$}), large pointed HST and subsequently JWST programmes may therefore remain the most
effective approach until similarly deep data become available from WFIRST.

\section*{Acknowledgements}
This work is based on observations made with the NASA/ESA {\it Hubble Space
  Telescope}, using imaging data from the SPT follow-up GO programmes
12246 (PI: C.~Stubbs) and 12477  (PI: F.~W.~High), as well as archival data from
GO programmes 9425, 9500, 9583, 10134, 12064, 12440, and 12757, obtained via the data archive at the Space
Telescope Science Institute,
and catalogues based on observations taken by the 3D-HST Treasury Program
(GO 12177 and 12328) and the UVUDF Project (GO 12534, also based on data
from GO programmes 9978,
10086, 11563, 12498).
STScI is operated by the Association of Universities for Research in Astronomy, Inc. under NASA contract NAS 5-26555.
It is also based on observations made with ESO Telescopes at the La Silla Paranal Observatory under programmes
086.A-0741, 088.A-0796, 088.A-0889, 089.A-0824.
The scientific results reported in this article are based in part on
observations made by the {\it Chandra} X-ray Observatory
(ObsIDs 9332, 9333, 9334, 9335, 9336, 9345, 10851, 10864, 11738, 11739, 11741, 11742, 11748, 11799, 11859, 11864, 11870, 11997, 12001, 12002, 12014, 12091, 12180, 12189, 12258, 12264, 13116, 13117, 14017, 14018, 14022, 14023, 14349, 14350, 14351, 14437, 15572, 15574, 15579, 15582, 15588, 15589, 18241).

It is a pleasure to thank Gabriel Brammer, Pieter van Dokkum, Mattia
Fumagalli, Ivelina Momcheva, Rosalind Skelton, and the 3D-HST team for
 helpful discussions and for
making their photometric and grism
redshift catalogues available to us prior to public release.
We  thank Matthew Becker and Andrey Kravtsov for making results from their N-body
simulation \citep{becker11} available to us, Raul Angulo for providing
data from the Millennium XXL Simulation, and Klaus Dolag for providing access to data from the Magneticum Pathfinder Simulation.

TS, DA, and FR acknowledge support from the German Federal Ministry of Economics and Technology (BMWi) provided through DLR under projects 50 OR 1210,  50 OR 1308,  50 OR 1407, and 50 OR 1610.
TS also acknowledges support  from NSF through grant
AST-0444059-001 and SAO through grant GO0-11147A.
This work was supported in part by the Kavli Institute for Cosmological Physics at the University of Chicago through grant NSF PHY-1125897 and an endowment from the Kavli Foundation and its founder Fred Kavli.
HH acknowledges support from NWO VIDI grant number 639.042.814 and ERC FP7
grant 279396.
Work at Argonne National Laboratory was supported under U.S. Department of Energy contract DE-AC02-06CH11357.
The Munich group acknowledges the support by the DFG Cluster of Excellence ``Origin and Structure of the Universe'' and the Transregio program TR33 ``The Dark Universe''.
R.J.F.\ is supported in part by fellowships from the Alfred P.\ Sloan Foundation and the David and
Lucile Packard Foundation.
TdH is supported by a Miller Research Fellowship.
PS acknowledges support by the European DUEL Research-Training Network
(MRTN-CT-2006-036133) and by the Deutsche Forschungsgemeinschaft under the
project SCHN 342/7-1.
CBM acknowledges the support of the DFG under Emmy Noether grant Hi
1495/2-1.
BB is supported by the Fermi Research Alliance, LLC under Contract No. De-AC02-07CH11359 with the United States Department of Energy.
CR  acknowledges support from the Australian Research Council's Discovery Projects scheme (DP150103208).
The Dark Cosmology Centre is funded by the Danish National Research
Foundation.

\bibliographystyle{mn2e}

\bibliography{oir}

\appendix

\section{Galaxy Ellipticity Dispersion and Shape Measurement Weights}
\label{app:shapes_candels}

\begin{figure*}
  \includegraphics[width=1\columnwidth]{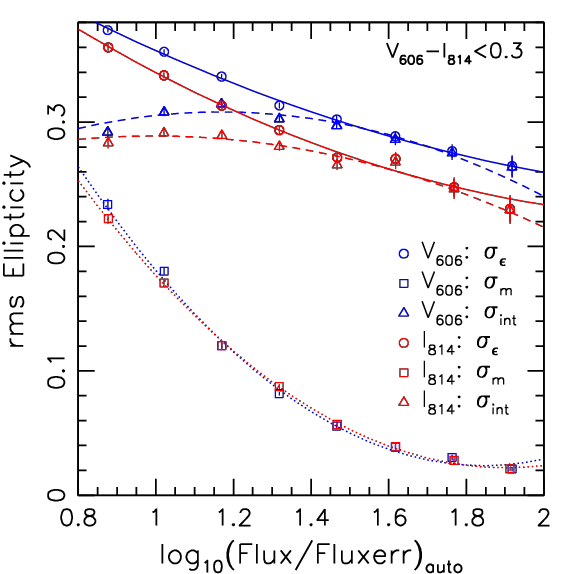}
  \includegraphics[width=1\columnwidth]{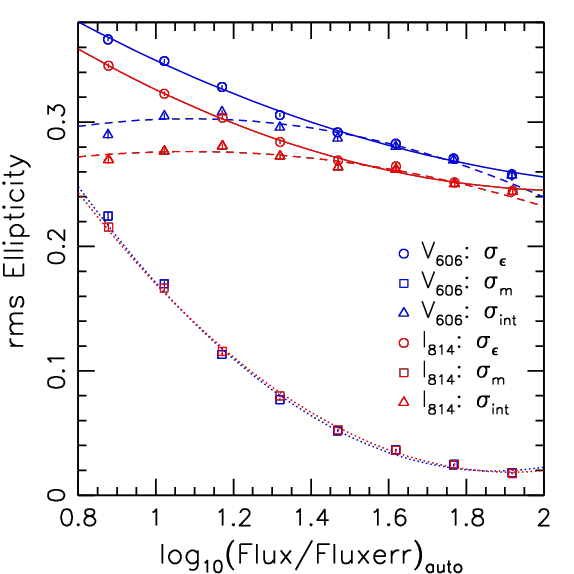}
\caption{Galaxy ellipticity dispersion per ellipticity component as function of the logarithmic flux
  signal-to-noise ratio (measured by \texttt{SExtractor} as
  \texttt{FLUX\_AUTO}/\texttt{FLUXERR\_AUTO}) with ({\it left}) and without
  colour selection ({\it right}), estimated from ACS F606W and F814W data in
  the CANDELS fields (see the text for details), and averaged over both
  ellipticity components.
The circles
show the
  r.m.s. of our KSB ellipticity estimates
  \mbox{$\sigma_\epsilon$}, with polynomial interpolations indicated by the
  solid curves. The squares
show the measurement noise $\sigma_\mathrm{m}$ estimated
from the difference between the ellipticity estimates
in
overlapping tiles, with polynomial interpolations indicated by the
  dotted curves.
The triangles show the estimate for intrinsic shape
  noise
$\sigma_\mathrm{int}=\sqrt{\sigma_\epsilon^2-\sigma_\mathrm{m}^2}$,
with polynomial interpolations indicated by the
  dashed curves.
The
  symbols mark the bin centres, and error-bars indicate the uncertainty
  estimated via bootstrapping.
\label{fig:erms_sn}}
\end{figure*}

\begin{figure*}
  \includegraphics[width=1\columnwidth]{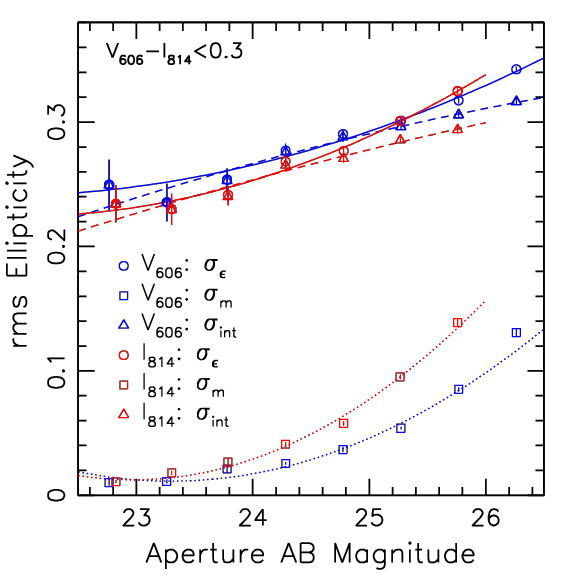}
  \includegraphics[width=1\columnwidth]{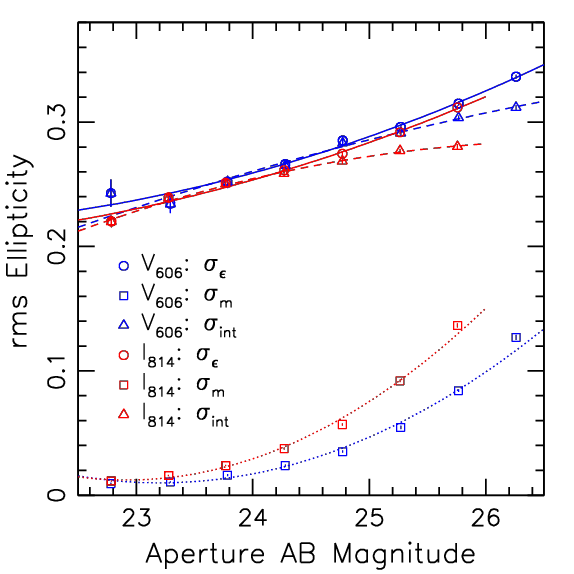}
\caption{Galaxy ellipticity dispersion per ellipticity component as function of AB
  magnitude $V_{606}$ or $I_{814}$. See the caption of Fig.\thinspace\ref{fig:erms_sn} for
  further details.
\label{fig:erms_mag}}
\end{figure*}

\begin{figure*}
  \includegraphics[width=1\columnwidth]{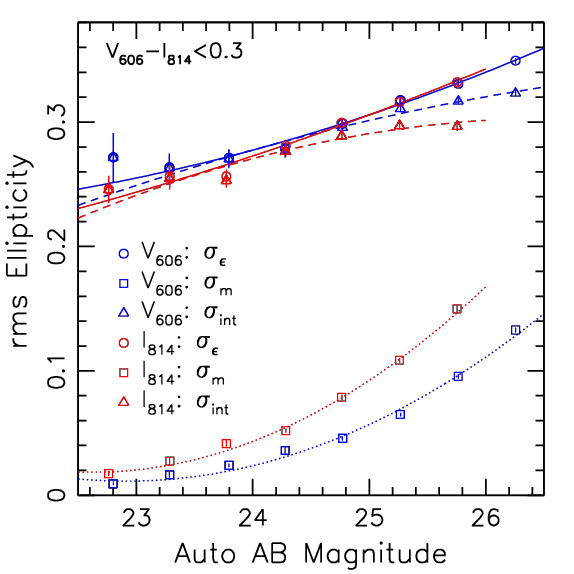}
  \includegraphics[width=1\columnwidth]{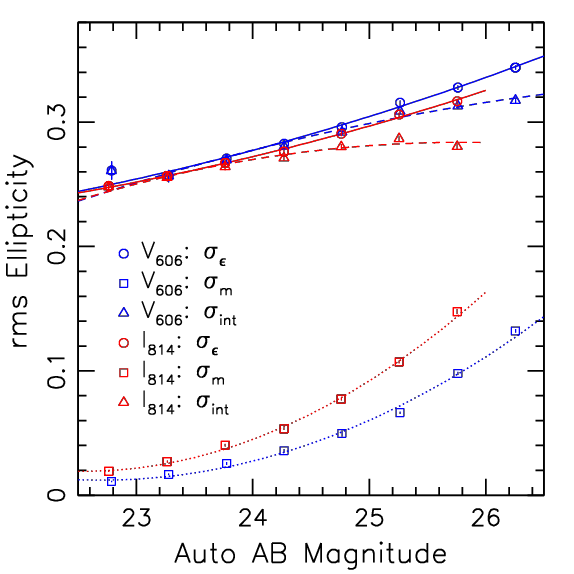}
\caption{Galaxy ellipticity dispersion per ellipticity component as function
  of AB auto
  magnitude from \texttt{SExtractor}.
See the caption of Fig.\thinspace\ref{fig:erms_sn} for
  further details.
\label{fig:erms_magauto}}
\end{figure*}

\begin{figure*}
  \includegraphics[width=1\columnwidth]{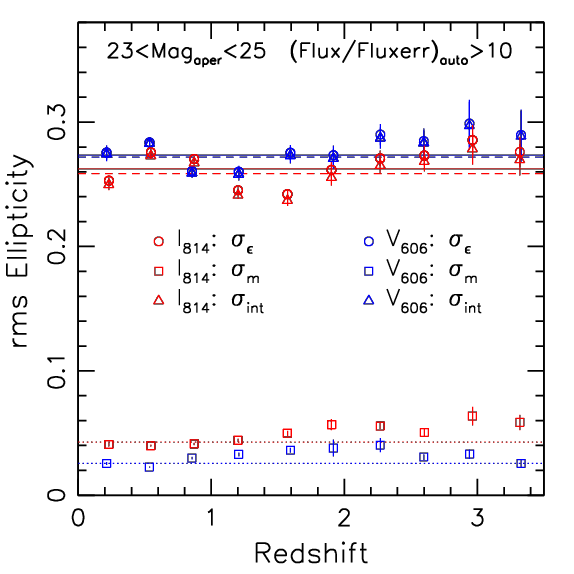}
  \includegraphics[width=1\columnwidth]{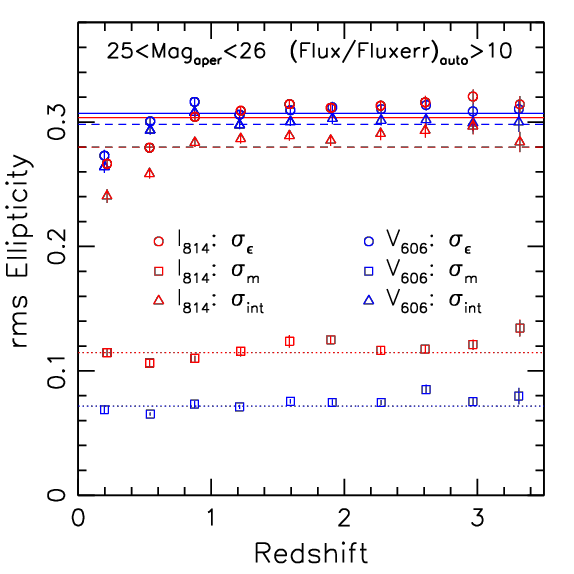}
\caption{Galaxy ellipticity dispersion per ellipticity component as function
  of photometric
  redshift
for bright \mbox{$23<\mathrm{mag}<25$} galaxies ({\it left}), and faint
\mbox{$25<\mathrm{mag}<26$} galaxies ({\it right}). The horizontal lines
show the weighted averages. See the caption of  Fig.\thinspace\ref{fig:erms_sn} for
  further details.
\label{fig:erms_redshift}}
\end{figure*}

As explained in Section \ref{sec:shear} we processed ACS  observations in the
CANDELS fields to be able to mimic our source selection in the photometric
redshift reference catalogues.
These blank field data also enable us to study the galaxy ellipticity distribution
as detailed in this Appendix.
On the one hand this allows us to optimise our weighting scheme for the
current study.
In addition, these estimates can be used to optimise future weak lensing observing
programmes and forecast their performance.
For the latter purpose we have studied shape estimates from both ACS
standard lensing filters F606W and F814W.
This also updates earlier results on the intrinsic ellipticity dispersion
estimated
by
\citet{leauthaud07} for F814W observations in the COSMOS Survey.

\subsection{Method}

Our ellipticity measurements $\epsilon$ provide estimates for the reduced shear
$g$.
We model the measured dispersion of the galaxy ellipticity
$\sigma_\epsilon$
with contributions from the intrinsic galaxy shapes
$\sigma_\mathrm{int}$ and measurement noise $\sigma_\mathrm{m}$
as
\begin{equation}
 \label{eq:sigmae2}
  \sigma_\epsilon^2=\sigma_\mathrm{int}^2+\sigma_\mathrm{m}^2 \,.
\end{equation}
The  contribution from the cosmological shear in CANDELS is small compared to
$\sigma_\epsilon$, and for the purpose of this study we regard it as part of $\sigma_\mathrm{int}$.
To estimate $\sigma_\mathrm{m}$ we
make use of the overlap region of neighbouring
ACS tiles (that have similar noise properties), where we have two
 estimates
 (a,b)
of the ellipticity of each galaxy with two independent realisations of the
measurement noise for identical $\epsilon_\mathrm{int}$.
After rotating the ellipticities to the same coordinate frame, the
dispersion of their
difference  \mbox{$\Delta\epsilon=\epsilon^\mathrm{a}-\epsilon^\mathrm{b}$}
allows us to estimate
\begin{equation}
\sigma_\mathrm{m}^2=\sigma_{\Delta\epsilon}^2/2\,,
\end{equation}
from which we compute $\sigma_\mathrm{int}$  according to (\ref{eq:sigmae2}).
Generally, we quote  r.m.s.  ellipticity  values {\it per ellipticity
  component},
where we compute the average from both components as
\begin{equation}
  \sigma_\epsilon^2= \left(\sigma_{\epsilon,1}^2+\sigma_{\epsilon,2}^2\right)/2\,.
\end{equation}

\subsection{Data}
\label{app:candels:data}
For this analysis we generated and analysed tile-wise F606W and F814W stacks of 4 ACS
exposures each.
We include the initial AEGIS ACS F606W and F814W observations\\
\citep[][Proposal ID 10134]{davis2007}.
Similar to \citet{schrabback07}
we generate F606W stacks in  GOODS-South and  GOODS-North that always combine two
epochs of the observations from \citet[][Proposal IDs 9425,
9583]{giavalisco2004}. In GOODS-South we also include F606W observations
from  GEMS  \citep[][Proposal ID
9500]{rix2004}, which provides some additional overlap with the S14
WFC3/IR-detected catalogues.
Generally, we limit our analysis to the overlap region with the  S14
catalogues to enable the colour selection and provide constraints as
function of photometric redshift.
For the COSMOS and UDS fields we use the  F606W and F814W observations from
CANDELS \citep[][Proposal IDs 12440, 12064]{grogin2011}. Here, the tile-wise
 F606W stacks have slightly shorter integration times of 1.3--1.7{\thinspace}ks compared
 to our targeted $\sim 2$ks depth.
For the constraints on the ellipticity dispersion we therefore
include these observations
only when studying the ellipticity
dispersion as function of flux signal-to-noise ratio, where the impact of the
shallower depth is minimal.

\subsection{Discussion}
We plot our estimates for the measured ellipticity dispersion
$\sigma_\epsilon$,
the intrinsic ellipticity dispersion $\sigma_\mathrm{int}$, and the
measurement noise  $\sigma_\mathrm{m}$ for both ACS filters in Figures
\ref{fig:erms_sn} to \ref{fig:erms_redshift}.

We investigate the dependencies on the logarithmic flux signal-to-noise ratio \mbox{$\log_{10}\left(\mathrm{Flux}/\mathrm{Fluxerr}\right)_\mathrm{auto}$},
defined via the ratio  \texttt{FLUX\_AUTO}/\texttt{FLUXERR\_AUTO}
from \texttt{SExtractor} in Fig.\thinspace\ref{fig:erms_sn},
 on the aperture magnitude in
Fig.\thinspace\ref{fig:erms_mag},
and on the auto magnitude from \texttt{SExtractor} in
Fig.\thinspace\ref{fig:erms_magauto},
in all cases with (left panels) and
without (right panels) applying our colour selection.
As expected, the measurement noise  $\sigma_\mathrm{m}$ increases steeply
towards low signal-to-noise and fainter magnitudes.
This is one of the reasons why $\sigma_\epsilon$
increases towards lower  signal-to-noise and
fainter magnitudes.
Interestingly, we find that $\sigma_\mathrm{int}$ also increases towards
fainter magnitudes.
The analysis of COSMOS data by \citet{leauthaud07} also hinted at this trend
with magnitude, but these authors discussed that it might be an artefact
from their  simplified estimator of the measurement error.
We expect that our estimate of the measurement noise from overlapping tiles
is fairly robust,
and therefore suggest that this indeed appears to be a real effect, showing that
intrinsically fainter galaxies have a broader ellipticity distribution.

As a function of the signal-to-noise ratio we largely observe  the corresponding trend
of an increasing $\sigma_\epsilon$ and $\sigma_\mathrm{int}$ towards lower
\mbox{$\log_{10}\left(\mathrm{Flux}/\mathrm{Fluxerr}\right)_\mathrm{auto}$}, but note that
our estimate for  $\sigma_\mathrm{int}$ flattens at
\mbox{$\log_{10}\left(\mathrm{Flux}/\mathrm{Fluxerr}\right)_\mathrm{auto}\sim 1-1.2$}
and eventually turns over to decreasing $\sigma_\mathrm{int}$.
Using  stacks of different depth we verified that this flattening
is not
intrinsic to the galaxies.
Instead, we expect that
 the validity of Eq.\thinspace\ref{eq:sigmae2} breaks down for large $\sigma_\mathrm{m}$.
In addition,
 selection effects may have some influence,
e.g.
the
cuts applied in size and \mbox{$\mathrm{Tr}[P^g]/2$},
as well as non-Gaussian tails in the measured ellipticity distribution at low
 signal-to-noise.

Comparing the left and right panels in Figures\thinspace\ref{fig:erms_sn} to \ref{fig:erms_magauto}
we find that the application of our colour selection to remove cluster
galaxies has only a relatively small impact on the ellipticity dispersion: Applying the
colour selection \mbox{$V_{606}-I_{814}<0.3$} (which preferentially selects
blue high-$z$ background galaxies) increases $\sigma_\epsilon$ by
\mbox{$0.004\pm 0.002$}
(\mbox{$0.009\pm 0.002$})
and  $\sigma_\mathrm{int}$ by \mbox{$0.004\pm 0.002$}
(\mbox{$0.008\pm 0.002$})
 at magnitudes
\mbox{$24\le\mathrm{mag}_\mathrm{aper}\le 26$}
in the F606W (F814W) filter.
This can be compared to the dependence of the ellipticity dispersion on
photometric redshift shown in Fig.\thinspace\ref{fig:erms_redshift}, where
we split the sample into bright (left panel) and faint (right panel) galaxies.
Over the broad redshift range covered by the HST data the redshift
dependence appears to be relatively weak. Most notably, the faint
galaxies show an increase in $\sigma_\epsilon$ and  $\sigma_\mathrm{int}$
between redshift 0 and \mbox{$\sim 1$}.
In principle, one expects such a trend,
as galaxies at higher redshifts are observed at bluer rest-frame
wavelengths, with stronger light contributions from sites of star formation.
However note that it is more challenging to robustly infer conclusions
on the redshift dependence of the shape distribution, as this is more
strongly affected by large-scale structure variations  \citep[compare
e.g.][]{kannawadi15}.
We therefore suggest to investigate these trends further in the future with larger data sets.

\subsection{Comparing the weak lensing efficiency of F606W and F814W}
\label{app:606vs814}
In Figures\thinspace\ref{fig:erms_sn} to \ref{fig:erms_magauto}
 $\sigma_\mathrm{int}$ is  typically
lower for the analysis of the F814W data than for the F606W images at a
given signal-to-noise ratio or magnitude.
However, when interpreting this one has to
keep in mind that the bins do not contain identical sets of
galaxies.
To facilitate  a fair direct comparison of the performance of both filters
for weak lensing measurements we limit the
analysis to the F606W and F814W AEGIS observations, which were taken under
very similar conditions with similar exposure times.
As a first test, we compare the ellipticity dispersions computed from those
galaxies that have robust shape estimates and
\mbox{$(\mathrm{Flux}/\mathrm{Fluxerr})_\mathrm{auto}>10$}
in {\it both} bands.
Including the matched galaxies
with \mbox{$24<V_{606}<26$} we
find that on average $\sigma_\mathrm{int}$  ($\sigma_\epsilon$)
is lower for the F814W shape estimates by $0.022\pm0.003$ ($0.019\pm0.003$)
compared to the F606W shapes
when no colour selection is applied, and by $0.016\pm0.006$ ($0.009\pm0.004$)
 when blue galaxies are selected with \mbox{$V_{606}-I_{814}<0.3$}.
Hence, we find that intrinsic galaxy shapes are slightly rounder when observed in the
redder filter, which reduces their weak lensing shape noise.
However, the quantity that actually sets the effective noise level for weak lensing studies
is the effective source density after colour selection, which we define as
\begin{equation}
n_\mathrm{eff}=\sum_\mathrm{mag} n(\mathrm{mag}) \times
\left(\frac{\sigma_\epsilon^\mathrm{ref}}{\sigma_\epsilon(\mathrm{mag})}\frac{\langle\beta\rangle(\mathrm{mag})}{\langle\beta\rangle^\mathrm{ref}}\right)^2 \,.
\end{equation}
For a cluster at \mbox{$z_\mathrm{l}=1.0$} we find from the AEGIS data that
$n_\mathrm{eff}$ is higher by a factor 1.28 (1.06) for F606W compared to
F814W when applying (when not applying) the colour selection with \mbox{$V_{606}-I_{814}<0.3$}.
Hence, if only a single band is observed with HST, F606W is slightly more
efficient for the shape measurements than F814W.
However, given that the ratio between the estimates is close to unity, we
expect that programmes which have observations in {\it both} F606W and
F814W can achieve a
higher effective source density when jointly estimating shapes
from both bands.
Our work has shown the necessity for depth-matched colours for the
cluster member removal.
Therefore, we suggest that future HST weak
lensing programmes of clusters at \mbox{$0.7\lesssim z_l\lesssim 1.1$} should
consider to split their observations between F606W and F814W to obtain both
colour estimates and joint shape measurements from both bands.

\subsection{Fitting functions and shape weights}
\label{app:shapeweights}
We compute second-order polynomial interpolations for the ellipticity dispersions
\mbox{$y\in \left\{\sigma_\epsilon, \sigma_\mathrm{int},
      \sigma_\mathrm{m}\right\}$}
as a function of logarithmic signal-to-noise and magnitude \mbox{$x\in
  \left\{\mathrm{log}_{10}(\mathrm{Flux}/\mathrm{Fluxerr})_\mathrm{auto},
      \mathrm{Mag}_\mathrm{aper}, \mathrm{Mag}_\mathrm{auto} \right\}$}
within limits \mbox{$x_\mathrm{min}<x<x_\mathrm{max}$}
as
\begin{equation}
y=a+b\hat{x}+c\hat{x}^2\,,
\end{equation}
where
\mbox{$\hat{x}=x-x_\mathrm{min}$}.
For our weak lensing analysis of SPT clusters we  compute empirical shape weights for galaxy $i$ as
\begin{equation}
\label{eq:shapeweights}
w_i=\left[ \sigma_\epsilon^\mathrm{fit}\left(\mathrm{log}_{10}(\mathrm{Flux}/\mathrm{Fluxerr})_{\mathrm{auto},i}\right) \right]^{-2}\,.
\end{equation}
from the interpolation of $\sigma_\epsilon$ as function of the logarithmic
signal-to-noise ratio for the \mbox{$V_{606}-I_{814}<0.3$} colour-selected CANDELS galaxies.
We plot the best-fit
interpolations in Figures \ref{fig:erms_sn} to \ref{fig:erms_magauto}
and list their
polynomial coefficients
in Table \ref{tab:sigmae_fits}.

\begin{table*}
\caption{Parameters and coefficients for the polynomial interpolation of the
  ellipticity dispersions $\sigma_\epsilon$, $\sigma_\mathrm{int}$, and
  $\sigma_\mathrm{m}$ in CANDELS as function of magnitude and logarithmic
  signal-to-noise ratio.
\label{tab:sigmae_fits}}
\begin{center}
\begin{tabular}{ccccccrrrr}
\hline
\hline
Band & Colour & $x$ & $x_\mathrm{min}$  & $x_\mathrm{max}$  & $y$ & \multicolumn{1}{c}{$a$}   &
\multicolumn{1}{c}{$b$}   & \multicolumn{1}{c}{$c$} \\
\hline
$I_{814}$ & all & $\mathrm{log}_{10}(\mathrm{Flux}/\mathrm{Fluxerr})_\mathrm{auto}$ & 0.75 & 2 & $\sigma_\epsilon$ & $0.36777$ & $-0.18359$ & $0.06843$\\
$I_{814}$ & all & $\mathrm{log}_{10}(\mathrm{Flux}/\mathrm{Fluxerr})_\mathrm{auto}$ & 0.75 & 2 & $\sigma_\mathrm{int}$ & $0.27050$ & $0.03504$ & $-0.05252$\\
$I_{814}$ & all & $\mathrm{log}_{10}(\mathrm{Flux}/\mathrm{Fluxerr})_\mathrm{auto}$ & 0.75 & 2 & $\sigma_\mathrm{m}$ & $0.26390$ & $-0.42101$ & $0.18058$\\
$I_{814}$ & all & $\mathrm{Mag}_\mathrm{aper}$ & 22.5 & 26 & $\sigma_\epsilon$ & $0.22123$ & $0.01644$ & $0.00340$\\
$I_{814}$ & all & $\mathrm{Mag}_\mathrm{aper}$ & 22.5 & 26 & $\sigma_\mathrm{int}$ & $0.21232$ & $0.03411$ & $-0.00402$\\
$I_{814}$ & all & $\mathrm{Mag}_\mathrm{aper}$ & 22.5 & 26 & $\sigma_\mathrm{m}$ & $0.01480$ & $-0.01211$ & $0.01453$\\
$I_{814}$ & all & $\mathrm{Mag}_\mathrm{auto}$ & 22.5 & 26 & $\sigma_\epsilon$ & $0.24301$ & $0.01649$ & $0.00201$\\
$I_{814}$ & all & $\mathrm{Mag}_\mathrm{auto}$ & 22.5 & 26 & $\sigma_\mathrm{int}$ & $0.23712$ & $0.02839$ & $-0.00433$\\
$I_{814}$ & all & $\mathrm{Mag}_\mathrm{auto}$ & 22.5 & 26 & $\sigma_\mathrm{m}$ & $0.01925$ & $-0.00090$ & $0.01200$\\
$I_{814}$ & $V_{606}-I_{814}<0.3$ & $\mathrm{log}_{10}(\mathrm{Flux}/\mathrm{Fluxerr})_\mathrm{auto}$ & 0.75 & 2 & $\sigma_\epsilon$ & $0.38420$ & $-0.19190$ & $0.05716$\\
$I_{814}$ & $V_{606}-I_{814}<0.3$ & $\mathrm{log}_{10}(\mathrm{Flux}/\mathrm{Fluxerr})_\mathrm{auto}$ & 0.75 & 2 & $\sigma_\mathrm{int}$ & $0.28447$ & $0.03555$ & $-0.07253$\\
$I_{814}$ & $V_{606}-I_{814}<0.3$ & $\mathrm{log}_{10}(\mathrm{Flux}/\mathrm{Fluxerr})_\mathrm{auto}$ & 0.75 & 2 & $\sigma_\mathrm{m}$ & $0.27431$ & $-0.43743$ & $0.18966$\\
$I_{814}$ & $V_{606}-I_{814}<0.3$ & $\mathrm{Mag}_\mathrm{aper}$ & 22.5 & 26 & $\sigma_\epsilon$ & $0.22602$ & $0.00757$ & $0.00698$\\
$I_{814}$ & $V_{606}-I_{814}<0.3$ & $\mathrm{Mag}_\mathrm{aper}$ & 22.5 & 26 & $\sigma_\mathrm{int}$ & $0.21238$ & $0.02943$ & $-0.00130$\\
$I_{814}$ & $V_{606}-I_{814}<0.3$ & $\mathrm{Mag}_\mathrm{aper}$ & 22.5 & 26 & $\sigma_\mathrm{m}$ & $0.01583$ & $-0.01478$ & $0.01571$\\
$I_{814}$ & $V_{606}-I_{814}<0.3$ & $\mathrm{Mag}_\mathrm{auto}$ & 22.5 & 26 & $\sigma_\epsilon$ & $0.23050$ & $0.02525$ & $0.00195$\\
$I_{814}$ & $V_{606}-I_{814}<0.3$ & $\mathrm{Mag}_\mathrm{auto}$ & 22.5 & 26 & $\sigma_\mathrm{int}$ & $0.22288$ & $0.03886$ & $-0.00469$\\
$I_{814}$ & $V_{606}-I_{814}<0.3$ & $\mathrm{Mag}_\mathrm{auto}$ & 22.5 & 26 & $\sigma_\mathrm{m}$ & $0.01869$ & $-0.00322$ & $0.01307$\\
$V_{606}$ & all & $\mathrm{log}_{10}(\mathrm{Flux}/\mathrm{Fluxerr})_\mathrm{auto}$ & 0.75 & 2 & $\sigma_\epsilon$ & $0.38882$ & $-0.16903$ & $0.05008$\\
$V_{606}$ & all & $\mathrm{log}_{10}(\mathrm{Flux}/\mathrm{Fluxerr})_\mathrm{auto}$ & 0.75 & 2 & $\sigma_\mathrm{int}$ & $0.29414$ & $0.05089$ & $-0.07555$\\
$V_{606}$ & all & $\mathrm{log}_{10}(\mathrm{Flux}/\mathrm{Fluxerr})_\mathrm{auto}$ & 0.75 & 2 & $\sigma_\mathrm{m}$ & $0.27001$ & $-0.44604$ & $0.19850$\\
$V_{606}$ & all & $\mathrm{Mag}_\mathrm{aper}$ & 22.5 & 26.5 & $\sigma_\epsilon$ & $0.22918$ & $0.01439$ & $0.00371$\\
$V_{606}$ & all & $\mathrm{Mag}_\mathrm{aper}$ & 22.5 & 26.5 & $\sigma_\mathrm{int}$ & $0.21549$ & $0.03276$ & $-0.00186$\\
$V_{606}$ & all & $\mathrm{Mag}_\mathrm{aper}$ & 22.5 & 26.5 & $\sigma_\mathrm{m}$ & $0.01564$ & $-0.01605$ & $0.01140$\\
$V_{606}$ & all & $\mathrm{Mag}_\mathrm{auto}$ & 22.5 & 26.5 & $\sigma_\epsilon$ & $0.24435$ & $0.01885$ & $0.00208$\\
$V_{606}$ & all & $\mathrm{Mag}_\mathrm{auto}$ & 22.5 & 26.5 & $\sigma_\mathrm{int}$ & $0.23647$ & $0.03082$ & $-0.00233$\\
$V_{606}$ & all & $\mathrm{Mag}_\mathrm{auto}$ & 22.5 & 26.5 & $\sigma_\mathrm{m}$ & $0.01257$ & $-0.00372$ & $0.00912$\\
$V_{606}$ & $V_{606}-I_{814}<0.3$ & $\mathrm{log}_{10}(\mathrm{Flux}/\mathrm{Fluxerr})_\mathrm{auto}$ & 0.75 & 2 & $\sigma_\epsilon$ & $0.39491$ & $-0.16019$ & $0.04158$\\
$V_{606}$ & $V_{606}-I_{814}<0.3$ & $\mathrm{log}_{10}(\mathrm{Flux}/\mathrm{Fluxerr})_\mathrm{auto}$ & 0.75 & 2 & $\sigma_\mathrm{int}$ & $0.29096$ & $0.08216$ & $-0.09812$\\
$V_{606}$ & $V_{606}-I_{814}<0.3$ & $\mathrm{log}_{10}(\mathrm{Flux}/\mathrm{Fluxerr})_\mathrm{auto}$ & 0.75 & 2 & $\sigma_\mathrm{m}$ & $0.28751$ & $-0.48200$ & $0.22022$\\
$V_{606}$ & $V_{606}-I_{814}<0.3$ & $\mathrm{Mag}_\mathrm{aper}$ & 22.5 & 26.5 & $\sigma_\epsilon$ & $0.24319$ & $0.00763$ & $0.00486$\\
$V_{606}$ & $V_{606}-I_{814}<0.3$ & $\mathrm{Mag}_\mathrm{aper}$ & 22.5 & 26.5 & $\sigma_\mathrm{int}$ & $0.22404$ & $0.03115$ & $-0.00180$\\
$V_{606}$ & $V_{606}-I_{814}<0.3$ & $\mathrm{Mag}_\mathrm{aper}$ & 22.5 & 26.5 & $\sigma_\mathrm{m}$ & $0.01884$ & $-0.01892$ & $0.01190$\\
$V_{606}$ & $V_{606}-I_{814}<0.3$ & $\mathrm{Mag}_\mathrm{auto}$ & 22.5 & 26.5 & $\sigma_\epsilon$ & $0.24607$ & $0.01653$ & $0.00295$\\
$V_{606}$ & $V_{606}-I_{814}<0.3$ & $\mathrm{Mag}_\mathrm{auto}$ & 22.5 & 26.5 & $\sigma_\mathrm{int}$ & $0.23311$ & $0.03313$ & $-0.00234$\\
$V_{606}$ & $V_{606}-I_{814}<0.3$ & $\mathrm{Mag}_\mathrm{auto}$ & 22.5 & 26.5 & $\sigma_\mathrm{m}$ & $0.01309$ & $-0.00857$ & $0.01044$\\
\hline
\end{tabular}
\end{center}
\vspace{0.2cm}

\end{table*}

\section{Non-matching galaxies in CANDELS}
\label{se:app:non_matches}
We have investigated the \mbox{$\sim 2.4\%$} of non-matching
galaxies between our CANDELS F606W shear catalogue and the S14 photo-$z$
catalogue (see Sect.\thinspace\ref{se:photo_ref_cats})
by visually inspecting a random subset.
Most of the non-matching
galaxies can be explained through differences in the object detection
or deblending given the different detection bands (optical F606W vs. NIR
F125W+F140W+F160W). For \mbox{$\sim 0.7\%$} of the total galaxies  centroid shifts prevent a
match. These should not affect the source redshift distribution.
For \mbox{$\sim 1.2\%$} the S14 catalogue contains a single object which is
associated with two deblended objects in our F606W shear catalogue.
If such differences in the  deblending would occur independent of redshift,
there would be no net effect on the source redshift distribution.
However, such differences might be more frequent for high-$z$ ($z\gtrsim 1$) galaxies,
where the F606W images probe rest-frame UV wavelengths and mostly detect
sites of star formation, while the IR imaging probes the stellar content of
the galaxies.
Finally, \mbox{$\sim 0.4\%$} of the total galaxies show clear isolated galaxies in our F606W shear
catalogue that are missing in the  S14 NIR-detected catalogue, possibly
because they are too faint and too blue.

To obtain a rough estimate for the resulting uncertainty of these effects on
our analysis,  we assume a  scenario where both the missing isolated galaxies
(\mbox{$\sim 0.4\%$}) plus the excess half of the differently deblended galaxies
(\mbox{$\sim 0.6\%$}) constitute an excess population of 100\% blue (\mbox{$V_{606}-I_{814}<0.3$}) galaxies
at high redshifts (\mbox{$z\simeq 2$}).
This scenario is pessimistic for the differently deblended galaxies as explained
above (no impact if the effect is redshift independent).
For the missing isolated galaxies the scenario is likely to be realistic, but we
note that it would also overestimate the
impact in case some of the galaxies are redder and removed by our
\mbox{$V_{606}-I_{814}<0.3$} colour selection.
At our median cluster redshift \mbox{$z_\mathrm{l}=0.88$} the scenario leads
to a {\it relative} increase in $\langle\beta\rangle$ by only +0.5\%, thanks
to our colour selection which already selects mostly  \mbox{$z>1$} galaxies.

\section{Cross-check for the redshift distribution using spatial
  cross-correlations}
\label{app:candels_x}

\begin{figure*}
  \includegraphics[width=0.69\columnwidth]{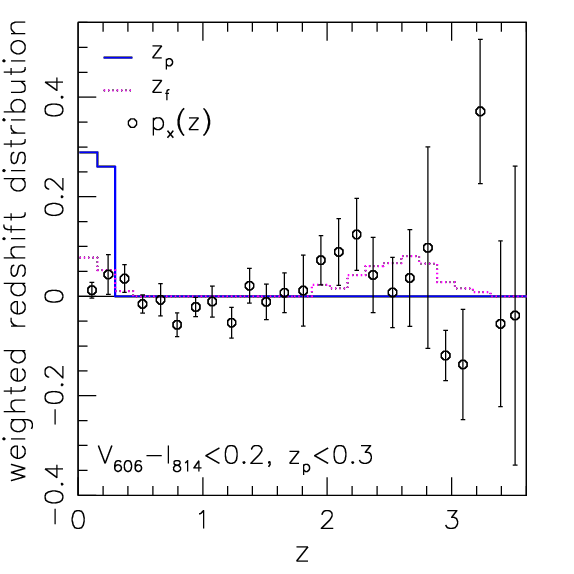}
 \includegraphics[width=0.69\columnwidth]{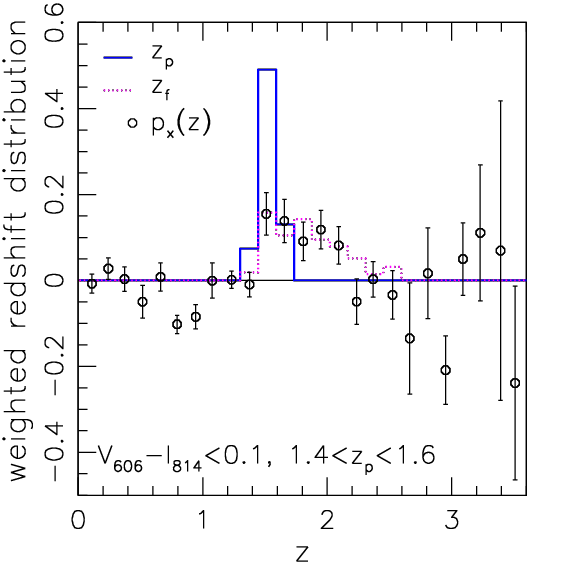}
 \includegraphics[width=0.69\columnwidth]{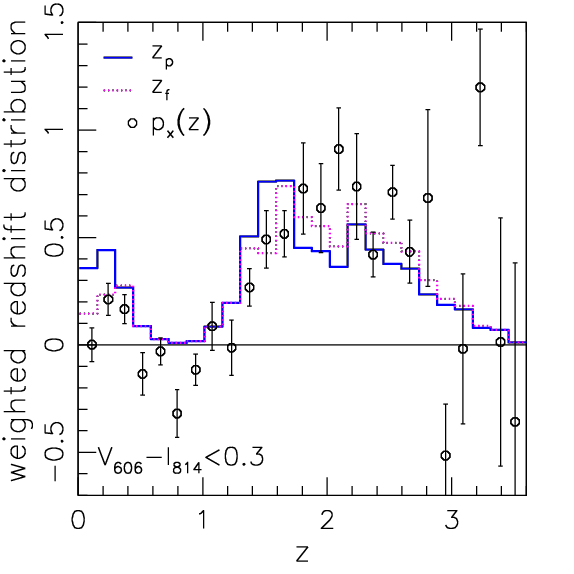}
\caption{Comparison of the histograms of the 3D-HST peak photometric redshift $z_\mathrm{p}$
   (blue solid) and the redshifts $z_\mathrm{f}$ that are statistically
   corrected based on the HUDF comparison
   (magenta dotted)
to the reconstructed redshift distribution \mbox{$p_\mathrm{x}(z)$} inferred
from the cross-correlation analysis (black circles) using
colour-selected CANDELS galaxies with \mbox{$24<V_{606}<26.5$} and applying shape weights
from our CANDELS shear catalogue.
The {\it left} and {\it middle} panels correspond to the galaxies for which we apply the
corrections for catastrophic outliers or redshift focusing, respectively,
while the
{\it right} panel includes the  full sample.
Error-bars show the dispersion of the \mbox{$p_\mathrm{x}(z)$} estimates
when splitting the total sample into ten sub-areas and bootstrapping the
contributing sub-areas. The large scatter at  \mbox{$z\gtrsim 2.8$}
is caused by the small  spectroscopic sample at these redshifts.
The negative peak at \mbox{$0.7\lesssim z \lesssim 1.0$}
is an artefact resulting from spatial density variations in the
spectroscopic sample and the colour selection applied to the photometric sample.  \label{fig:zdist_cross}}
\end{figure*}

A number of  studies have explored the use of spatial
cross-correlation techniques
to constrain
source redshift
distributions
\citep[e.g.][]{newman08,matthews10,benjamin13}.
In particular, \citet{newman08,matthews10,schmidt13,rahman15,rahman16,scottez16}
aim at reconstructing the redshift distribution of a sample with an unknown
redshift distribution  (``photometric sample'')
via its
 spatial cross-correlation with galaxies in redshift slices of an incomplete spectroscopic reference
 sample.
The cross-correlation
amplitude increases if a larger fraction of the photometric sample is located
within the redshift range of the corresponding slice. As a result,
information on the redshift distribution  of
the photometric
sample
can be inferred.
When using photometric samples with a broad redshift distribution
the accuracy of the method  is limited by how well a potential
redshift evolution of the relative galaxy bias
between the populations
can be accounted for \citep[e.g.][]{rahman15}.
However, the impact of this limitation can be reduced if the photometric sample can
be split into sub-samples with relatively narrow individual redshift
distributions, as suggested by \cite{schmidt13,menard13} and applied to SDSS
data in \cite{rahman16}.
The CANDELS data are well suited to employ this technique, as
considerable spectroscopic (or grism) redshift samples are available (Sect.\thinspace\ref{se:zdist_candels_checks}), and
given
that the 3D-HST photo-$z$s allow for a relatively clean subdivision into
narrower redshift slice for most of the galaxies.

We employ the \textsc{The-wiZZ}\footnote{\url{https://github.com/morriscb/The-wiZZ}} implementation \citep{morrison17} of the cross-correlation technique
described in \citet{schmidt13,menard13} to obtain an independent cross-check for our
estimate of the colour-selected CANDELS redshift distribution.
For this we use the combined sample of high-fidelity spectroscopic and
high-quality grism redshifts
(see Sect.\thinspace\ref{se:zdist_candels_checks}) as spectroscopic reference
sample (without colour selection)
and the colour-selected photo-$z$ sample as photometric sample,
splitting galaxies into 25 linear bins in $z_\mathrm{s/g}$ or
$z_\mathrm{p}$, respectively, between \mbox{$z=0.01$} and \mbox{$z=3.6$}.
We compare the estimate for the redshift probability distribution \mbox{$p_\mathrm{x}(z)$} obtained from the cross-correlation analysis using physical separations between 30 kpc and 300 kpc
to the $z_\mathrm{p}$ and $z_\mathrm{f}$  histograms in
Fig.\thinspace\ref{fig:zdist_cross} using galaxies with
\mbox{$24<V_{606}<26.5$} and the actual shape weights from our CANDELS shear
catalogue.

The left and the middle panels
of Fig.\thinspace\ref{fig:zdist_cross}
correspond to the subset of CANDELS galaxies
for which we implemented statistical corrections
(see Sect.\thinspace\ref{sec:correct_sys_photoz}) for catastrophic redshift
outliers (\mbox{$V_{606}-I_{814}<0.2$}, \mbox{$z_\mathrm{p}<0.3$}) or
redshift focusing (\mbox{$V_{606}-I_{814}<0.1$},
\mbox{$1.4<z_\mathrm{p}<1.6$}), respectively.
In both cases we find that the redshift distribution inferred from the
 cross-correlation analysis is largely consistent with the statistically
 corrected distribution based on the HUDF analysis ($z_\mathrm{f}$), while it is clearly incompatible with the
uncorrected distribution in the selected $z_\mathrm{p}$ ranges,
providing an independent confirmation for the HUDF-based correction scheme.
The right panel of Fig.\thinspace\ref{fig:zdist_cross}
shows the  combined  \mbox{$p_\mathrm{x}(z)$} reconstruction for the full
colour selected sample (\mbox{$V_{606}-I_{814}<0.3$}).
Consistent with the other panels the reconstruction  describes the $z_\mathrm{f}$ histogram better
than
the $z_\mathrm{p}$  histogram, both at low redshifts  (\mbox{$z< 0.3$}) and
around the
broad peak at \mbox{$z\sim 2$}.

The statistical error-bars shown in Fig.\thinspace\ref{fig:zdist_cross}
indicate the dispersion of the  \mbox{$p_\mathrm{x}(z)$} reconstruction
when  splitting the combined CANDELS data set into 10 subareas of
equal area and obtaining 1000 bootstrap resamples of the subareas included
in the analysis.
We expect that this yields a good approximation for the statistical
uncertainty for most of the redshift range of interest. However,
at the highest redshifts (\mbox{$z\gtrsim 2.8$}) the spectroscopic samples become very small
(compare Fig.\thinspace\ref{fig:zdist_f814w}), likely introducing
additional uncertainties that are not fully captured by the error-bars.
This is also suggested by the large fluctuations of both the recovered
\mbox{$p_\mathrm{x}(z)$}  and the error-bars between neighbouring high-$z$
bins.

We note the substantially negative \mbox{$p_\mathrm{x}(z)$} reconstructions
at \mbox{$0.7\lesssim z \lesssim 1.0$}
in the middle and right panel of  Fig.\thinspace\ref{fig:zdist_cross}.
 At these redshifts the full spectroscopic sample contains a large number of
 galaxies (no colour selection applied to the spectroscopic sample). We therefore expect that the error-bars are robust and that the
 negative \mbox{$p_\mathrm{x}(z)$} estimates are indeed significant.
We interpret these negative \mbox{$p_\mathrm{x}(z)$} values as a spurious effect
caused by our colour selection, which explicitly
removes galaxies at these redshifts from the photometric sample.
Therefore, the photometric sample is spatially underdense in regions
that are physically overdense at these redshifts.
In contrast, the spectroscopic sample is
spatially
over-represented in regions of physical overdensities at these redshifts.
This results in a net anti-correlation between the samples and negative
\mbox{$p_\mathrm{x}(z)$} estimates.
As a possible solution to this problem \cite{rahman15} suggest to homogenise the spatial density of the
spectroscopic sample by removing galaxies in overdense regions.
However, as the spectroscopic sample employed in our analysis (14,472
galaxies)
is already much smaller than the sample  employed by \cite{rahman15} (791,546
galaxies)
we do not follow this approach.
As an approximate solution for this systematic effect
we instead set the \mbox{$p_\mathrm{x}(z)$} values of the two bins in Fig.\thinspace\ref{fig:zdist_cross} at
\mbox{$0.7< z < 1.0$} to zero when computing $\langle\beta \rangle$ as
described in the next paragraph.
This is justified by multiple tests presented in this work that suggest that
the residual contamination by galaxies at these redshifts should be very low
and close to zero (Sections \ref{se:photo_color_select_acs}, \ref{se:zdist_candels_checks}, \ref{se:source_sel_scatter}, \ref{se:photo_number_density_tests}).
However, outside this redshift range we treat bins with negative   \mbox{$p_\mathrm{x}(z)$}
     as negative contributions in the computation of $\langle\beta
\rangle$.
This is needed in order to achieve unbiased results
in the case of purely  statistical scatter that has equal chance to be
positive or negative.

For a quantitative comparison of the  \mbox{$p_\mathrm{x}(z)$} distribution
and the histograms shown in the right panel of
Fig.\thinspace\ref{fig:zdist_cross} we compute $\langle\beta \rangle$
for our median cluster redshift \mbox{$z_\mathrm{l}=0.88$},
dealing with negative \mbox{$p_\mathrm{x}(z)$} as explained in the previous
paragraph, and generally limiting the considered redshift range to
\mbox{$z<3.2$}
to minimise the impact of the highest-$z$ data points for which the
\mbox{$p_\mathrm{x}(z)$} recovery suffers the strongest from the small
spectroscopic sample.
The resulting  \mbox{$\langle\beta (p_\mathrm{x}(z))\rangle=0.403\pm 0.017$} from the
cross-correlation analysis
(the error indicates the statistical scatter from the bootstrap resamples) is consistent with
\mbox{$\langle\beta  (z_\mathrm{f})\rangle=0.366\pm 0.008$}
 from the HUDF-corrected catalogues
within $2\sigma$.
We conclude that the cross-correlation analysis independently supports the
results from the HUDF analysis, but note that the spectroscopic samples
within the relatively small CANDELS areas are not yet sufficiently large to
constrain the redshift distribution with very high precision.

\section{Details of the ACS+FORS2 colour measurements and the accounting for photometric scatter }
\label{app:details_fors2color_scatter}

\subsection{ACS+FORS2 colour measurement}
\label{app:details_fors2color}
To measure colours between the
F606W and FORS2 I-band images
 images we
convolve each mosaic
F606W  image with a Gaussian kernel such that
the resulting PSF has the same \texttt{FLUX\_RADIUS} measured by
\texttt{Source Extractor}
as the corresponding FORS2  I-band
 image (we empirically account
for the impact of non-Gaussian VLT PSF profiles in
Appendix \ref{app:tiecolor}).
For some of the FORS2 stacks we found small residual systematic offsets of  object
positions in some image regions with respect to their location in the
corresponding ACS mosaic  (typically \mbox{$\lesssim 0\farcs3$}).
To not bias the colour measurement, we therefore fit and subtract a smooth
5th-order 2D-polynomial interpolation of the measured positional offsets to the
catalogue positions. We overlayed and visually inspected these corrections
on all images to ensure that they are robust.
We then measure object fluxes in circular  apertures with diameter 1\farcs5
both in the VLT and the convolved ACS image.
We transform them into
magnitudes, correct these for galactic extinction,
 and compute the colour estimate \mbox{$V_{606,\mathrm{con}}-I_\mathrm{FORS2}$}.

\begin{figure*}
  \includegraphics[width=0.9\columnwidth]{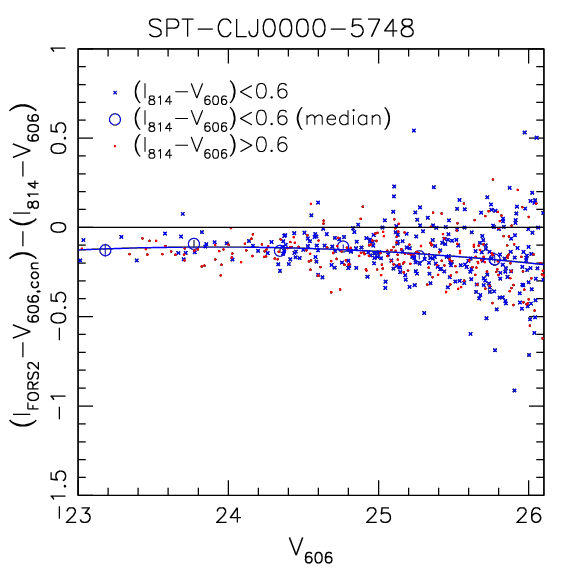}
  \includegraphics[width=0.9\columnwidth]{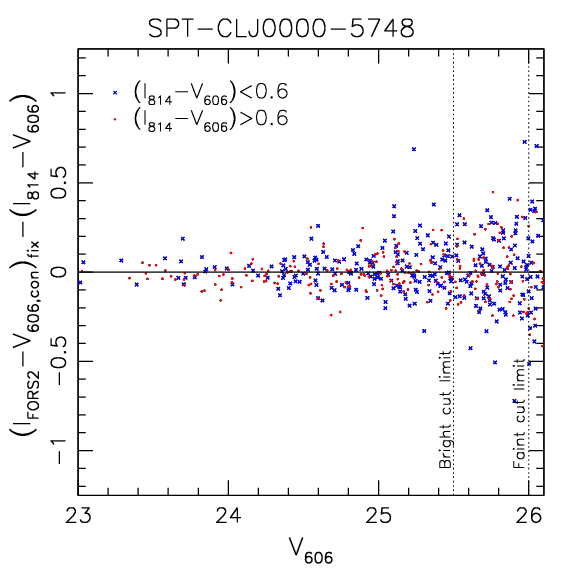}
  \includegraphics[width=0.9\columnwidth]{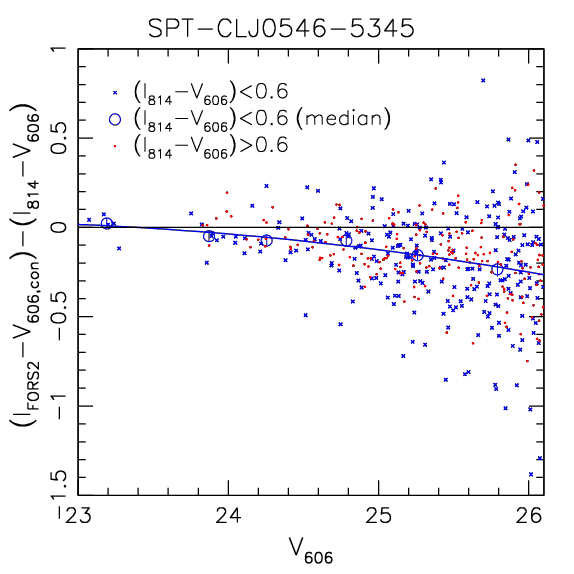}
  \includegraphics[width=0.9\columnwidth]{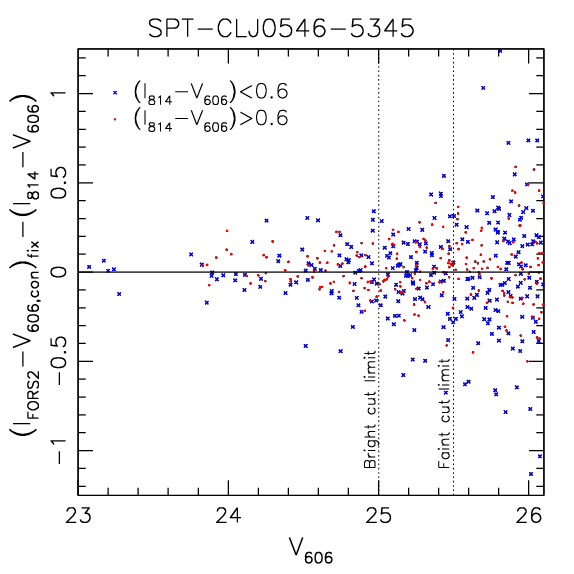}
\caption{Details on the colour selection for
SPT-CL{\thinspace}$J$0000$-$5748 ({\it top}) and
SPT-CL{\thinspace}$J$0546$-$5345 ({\it bottom}):
{\it Left:} Difference between the colour
  \mbox{$(I_\mathrm{FORS2}-V_{606,\mathrm{con}})$} measured with the FORS2
I-band and
 the convolved HST/ACS
F606W images,
and the colour
  \mbox{$(I_{814}-V_{606})$} measured from the unconvolved HST/ACS data in
  the inner cluster region, as function of \mbox{$V_{606}$}. Small blue crosses indicate blue galaxies
  with \mbox{$(I_{814}-V_{606})<0.6$}, while red points show red galaxies
  with \mbox{$(I_{814}-V_{606})>0.6$}. The open circles
  mark the median values for the blue galaxies within 0.5\thinspace mag wide magnitude
  bins, with error-bars indicating the uncertainty on the mean for a
  Gaussian distribution, and the curve showing their best-fit second-order
  polynomial interpolation. The {\it right} panels show the same data after
  subtraction of this function. We sample the photometric scatter
  distribution for the ACS-FORS2 selection from this distribution of offsets.
Because of the lower scatter in the
deeper FORS2 data of SPT-CL{\thinspace}$J$0000$-$5748 we can include fainter
galaxies in the  ACS-FORS2 selection than for SPT-CL{\thinspace}$J$0546$-$5345 (see Table \ref{tab:vltdata} and the indicated
bright/faint cut limits).
\label{fig:scatter}}
\end{figure*}

\subsection{Tying the ACS+FORS2 colours to the ACS-only colours}
\label{app:tiecolor}
We have ACS-based \mbox{$V_{606} - I_{814}$}
and ACS+FORS2-based \mbox{$V_{606,\mathrm{con}}-I_\mathrm{FORS2}$}
colour estimates for the galaxies in the inner cluster regions.
We use these galaxies to refine the calibration of the
\mbox{$V_{606,\mathrm{con}}-I_\mathrm{FORS2}$}
colours for all galaxies and tie them to the \mbox{$V_{606} - I_{814}$} colour selection
available in the 3D-HST CANDELS catalogues.
The left panels of Fig.\thinspace\ref{fig:scatter} shows the difference
of these colour estimates as function of \mbox{$V_{606}$} for two example
clusters.
The top row corresponds to SPT-CL{\thinspace}$J$0000$-$5748, which has one of
the deepest and best-seeing FORS2 I-band
stacks in our sample,
resulting in relatively moderate photometric scatter. Here the analysis
reveals a \mbox{$\sim 0.11$}\thinspace mag colour offset for bright
galaxies.
We expect that this offset is in part caused by the offset in
Eq.\thinspace(\ref{eq:picklesoffset}). Further contributions might come from
uncertainties in the $I_\mathrm{FORS2}$
zero-point calibration due to  the small number of stars available for its
determination, or inaccuracies in the PSF homogenisation.
In comparison, the bottom row reveals a larger photometric
scatter for SPT-CL{\thinspace}$J$0546$-$5345, which has a shallower magnitude
limit and worse image quality (see Table \ref{tab:vltdata}).
For such VLT data
we typically detect a shift of the median colour
difference (indicated through the open circles) at faint magnitudes towards
negative values.
In part this is caused by the
asymmetric and biased scatter in
logarithmic magnitude space.
However, further effects could lead to a
magnitude-dependent colour offset: for example, we acknowledge that our PSF
homogenisation
only ensures equal flux radii between the bands. However, residual
differences in the actual PSF shapes might lead to slightly different
fractions in the total PSF flux lost outside the aperture.
This would lead to a magnitude-dependent colour offset given that fainter
objects are typically less resolved.
Understanding the exact
combination of these
effects
 for each
cluster
is not necessary
given that we directly tie the
\mbox{$V_{606,\mathrm{con}}-I_\mathrm{FORS2}$} colours to the \mbox{$V_{606}
  - I_{814}$} colours empirically:
To do so, we fit the median values of the colour offsets determined in
0.5\thinspace mag-wide bins between \mbox{$23<V_{606}<26$} with a second
order polynomial in $V_{606}$ and subtract this model from all
\mbox{$V_{606,\mathrm{con}}-I_\mathrm{FORS2}$}
colour estimates in the cluster field to obtain
\mbox{$(V_{606,\mathrm{con}}-I_\mathrm{FORS2})_\mathrm{fix}$} (see Fig.\thinspace\ref{fig:scatter}).
We only use relatively blue galaxies with \mbox{$V_{606}
  - I_{814}<0.6$} to derive this fit. This is motivated by small
differences in the effective filter curves of $I_\mathrm{FORS2}$ and
$I_{814}$.
In particular, $I_\mathrm{FORS2}$ cuts off transmission red-wards of \mbox{$\sim 870$ nm},
while  $I_{814}$ has a transmission tail out to \mbox{$\sim 960$ nm}. Thus,
we expect non-negligible colour differences for very red objects.
Given that we generally apply fairly blue cuts in colour this is
not a problem for our analysis.
However, we exclude red galaxies when deriving the fit as they are over-represented compared to CANDELS
in the cluster fields.

\subsection{Accounting for photometric scatter}
\label{app:scatter}

\subsubsection{ACS-only colour selection}
\label{sec:noise:acsonly}

In the inner cluster regions covered by the
F606W and
F814W ACS
images we include galaxies in the magnitude range
\mbox{$24<V_{606}<26.5$}.
The brighter magnitude limit has been chosen as galaxies passing our
colour selection at even brighter magnitudes are dominated by foreground
galaxies.
The fainter magnitude limit approximately matches the
\mbox{$S/N$} cut applied in the weak lensing shape analysis (see
Sect.\thinspace\ref{sec:shear}).
Our ACS images have typical $5\sigma$ limits for the adopted 0\farcs7
apertures of \mbox{$V_{606,\mathrm{lim}}=27.15$} and
\mbox{$I_{814,\mathrm{lim}}=26.60$}.
Therefore, the faintest galaxies included at the colour cut
(\mbox{$V_{606}=26.5$}, \mbox{$V_{606}
  - I_{814}=0.3$})    still have fairly high
photometric signal-to-noise \mbox{$(S/N)_{606}=9.1$} and
\mbox{$(S/N)_{814}=7.2$}.
Accordingly, photometric noise has only minor impact on the colour selection
for these galaxies.
Nonetheless, we account for it by adding random Gaussian scatter
to the  \citetalias{skelton14} catalogues, which are typically based on
deeper ACS mosaic stacks compared to the ones used for our shape analysis, prior to
the colour selection,
such that they have the same limiting magnitudes in $V_{606}$ and $I_{814}$
as our cluster field observations.
Also, we apply a slightly bluer
colour selection for the galaxies in the faintest magnitude bin (see Table \ref{tab:app:colourcuts} and Sect.\thinspace\ref{se:photo_color_select_acs}).

\begin{table*}
\caption{Overview of \mbox{$V-I$} colour cut limits applied in our analysis.
\label{tab:app:colourcuts}}
\begin{center}
\begin{tabular}{ccccc}
\hline
\hline
$z_\mathrm{l}$  & \multicolumn{2}{c}{\mbox{$V_{606}-I_{814}$}} & \multicolumn{2}{c}{\mbox{$(V_{606,\mathrm{con}}-I_\mathrm{FORS2})_\mathrm{fix}$}} \\
& \mbox{$24<V_{606}<26$}  & \mbox{$26<V_{606}<26.5$}  & bright & faint\\
\hline
$<1.01$ & 0.3 & 0.2 & 0.2 & 0.0\\
$>1.01$ & 0.2 & 0.1 & 0.1 & $-0.1$\\
\hline
\end{tabular}
\end{center}
{\flushleft
Note. --- Colour cut limits applied in our analysis.
{\it Column 1:} Cluster redshift range.
{\it Column 2:} Colour-cut in ACS-only colour \mbox{$V_{606}-I_{814}$} for galaxies with \mbox{$24<V_{606}<26$}.
{\it Column 3:} Colour-cut in ACS-only colour \mbox{$V_{606}-I_{814}$} for galaxies with \mbox{$26<V_{606}<26.5$}.
{\it Column 4:} Colour cut in the ACS+FORS2 colour
\mbox{$(V_{606,\mathrm{con}}-I_\mathrm{FORS2})_\mathrm{fix}$} after
tying it to the \mbox{$V_{606}-I_{814}$} colour (see
Appendix \ref{app:tiecolor}),
as employed for ``bright cut'' magnitude bins
with low photometric scatter \mbox{$\sigma_{\Delta(V-I)}<0.2$}.
{\it Column 5:} As column 4, but for the ``faint cut'' magnitude bins with
increased photometric scatter \mbox{$0.2<\sigma_{\Delta(V-I)}<0.3$}.\\
}
\end{table*}

\subsubsection{ACS+FORS2 colour selection}
\label{se:scatter_acsfors2}
The
colour estimates
\mbox{$(V_{606,\mathrm{con}}-I_\mathrm{FORS2})_\mathrm{fix}$} that include
the FORS2 data
are more strongly affected by photometric scatter than the \mbox{$V_{606} - I_{814}$}
colours obtained from the high-resolution ACS data only
(see Fig.\thinspace\ref{fig:scatter}).
To ensure that we can still apply a consistent colour selection to the
\citetalias{skelton14} catalogues
we do the following:

First, we limit the analysis to relatively bright $V_{606}$  magnitudes, to
ensure that the scatter is small enough to not compromise the exclusion of
galaxies at the cluster redshift considerably.
For this we compute the r.m.s. scatter $\sigma_{\Delta(V-I)}$ in the colour difference
\mbox{$\Delta(V-I)\equiv (V_{606,\mathrm{con}}-I_\mathrm{FORS2})_\mathrm{fix}-(V_{606} - I_{814})$} of blue
galaxies (\mbox{$V_{606} - I_{814}<0.6$}) in 0.5\thinspace mag-wide bins in $V_{606}$.
For the ACS+FORS2 colour selection we only include magnitudes bins
with
scatter \mbox{$\sigma_{\Delta(V-I)}<0.3$}.
Here we employ our standard (``bright'') colour cut for the magnitude bins
with low scatter \mbox{$\sigma_{\Delta(V-I)}<0.2$}, and a more
conservative (``faint'') colour cut for magnitude bins with slightly larger
scatter \mbox{$0.2<\sigma_{\Delta(V-I)}<0.3$}, see columns 5 and 6 in
Table \ref{tab:vltdata} for the corresponding magnitude bins in each
cluster
and Table \ref{tab:app:colourcuts} for the colour cuts as function of
cluster redshift.

Second, we
add noise to the \mbox{$V_{606} - I_{814}$} colour
estimates in the CANDELS catalogue prior to the colour cut, similarly to our
approach for the ACS-only selection.
However, in contrast to Appendix\thinspace\ref{sec:noise:acsonly} we do not
assume a Gaussian noise distribution here, but
randomly sample the noise from the actual distribution of the colour
differences
\mbox{$(V_{606,\mathrm{con}}-I_\mathrm{FORS2})_\mathrm{fix}-(V_{606} -
  I_{814})$} shown in the right panels of Fig.\thinspace\ref{fig:scatter}.
 The motivation for not using a Gaussian approximation is given by the
 skewness in the distribution and presence of outliers.
In practice, we again divide the galaxies into \thinspace 0.5\thinspace mag-wide bins
in $V_{606}$.
We further subdivide these galaxies into  sub-bins according to their
\mbox{$V_{606} - I_{814}$}  colour
if sufficiently many galaxies are available to provide sub-bins containing
at least 30 galaxies each.
For each galaxy in the CANDELS catalogue we then identify the corresponding
bin/sub-bin and randomly assign a colour difference drawn from this
bin/sub-bin.
Note that we introduce the further colour subdivision as red galaxies (which are later removed by the colour cut)
show a lower scatter at a given $V_{606}$ magnitude\footnote{This is expected
  since \mbox{$(V_{606,\mathrm{con}}-I_\mathrm{FORS2})$} receives
 roughly comparable scatter contributions from $V_{606,\mathrm{con}}$ and
  $I_\mathrm{FORS2}$, with a reduced scatter in  $I_\mathrm{FORS2}$
  for red objects.}.

\section{Limitations of a statistical correction for cluster member contamination}
\label{app:why_not_boost}

Weak lensing studies that use wide-field imaging data and do not have
sufficient colour information for a robust removal of cluster galaxies can attempt to
statistically correct their shear profiles for the dilution effect of cluster members in the source
samples \citep[see e.g.][]{hoekstra15}.
For this, they need to estimate the relative excess counts as function of
cluster-centric distance,
ideally accounting for the impact of masks, obscuration by cluster members,
and magnification,
and fit it with a model, typically in the form
\begin{equation}
  n_\mathrm{measure}(r)=\frac{n_\mathrm{bg}}{1-f(r)}\,,
\end{equation}
and scale the shear profile as
\begin{equation}
  \langle g_\mathrm{t} \rangle^\mathrm{boosted}(r)=\langle g_\mathrm{t} \rangle(r) \frac{1}{1-f(r)}\,.
\end{equation}
Here we consider two previously employed
models for the projected density profiles of cluster galaxies, namely the projected singular isothermal sphere (SIS) model
\begin{equation}
f(r)=f_{500} \frac{r_\mathrm{500c}}{r}
\end{equation}
\citep[e.g.][]{hoekstra07}
and an exponential model
\begin{equation}
f(r)=f_{500} e^{1-r/r_\mathrm{500c}}
\end{equation}
\citep[e.g.][]{applegate14}, where $f_\mathrm{500}$ corresponds to the contamination
at $r_\mathrm{500c}$.

\begin{figure*}
\includegraphics[width=0.99\columnwidth]{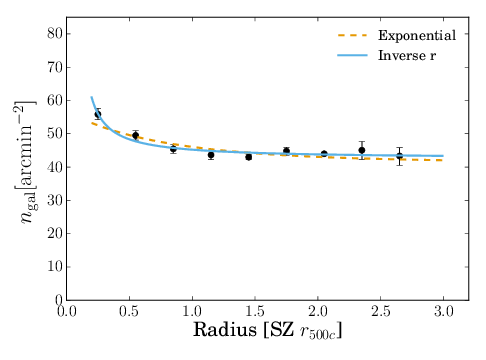}
 \includegraphics[width=0.99\columnwidth]{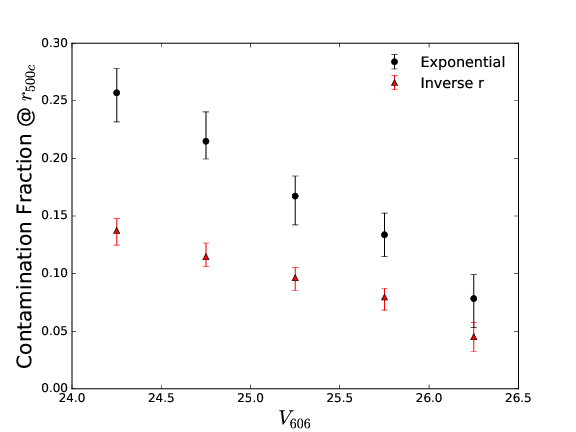}
 \caption{Cluster member contamination for the analysis centred around the X-ray centroid when no colour selection is
   applied. {\it Left:} Number density profile combining all magnitude bins,
   where the curves indicate the best fit exponential and $1/r$ (SIS) model
   for the cluster member contamination. {\it Right:} Comparison of the
   estimated contamination fraction $f_{500}$ for the two models as function
   of magnitude.
\label{fig:ngal_nocc}}
\end{figure*}

We do not use this approach for our HST analysis as the
\mbox{$2\times 2$} ACS mosaics are too small to derive a robust estimate of
the background source density directly.
To test this, we use our source catalogues without colour selection,
estimate the mask- and obscuration-corrected source density profiles in
magnitude bins, and
fit them with both $f(r)$ profiles.
Combining the analysis from all clusters we find that both profiles provide
\reva{acceptable} fits for most of the radial range covered by the ACS data.
For example, when using only a single broad magnitude bin, the SIS model
returns
\reva{\mbox{$\chi^2=6.0$}}
for 7 degrees of freedom, whereas the exponential model returns
\reva{\mbox{$\chi^2=14.3$}}.
 \reva{The SIS model is clearly a better fit at small radii (see the
   \textit{left} panel of
   Fig.\thinspace\ref{fig:ngal_nocc}), but the exponential profile is not ruled out at high significance.}
Yet, the two models yield
\reva{uncomfortably}
different contamination fractions
\reva{(shown in the \textit{right} panel of
  Fig.\thinspace\ref{fig:ngal_nocc} as function of $V_{606}$)}.
As a test for the impact of these differences we artificially apply the two
different boost correction schemes
\reva{(taking their
 magnitude dependencies into
account)}
to our colour-selected shear
profiles and compare the resulting mass estimates.
Here we find that the exponential model leads to mass estimates which are
higher compared to those from the isothermal model by
\reva{\mbox{$\sim 14\%$}}.
As it is currently not clear what the correct functional form would be,
we conclude that the application of such a
contamination correction would introduce substantial systematic uncertainty.

One could consider to reduce this uncertainty by using external blank fields
to constrain the background source density.
Using our colour-selected catalogues we have demonstrated that the careful
matching of source selection criteria and noise properties between the
cluster and reference fields, which would be required for such an approach,
is in principle possible.
However, instead of providing an important validation test as in our
study, the information in the number density would then be used to correct
the signal, assuming that all other related  analyses steps were done
correctly.
Large-scale structure variations also introduce significant variations in the  source densities
between the five CANDELS fields.
Without colour selection we find that they lead to uncertainties in the estimated mean background density of
\mbox{$\sim 6\%$} at \mbox{$V\sim 24$} to  \mbox{$\sim 3\%$} at \mbox{$V\sim
  26$}.

In addition to the increased systematic uncertainty,  the use
of a contamination correction also  increases the statistical uncertainty compared
to a robust colour selection that adequately removes cluster galaxies.
First, the cluster members dilute the small-scale signal, which is the
regime providing the highest signal-to-noise contribution for our analysis.
Second, source density profiles are typically too noisy to measure the
contamination for individual clusters.
On the other hand, if an average contamination model is applied, extra
scatter in the mass constraints is introduced.

\section{Impact of contamination by very blue
  cluster members}
\label{se:test_extremely_blue}
The tests presented in Sect.\thinspace\ref{se:photo_number_density_tests}
show no indication for a significant residual contamination by cluster
members.
However, our estimates from
Sect.\thinspace\ref{se:source_sel_scatter} suggest that, in the
presence of noise and averaged over our cluster sample, our ACS-only (ACS+FORS2)
 colour selection should leave a
residual contamination of \mbox{$\sim 1.9\%$} (\mbox{$\sim 1.1\%$})
of very blue field galaxies at the corresponding
cluster redshifts.
Whether or not this can introduce a residual excess contamination by cluster
members depends on the relative properties of the galaxy distributions in
the field and cluster environment.

Luminous Compact Blue Galaxies
\citep[LCBGs, e.g.][]{koo94} represent an extreme star-bursting population of galaxies
 with very blue colours and compact sizes.
Such galaxies were also identified in cluster environments \citep{koo97},
making them the most relevant
potential contaminant
 for our colour-selected weak lensing source sample.
\citet{crawfords11,crawfords14} and \citet{crawfords16} identify and study LCBGs in five massive clusters at
\mbox{$0.5<z<0.9$} using a photometric preselection, Keck/DEIMOS
spectroscopy, and HST morphological measurements.
 For the \mbox{$z>0.6$} clusters in their sample \citet{crawfords11} find that the
 number density enhancement of the cluster LCBG population compared to the LCBG field density is
 comparable to or lower than the corresponding  enhancement of the total
 cluster population compared to the total field population.
In addition, \citet{crawfords16}
find that the relevant properties of the cluster  LCBGs
(star-formation rate, dynamical
mass, size, luminosity, and metallicity)  are indistinguishable from the
properties of field
LCBGs  at the same redshift.

Accordingly, we can make the conservative assumption that the relative fraction of cluster members that pass our
colour selection is equal to or lower than the fraction of field galaxies passing
the selection \mbox{$f_\mathrm{pass,field}\sim 1.9\%\thinspace (1.1\%)$} for the
ACS-only (ACS+FORS2) selection,  accounting for noise (see Sect.\thinspace\ref{se:source_sel_scatter}).
We then estimate the approximately expected average fraction of cluster galaxies in our
colour-selected source sample at \mbox{$r_\mathrm{500c}$} as
\begin{eqnarray}
f_\mathrm{500,expected}&=& f_\mathrm{pass,field} f_\mathrm{500,no-cc}
\left(\frac{n_\mathrm{gal,cc}}{n_\mathrm{gal,no-cc}}\right)^{-1} \\
\nonumber
&=&0.009\thinspace (0.008) \,,
\end{eqnarray}
where
\reva{\mbox{$f_\mathrm{500,no-cc}\simeq 0.15$}}
indicates an estimate for the average contamination at
\mbox{$r_\mathrm{500c}$} based on a number density profile analysis
when the colour  selection is
 {\it not} applied (see
\reva{the right panel of Fig.\thinspace\ref{fig:ngal_nocc},
averaging the values for the more conservative exponential model
according to the relative weight of the
corresponding
magnitude bin
in the
reduced shear profile fits}),
and \mbox{$n_\mathrm{gal,cc}/n_\mathrm{gal,no-cc}=0.33$ $(0.22)$}
 corresponds to the fraction of galaxies in the cluster fields passing the colour selection within
 the magnitude range of the ACS-only (ACS+FORS2) analysis.
We do not attempt to model the radial distribution of the expected
contaminating cluster galaxies, as LCBGs appear to follow a rather
shell-like distribution with a depletion in the cluster core
\citep{crawfords06}.
Instead, we assume that $f_\mathrm{500,expected}$ provides a reasonable
approximation for the typical contamination, which is likely conservative
given that the average \mbox{$\langle r_\mathrm{500c} \rangle=770$ kpc} of
our cluster sample \citep[based on the lensing analysis and assuming the
\reva{concentration--mass} relation from][see Sect.\thinspace\ref{sec:wlmasses}]{diemer15}
is more representative for the inner ($500$ kpc) than the outer ($1.5$ Mpc)
limit of the fit range for our default analysis.
With these conservative assumptions, the relative bias
for the average lensing efficiency
caused by the expected
cluster contamination is
\mbox{$\Delta\langle\beta\rangle/\langle\beta\rangle=-f_\mathrm{500,expected}=-0.009
  (-0.008)$}.
Given that this is even smaller than the uncertainty on
\mbox{$\langle\beta\rangle$} from line-of-sight variations between the
CANDELS fields (Sect.\thinspace\ref{sec:beta_los_variation}),
this bias
could well be ignored \reva{(we still include it in the systematic error budget in Table \ref{tab:sys})}.
But we note that future studies could attempt to model the contamination
more accurately and apply a correction.

\section{Additional figures}
\label{se:additional_figs}

\begin{figure*}
 \includegraphics[width=6.cm]{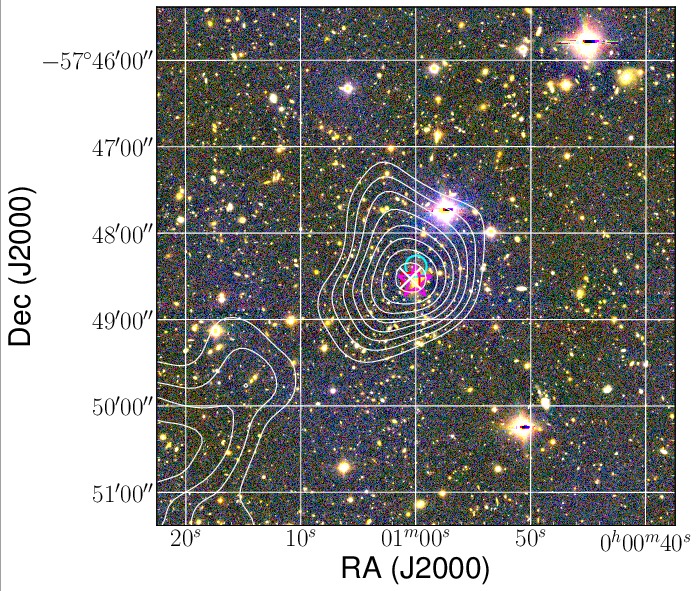}
 \includegraphics[width=6.cm]{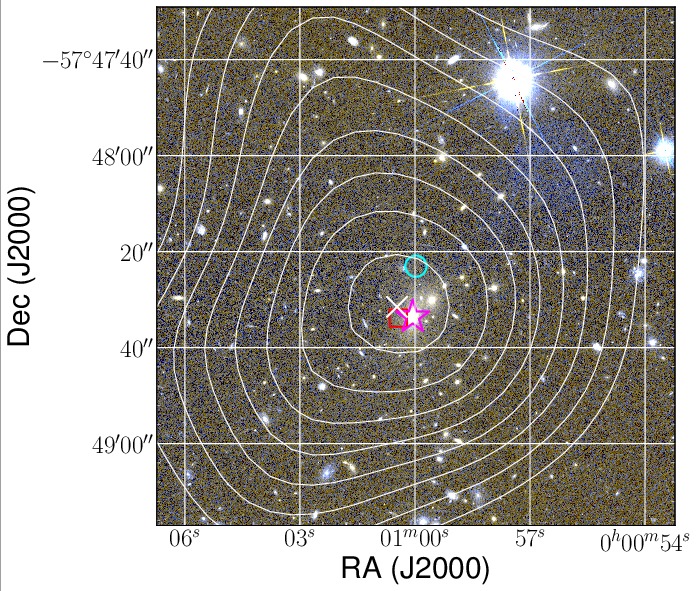}
\includegraphics[width=5.4cm]{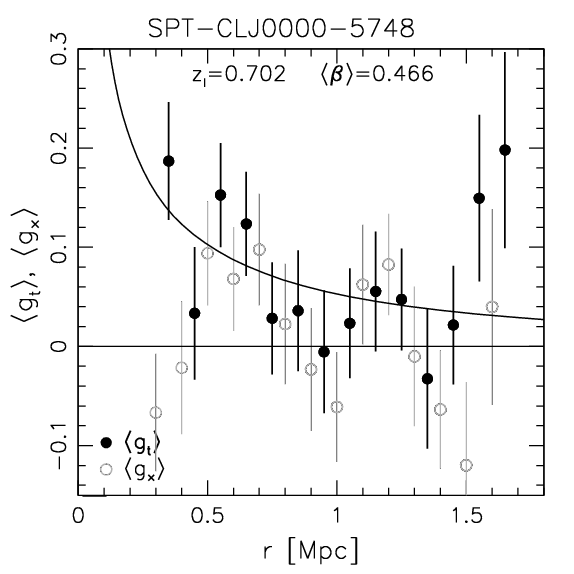}
\caption{
Weak lensing results for \clustera. See the descriptions in the captions of
Fig.\thinspace\ref{fig:massplotdemo_map} and Fig.\thinspace\ref{fig:massplotdemo_profile} for details.
\label{fig:massplota}}
\end{figure*}

\begin{figure*}
 \includegraphics[width=6.cm]{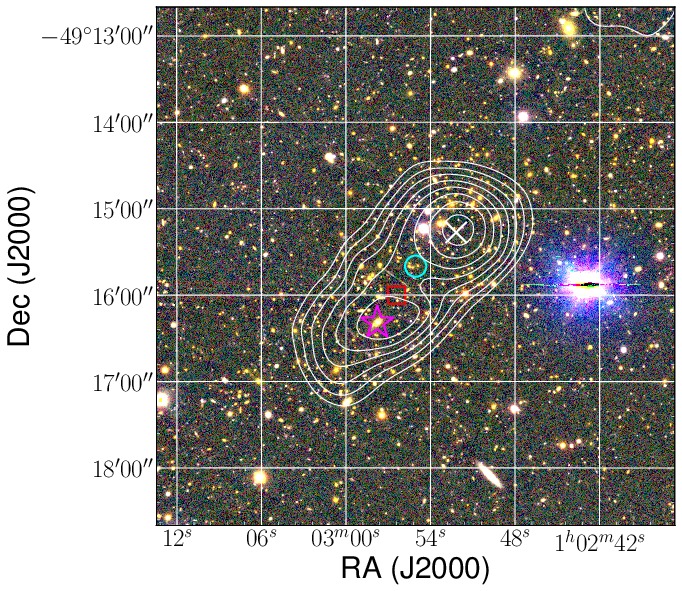}
 \includegraphics[width=6.cm]{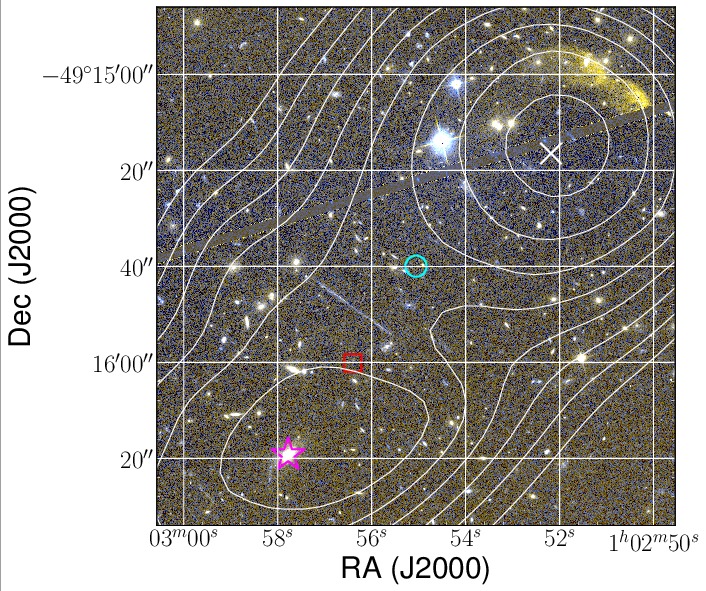}
\includegraphics[width=5.4cm]{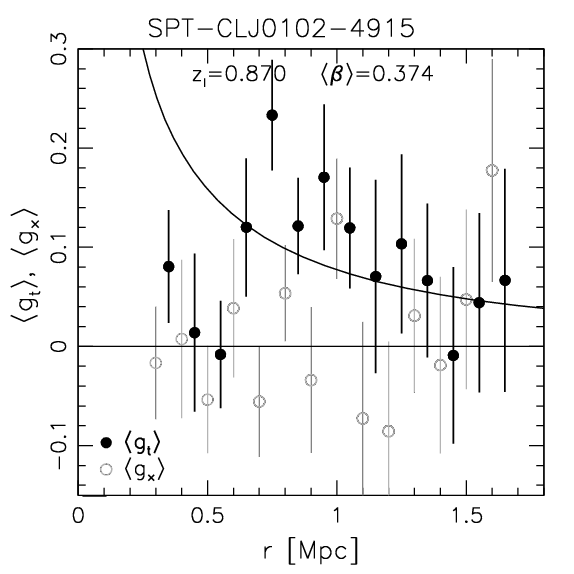}
\caption{Weak lensing results for \clusterb. See the descriptions in the captions of
Fig.\thinspace\ref{fig:massplotdemo_map} and Fig.\thinspace\ref{fig:massplotdemo_profile} for details.
\label{fig:massplotb}}
\end{figure*}

\begin{figure*}
 \includegraphics[width=6.cm]{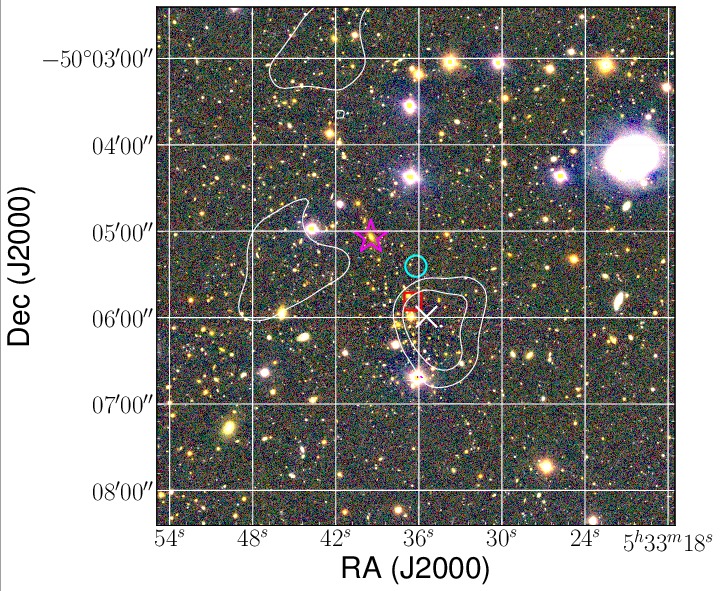}
 \includegraphics[width=6.cm]{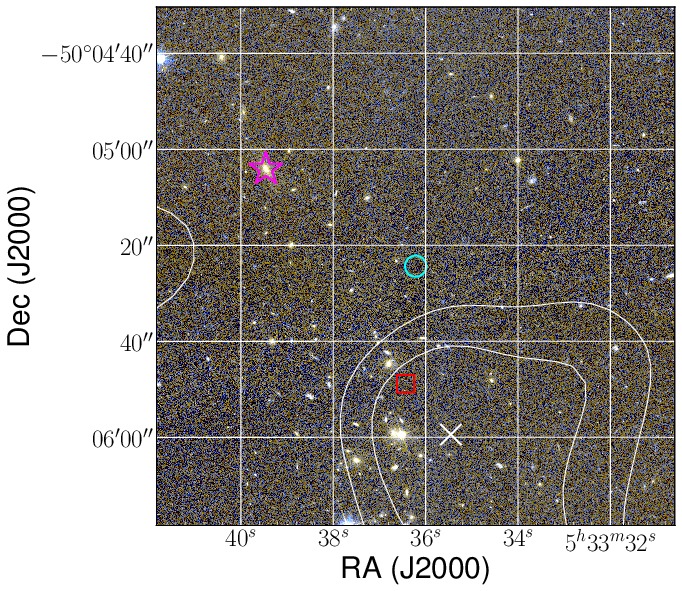}
\includegraphics[width=5.4cm]{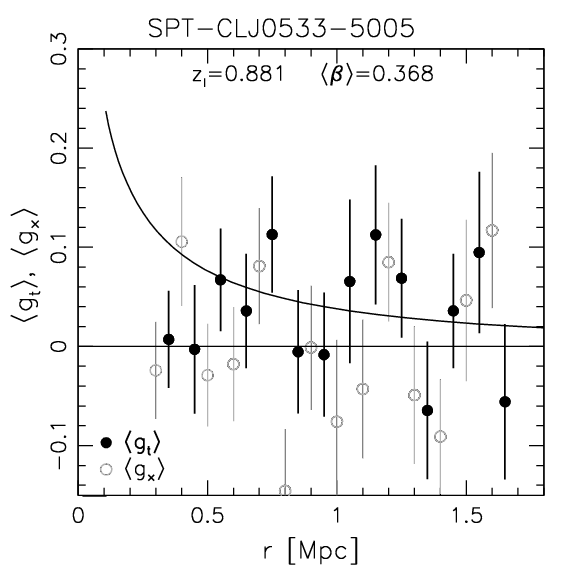}
\caption{Weak lensing results for \clusterc.  See the descriptions in the captions of
Fig.\thinspace\ref{fig:massplotdemo_map} and
Fig.\thinspace\ref{fig:massplotdemo_profile} for details.
\label{fig:massplotc}}
\end{figure*}

\begin{figure*}
 \includegraphics[width=6.cm]{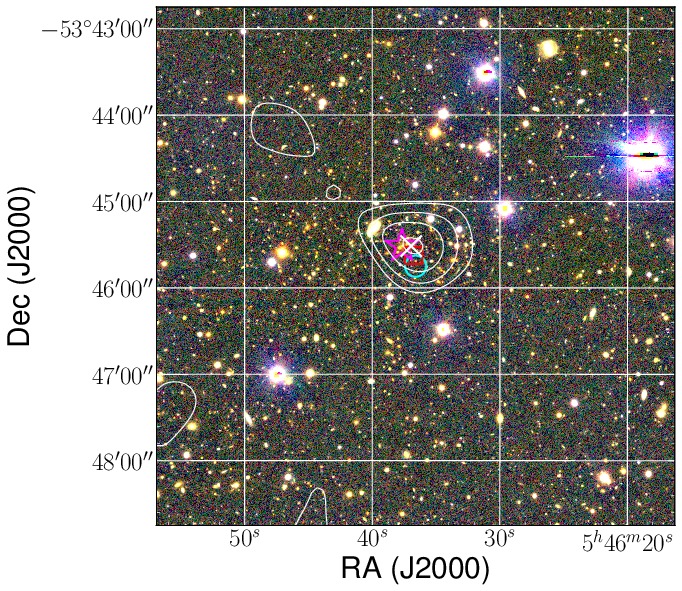}
 \includegraphics[width=6.cm]{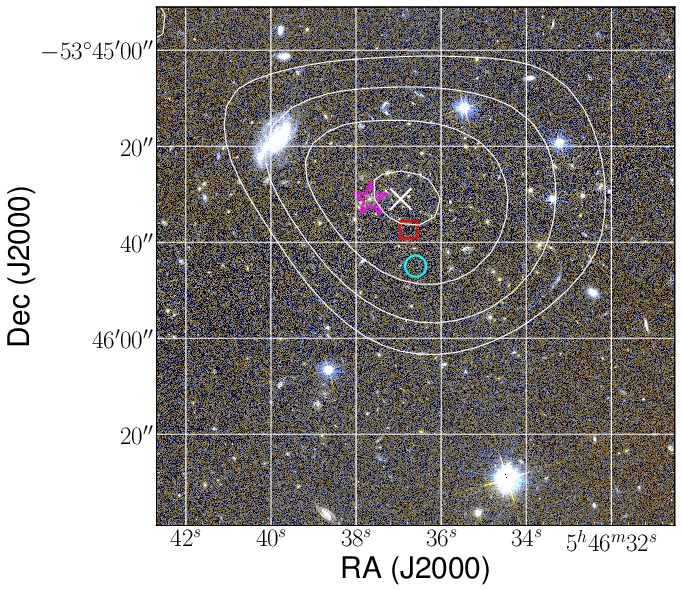}
\includegraphics[width=5.4cm]{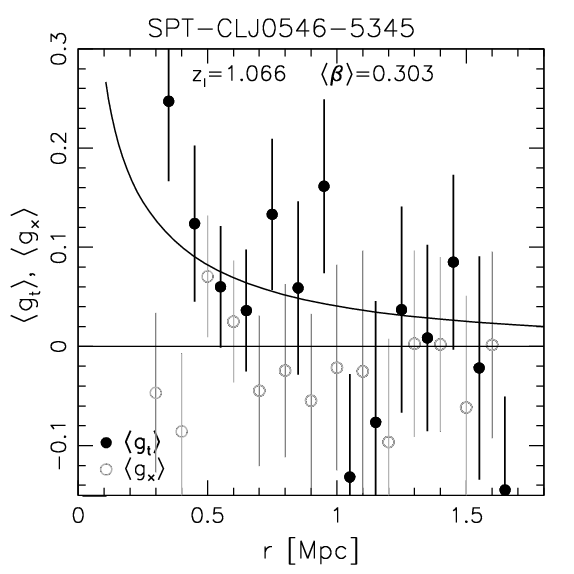}
\caption{
Weak lensing results for \clusterd.
 See the descriptions in the captions of
Fig.\thinspace\ref{fig:massplotdemo_map} and Fig.\thinspace\ref{fig:massplotdemo_profile} for details.
\label{fig:massplotd}}
\end{figure*}

\begin{figure*}
 \includegraphics[width=6.cm]{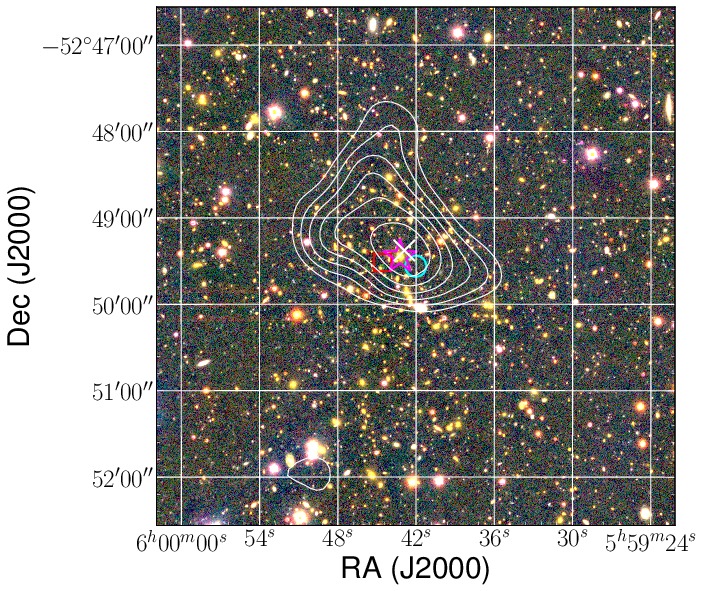}
 \includegraphics[width=6.cm]{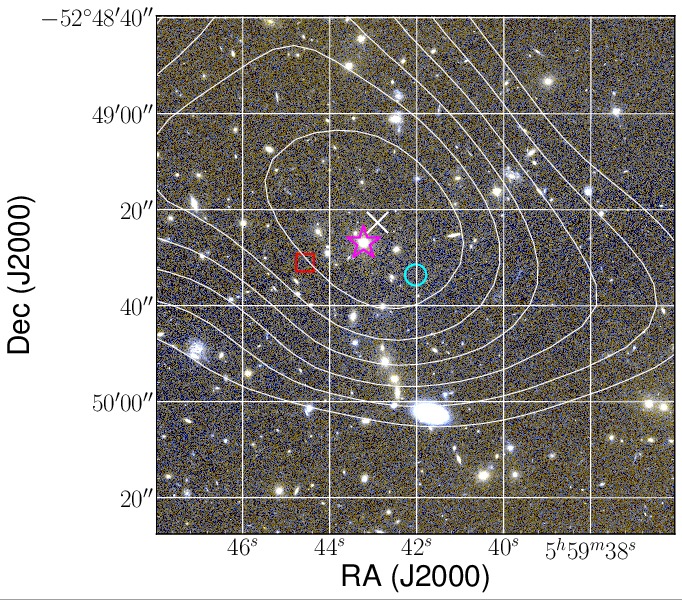}
\includegraphics[width=5.4cm]{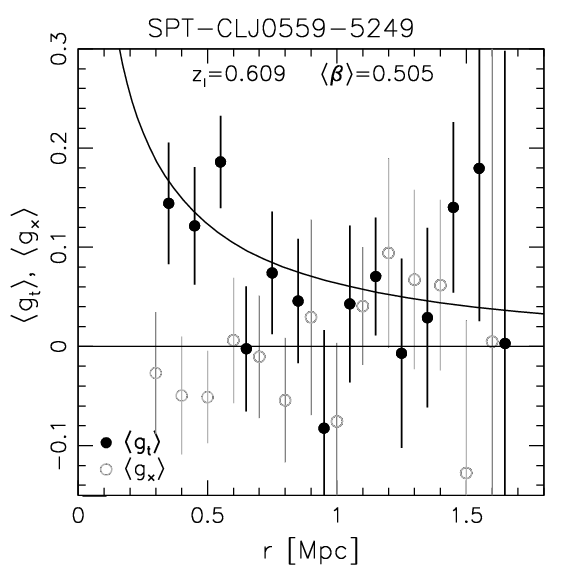}
\caption{Weak lensing results for \clustere.
 See the descriptions in the captions of
Fig.\thinspace\ref{fig:massplotdemo_map} and Fig.\thinspace\ref{fig:massplotdemo_profile} for details.
\label{fig:massplote}}
\end{figure*}

\begin{figure*}
 \includegraphics[width=6.cm]{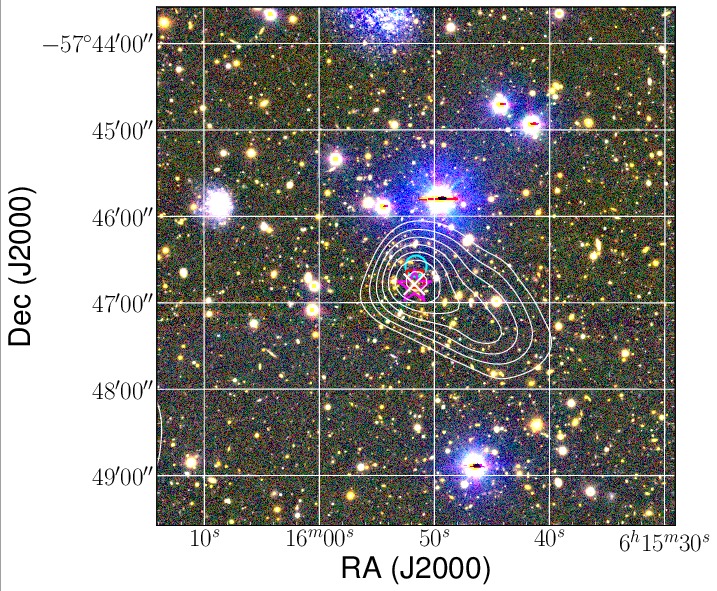}
 \includegraphics[width=6.cm]{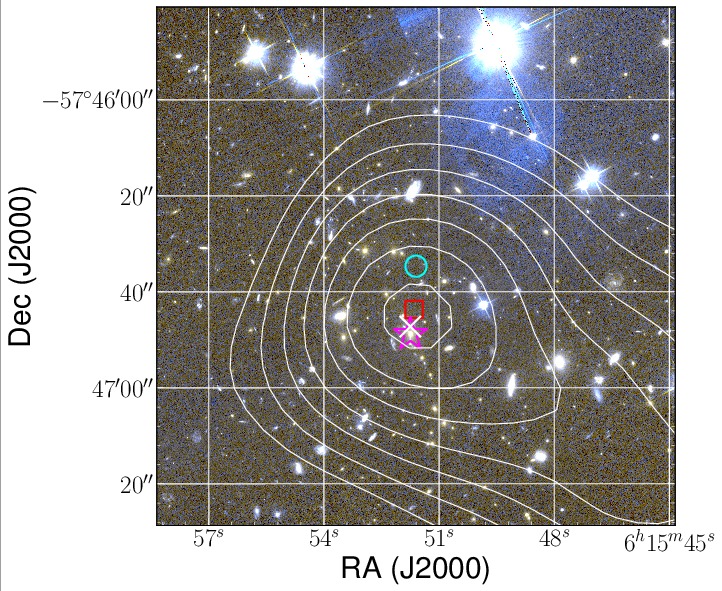}
\includegraphics[width=5.4cm]{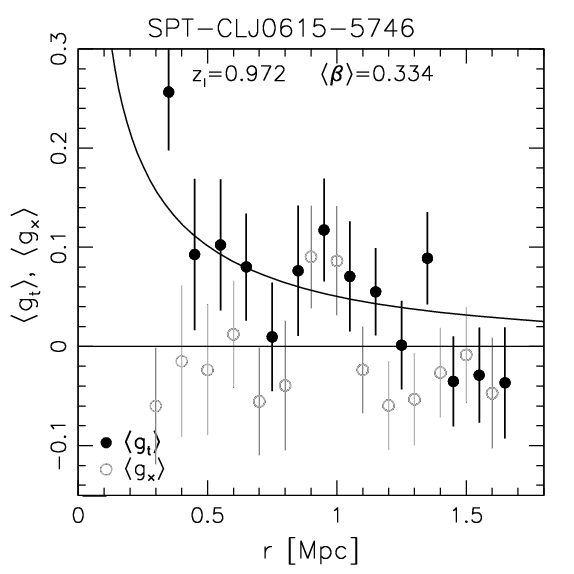}
\caption{
Weak lensing results for \clusterf.
 See the descriptions in the captions of
Fig.\thinspace\ref{fig:massplotdemo_map} and Fig.\thinspace\ref{fig:massplotdemo_profile} for details.
\label{fig:massplotf}}
\end{figure*}

\begin{figure*}
 \includegraphics[width=6.cm]{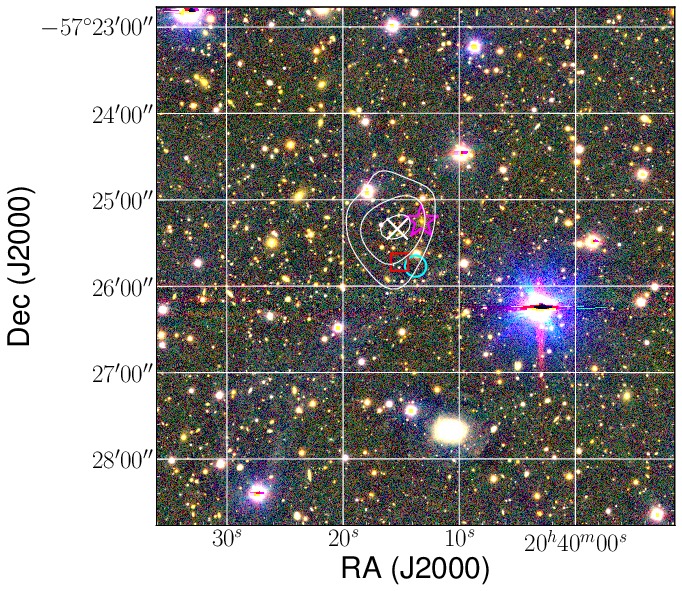}
 \includegraphics[width=6.cm]{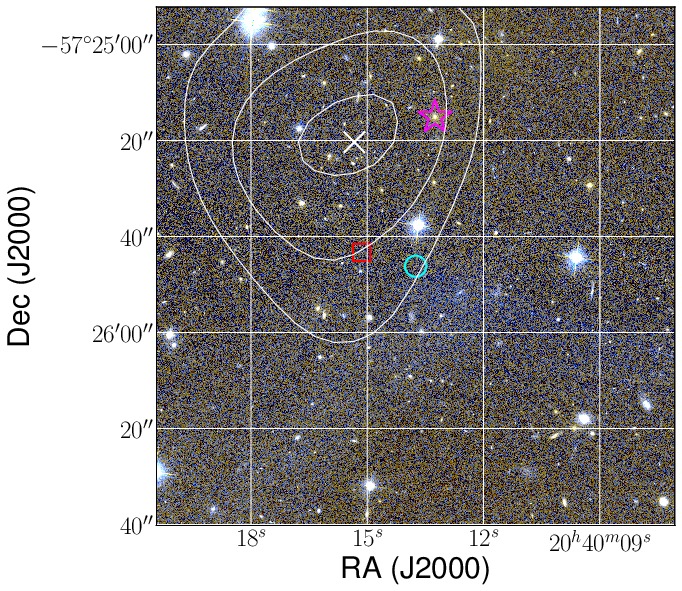}
\includegraphics[width=5.4cm]{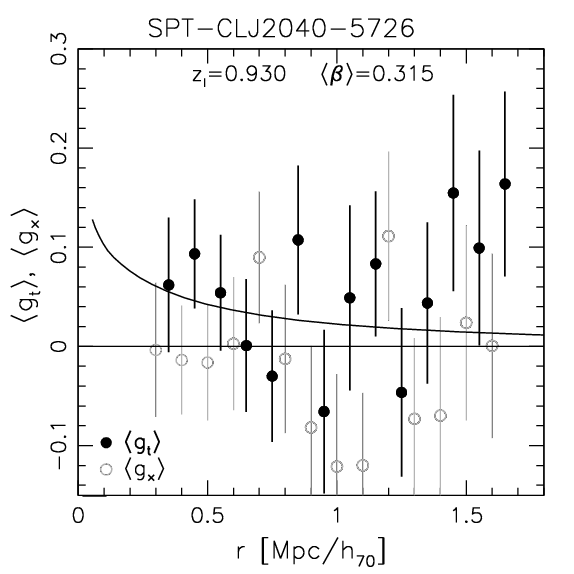}
\caption{Weak lensing results for \clusterg.
 See the descriptions in the captions of
Fig.\thinspace\ref{fig:massplotdemo_map} and Fig.\thinspace\ref{fig:massplotdemo_profile} for details.
\label{fig:massplotg}}
\end{figure*}

\begin{figure*}
 \includegraphics[width=6.cm]{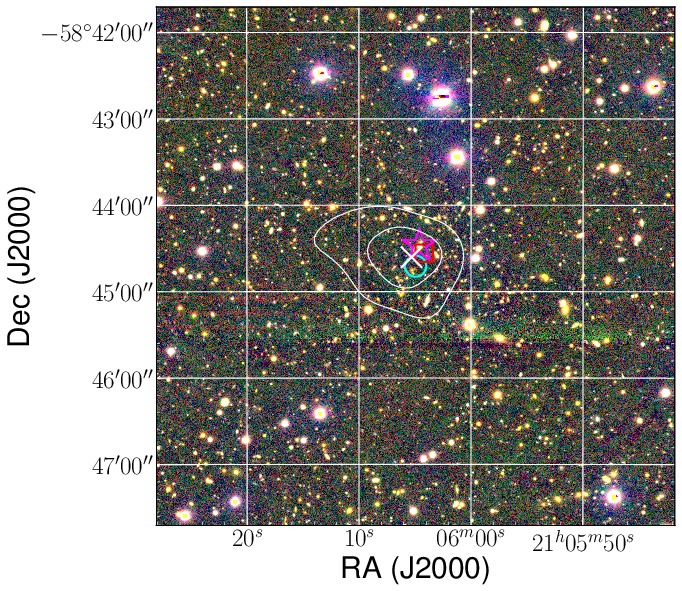}
 \includegraphics[width=6.cm]{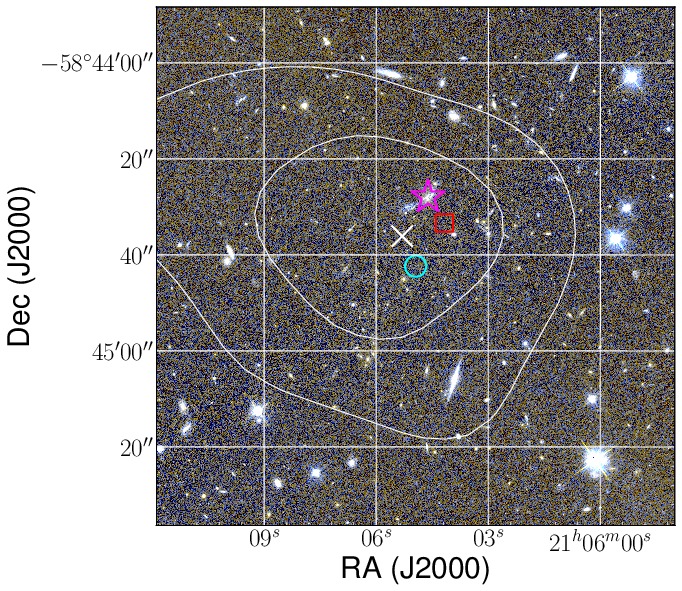}
\includegraphics[width=5.4cm]{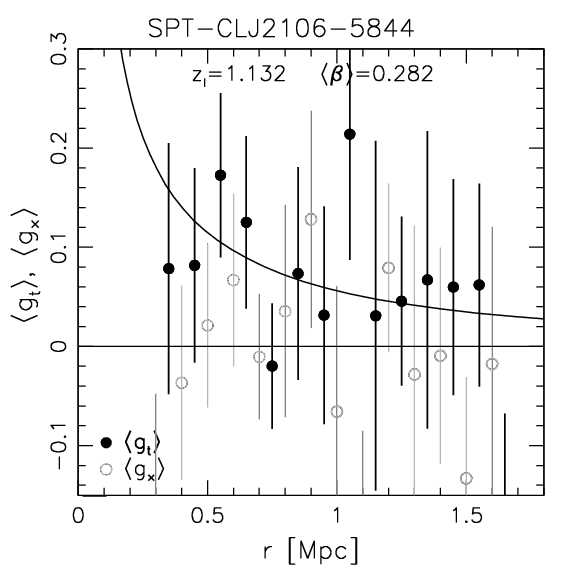}
\caption{
Weak lensing results for \clusterh.
 See the descriptions in the captions of
Fig.\thinspace\ref{fig:massplotdemo_map} and Fig.\thinspace\ref{fig:massplotdemo_profile} for details.
\label{fig:massploth}}
\end{figure*}

\begin{figure*}
 \includegraphics[width=6.cm]{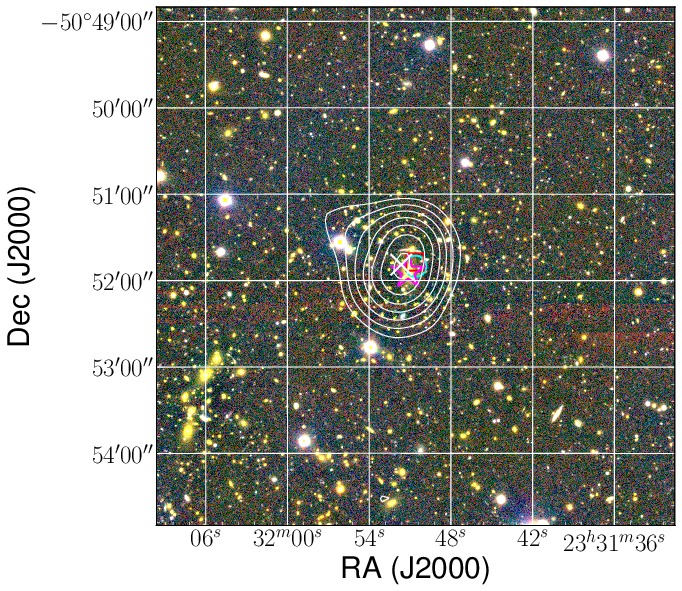}
 \includegraphics[width=6.cm]{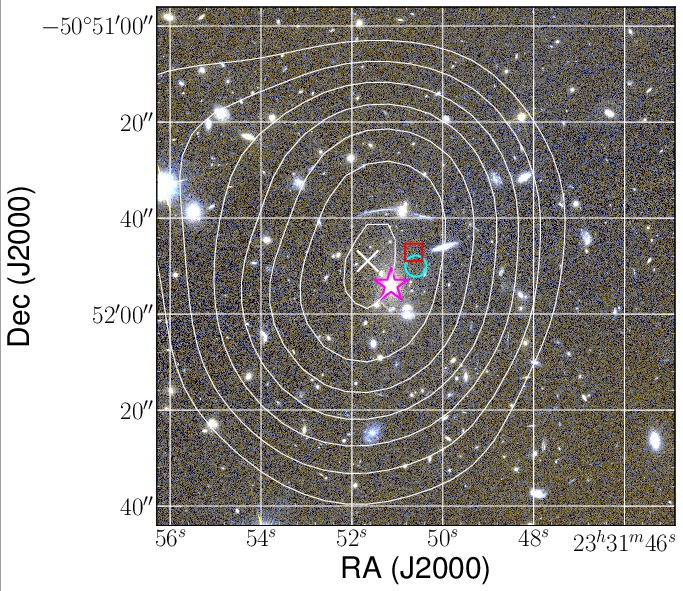}
\includegraphics[width=5.4cm]{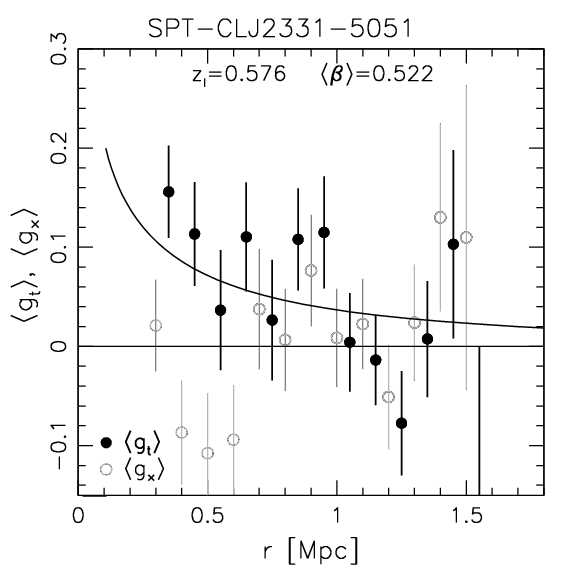}
\caption{Weak lensing results for \clusteri.
 See the descriptions in the captions of
Fig.\thinspace\ref{fig:massplotdemo_map} and Fig.\thinspace\ref{fig:massplotdemo_profile} for details.
\label{fig:massploti}}
\end{figure*}

\begin{figure*}
 \includegraphics[width=6.cm]{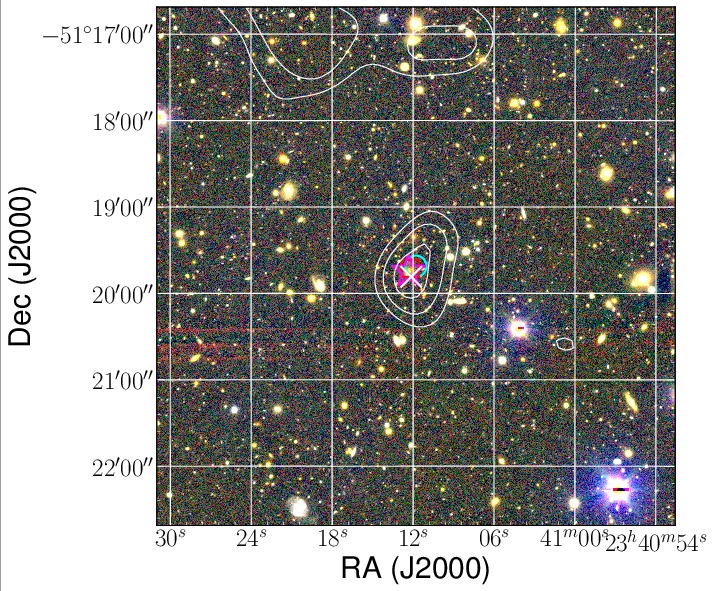}
 \includegraphics[width=6.cm]{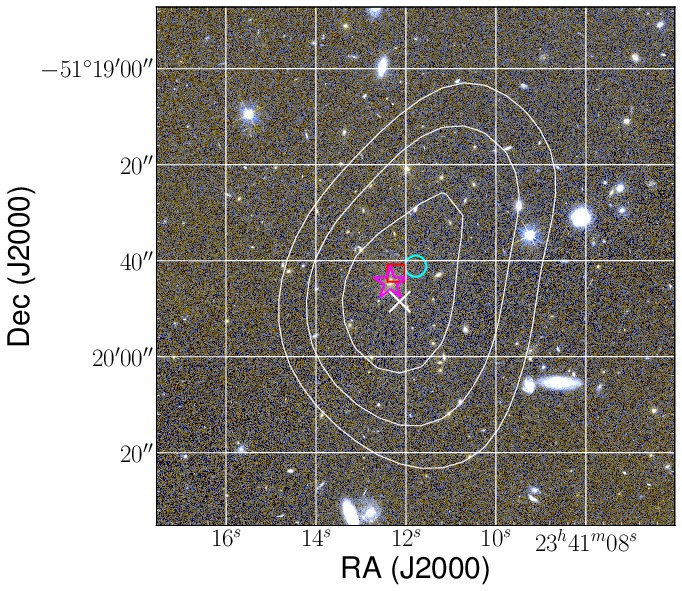}
\includegraphics[width=5.4cm]{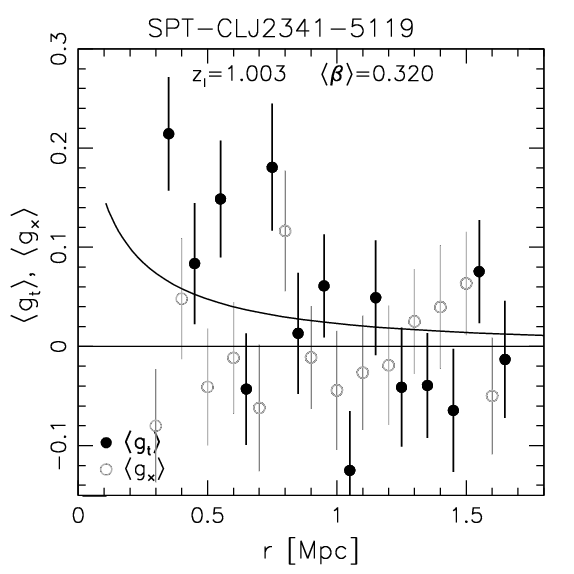}
\caption{Weak lensing results for \clusterk.
 See the descriptions in the captions of
Fig.\thinspace\ref{fig:massplotdemo_map} and Fig.\thinspace\ref{fig:massplotdemo_profile} for details.
\label{fig:massplotk}}
\end{figure*}

\begin{figure*}
 \includegraphics[width=6.cm]{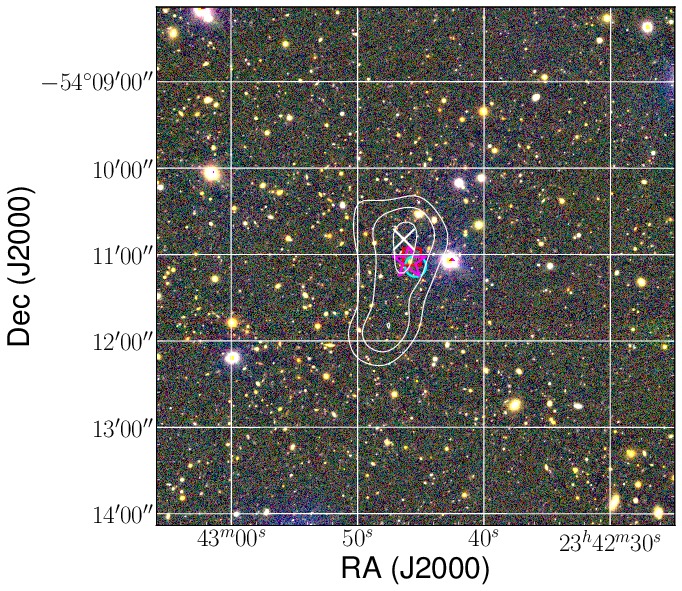}
 \includegraphics[width=6.cm]{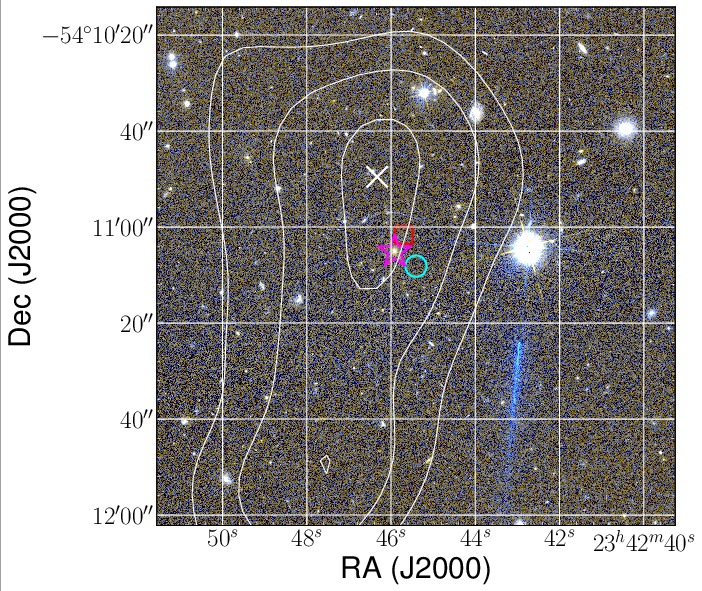}
\includegraphics[width=5.4cm]{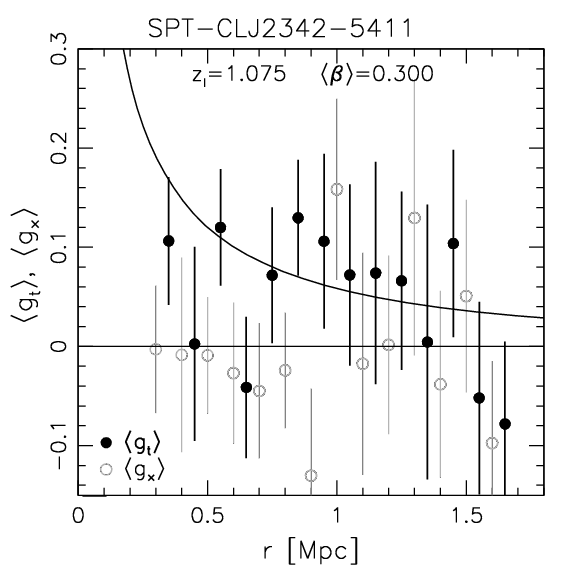}
\caption{
Weak lensing results for \clusterl.
 See the descriptions in the captions of
Fig.\thinspace\ref{fig:massplotdemo_map} and Fig.\thinspace\ref{fig:massplotdemo_profile} for details.
\label{fig:massplotl}}
\end{figure*}

\begin{figure*}
 \includegraphics[width=6.cm]{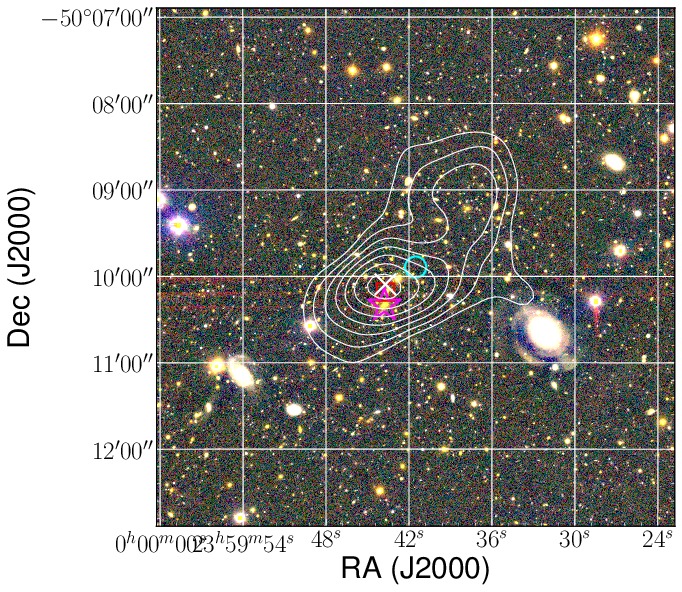}
 \includegraphics[width=6.cm]{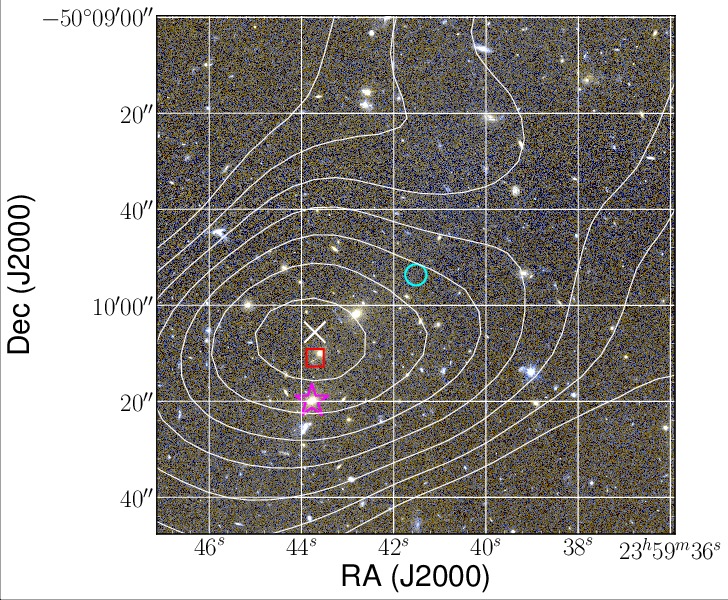}
\includegraphics[width=5.4cm]{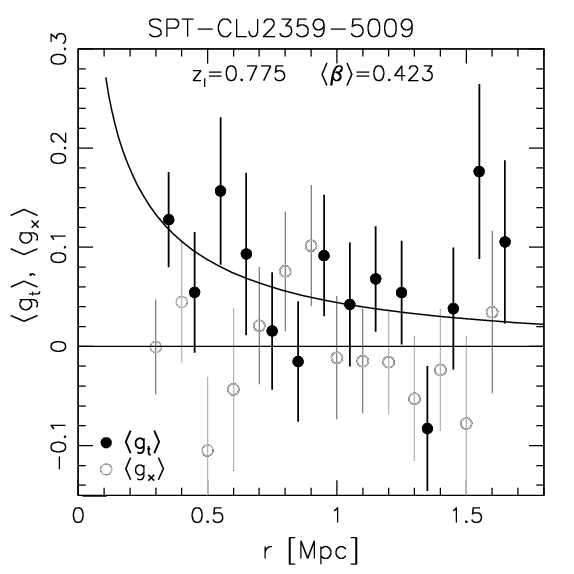}
\caption{Weak lensing results for \clusterm.
 See the descriptions in the captions of
Fig.\thinspace\ref{fig:massplotdemo_map} and Fig.\thinspace\ref{fig:massplotdemo_profile} for details.
\label{fig:massplotm}}
\end{figure*}

Figures \ref{fig:massplota} to \ref{fig:massplotm} complement
Fig.\thinspace\ref{fig:massplotdemo_map} and Fig.\thinspace\ref{fig:massplotdemo_profile}, showing the corresponding results for the other clusters.
In particular,  the
left
 and middle
panels
 show the weak lensing  \mbox{$S/N$} mass
reconstructions
overlaid onto
the corresponding VLT/FORS2 $BIz$
and central ACS
colour images, as well as the locations of the different cluster centres used in our analysis.
In the corresponding
 right
 panels we show the weak lensing
shear profiles centred onto the X-ray centroids.

\section*{Affiliations}
\altaffilmark{\Bonn} Argelander-Institut f\"{u}r Astronomie, Universit\"{a}t Bonn, Auf dem
H\"{u}gel 71, 53121, Bonn, Germany\\
\altaffilmark{\StanfordKIPAC}  Kavli Institute for Particle Astrophysics and
Cosmology, Stanford University, 382 Via Pueblo Mall, Stanford, CA
94305-4060, USA\\
\altaffilmark{\StanfordPhysics} Department of Physics, Stanford University, 382 Via Pueblo Mall, Stanford,
CA 94305-4060, USA\\
\altaffilmark{\KICPChicago} Kavli Institute for Cosmological Physics, University of Chicago, 5640 South Ellis Avenue, Chicago, IL 60637 \\
\altaffilmark{\Munich} Faculty of Physics, Ludwig-Maximilians University, Scheinerstr.\ 1, 81679 M\"{u}nchen, Germany\\
\altaffilmark{\ExcellenceCluster} Excellence Cluster Universe,
Boltzmannstr. 2, 85748 Garching, Germany\\
\altaffilmark{\Leiden} Leiden Observatory, Leiden University, Niels Bohrweg
2, NL-2300 CA Leiden, The Netherlands\\
\altaffilmark{\ANL} Argonne National Laboratory, 9700 S. Cass Avenue,
Argonne, IL, USA 60439\\
\altaffilmark{\UFlorida} Department of Astronomy, University of Florida,
Gainesville, FL 3261\\
\altaffilmark{\DARK}  Dark Cosmology Centre, Niels Bohr Institute,
University of Copenhagen, Juliane Maries Vej 30, DK-2100 Copenhagen,
Denmark\\
\altaffilmark{\StonyBrook} Department of Physics and Astronomy, Stony Brook University, Stony Brook, NY 11794, USA\\
\altaffilmark{\MIT} MIT Kavli Institute for Astrophysics and Space
Research, Massachusetts Institute of Technology, 77 Massachusetts
Avenue, Cambridge, MA 02139 \\
\altaffilmark{\Washington} Department of Astronomy, University of Washington, Box 351580, Seattle, WA 98195, USA\\
\altaffilmark{\SLAC} SLAC National Accelerator Laboratory, 2575 Sand Hill Road, Menlo Park, CA 94025, USA\\
\altaffilmark{\Harvard} Department of Physics, Harvard University, 17 Oxford Street, Cambridge, MA 02138\\
\altaffilmark{\CfA} Harvard-Smithsonian Center for Astrophysics,
60 Garden Street, Cambridge, MA 02138 \\
\altaffilmark{\Colby}{Department of Physics \& Astronomy, Colby College,
  5800 Mayflower Hill, Waterville, Maine 04901}\\
\altaffilmark{\FNAL}{Fermi National Accelerator Laboratory, Batavia, IL 60510-0500, USA}\\
\altaffilmark{\AAUChicago} Department of Astronomy and Astrophysics, University of Chicago, 5640 South Ellis Avenue, Chicago, IL 60637 \\
\altaffilmark{\PhysicsUChicago} Department of Physics, University of Chicago,
  5640 South Ellis Avenue, Chicago, IL 60637 \\
\altaffilmark{\ASIAA} Academia Sinica Institute of Astronomy and Astrophysics (ASIAA)
11F of AS/NTU Astronomy-Mathematics Building,
No.1, Sec. 4, Roosevelt Rd, Taipei 10617, Taiwan\\
\altaffilmark{\Hyderabad} Department of Physics, IIT Hyderabad, Kandi, Telangana 502285, India\\
\altaffilmark{\UCStCruz} Department of Astronomy and Astrophysics, University of California, Santa Cruz, CA 95064, USA\\
\altaffilmark{\McGill} Department of Physics, McGill University, 3600 Rue
University, Montreal, Quebec H3A 2T8, Canada\\
\altaffilmark{\Berkeley} Department of Physics, University of California,
Berkeley, CA 94720\\
\altaffilmark{\Durham} Institute for Computational Cosmology, Durham University, South Road, Durham DH1 3LE, UK\\
\altaffilmark{\MPE} Max Planck Institute for Extraterrestrial Physics,
Giessenbachstrasse 1, 85748 Garching, Germany\\
\altaffilmark{\Melbourne} School of Physics, University of Melbourne, Parkville, VIC
3010, Australia\\
\altaffilmark{\CTIO} Cerro Tololo Inter-American Observatory, Casilla 603, La Serena, Chile\\

\end{document}